%% file: 2d-duality.tex
\documentclass{article}
\pdfoutput=1

\usepackage{macros}

\preprint{ }
\title{2d dualities from 4d}

\author{Jiaqun Jiang,}
\author{Satoshi Nawata,}
\author{Jiahao Zheng}
\affiliation{Department of Physics and Center for Field Theory \& Particle Physics, Fudan University, 20005, Songhu Road, 200438 Shanghai, China}
\emailAdd{jiangjiaqun@gmail.com}
\emailAdd{snawata@gmail.com}
\emailAdd{azjh1997@gmail.com}
\abstract{We find new $\cN=(2,2)$ and $\cN=(0,2)$ dualities through the twisted compactifications of 4d supersymmetric theories on $S^2$. Our findings include dualities for both $\cN=(2,2)$ and $\cN=(0,2)$ non-Abelian gauge theories, as well as $\cN=(0,2)$ Gauge/Landau-Ginzburg duality. }

\begin{document}
\Yboxdim5pt
\allowdisplaybreaks
\maketitle

\section{Introduction}

Two-dimensional (2d) supersymmetric theories serve as simplified models that capture the essential features of broader quantum field theories. Additionally, many 2d supersymmetric theories exhibit exact solvability, allowing exact computation of physical observables such as correlation and partition functions. These exact results provide profound insights into the non-perturbative aspects of quantum field theory. They not only shed light on infrared physics but also reveal rich mathematical structures, including the geometry of the target space in non-linear sigma models and representations of infinite-dimensional symmetries.

The foundational groundwork for 2d $\mathcal{N}=(2,2)$ and $\mathcal{N}=(0,2)$ supersymmetric gauge theories was laid in \cite{Witten:1993yc}. Building on this foundation, the study of dualities in 2d supersymmetric theories has advanced significantly. In particular, $\mathcal{N}=(2,2)$ gauged linear sigma models have been instrumental in the study of Calabi-Yau sigma models and mirror symmetry. Most research has focused on Calabi-Yau manifolds in toric varieties, leading to the discovery of dualities and the understanding of phases in $\mathcal{N}=(2,2)$ Abelian gauge theories. Although progress on non-Abelian gauge theories has been limited, dualities for $\mathcal{N}=(2,2)$ non-Abelian gauge theories were proposed in \cite{Hori:2006dk, Hori:2011pd}. These dualities are scrutinized from the viewpoint of exact partition functions (see \cite{Benini:2016qnm} and references therein), and mathematics \cite{Rennemo:2016oiu}.

The $\cN=(0,2)$ triality proposed in \cite{Gadde:2013lxa} is the first crucial progress along this line of investigation in 2d $\cN=(0,2)$ gauge theories. Recently, a large class of dualities of $\cN=(0,2)$ quiver gauge theories has been proposed in \cite{Nawata:2023aoq} through the twisted compactification of Lagrangian class $\cS$ theories \cite{Gaiotto:2009we}. In this paper,  we further explore this direction, finding new 2d $\cN=(2,2)$ and $\cN=(0,2)$ dualities from 4d supersymmetric theories.

The structure of this paper is as follows. In Section \ref{sec:(2,2)}, we examine the twisted compactification of Lagrangian class $\mathcal{S}$ theories of type $A$, and generalize the construction to a wider family of 2d $\mathcal{N}=(2,2)$ quiver gauge theories. Through the computation of elliptic genera, we explicitly demonstrate that these theories are independent of the duality frame. In Section \ref{sec:(0,2)-dual}, we provide a detailed study of various 2d \(\mathcal{N}=(0,2)\) dualities, beginning with a discussion on 2d (0,2) Seiberg-like dualities for SU gauge groups (\S\ref{sec:SU-Seiberg}). We then extend 2d (0,2) Seiberg-like dualities to new trialities in \S\ref{sec:triality}. Building on these results, we will uncover various new (0,2) dualities. First, we derive dualities between SU and Sp gauge theories in \S\ref{sec:SU-Sp}. The next two subsections (\S\ref{sec:AS} and \ref{sec:Sym}) systematically examine dualities in SU gauge theories with different types of chiral matter, including anti-symmetric, and symmetric representations. In \S\ref{sec:adj}, we investigate SO and Sp gauge theories with adjoint chiral matter and their dualities to free chiral theories.  Finally, section \ref{sec:AS-Sym} investigates theories with both symmetric and anti-symmetric chiral multiplets, providing a duality to a Landau-Ginzburg model.
Various appendices are provided to supplement the main text.

In a sense, this paper serves as a sequel to \cite{Nawata:2023aoq}. Interested readers are encouraged to refer to \cite{Nawata:2023aoq} alongside this paper.

\section{\texorpdfstring{$\cN=(2,2)$}{N=(2,2)} dualities}\label{sec:(2,2)}
In this section, we consider the twisted compactification of Lagrangian class $\cS$ theories \cite{Gaiotto:2009we} on $S^2$, which leads to 2d $\cN=(2,2)$ quiver theories. Specifically, we focus on a particular topological twist referred to as ``flavored'' reduction in \cite{Gadde:2015wta}.

We will perform a topological twist of the holonomy $\U(1)_{S^2}$ of $S^2$ with a particular $\U(1)_\frakR$ $\frakR$-symmetry of 4d $\cN=2$ SCFT. 
Treating the theory as a 4d $\cN=1$ theory,  for the twisted compactification to be well-defined, the $\U(1)_\frakR$ charge $\frakr$ of a 4d $\cN=1$ chiral multiplet must be integral \cite{Closset:2013sxa}. Furthermore, if \emph{all} charges are non-negative integers, we can focus on the vanishing sector of the gauge magnetic flux on $S^2$ \cite{Gadde:2015wta}. In this case, a  4d $\mathcal{N}=1$ chiral multiplet with $\mathrm{U}(1)_{\mathfrak{R}}$-charge $\mathfrak{r}$ becomes $(1-\mathfrak{r})$ (0,2) chiral multiplets if $\mathfrak{r}<1$, or $(\mathfrak{r}-1)$ (0,2) Fermi multiplets if $\mathfrak{r}>1$. However, the 4d chiral multiplet with $\mathfrak{r}=1$ does not contribute to the 2d theory.

A 4d $\cN=2$ SCFT is endowed with $\SU(2)_R\times \U(1)_r$ $R$-symmetry. 
Here, we pick $\frakR=2R-f$ for the topological twist where $\U(1)_R\subset \SU(2)_R$ and $\U(1)_f$ is a flavor symmetry that distinguishes the half-hypermultiplets $(q,\tilde q)$. The symmetry analysis for supercharges and the fundamental fields is given in Tab.~\ref{tab:top-twist}. Since the 4d $\cN=2$ supercharges are uncharged under the $\U(1)_f$ flavor symmetry, the four supercharges $Q_{-}^1,Q_{+}^2,\tilde{Q}_{\dot+}^1,\tilde{Q}_{\dot-}^2$ are neutral under this twist. They have two opposite charges under the $\U(1)_{T^2}$ rotation group of $\bR^2$ (or $T^2$ in this paper) where the 2d theory lives, resulting in $\cN=(2,2)$ supersymmetry. 
 Note that $\U(1)_{-R+f/2}$ corresponds to the $\U(1)_V$ vector $R$-symmetry in 2d while $\U(1)_r$  is identified with the $U(1)_A$ axial $R$-symmetry in 2d.

The adjoint $\Phi$ in 4d $\cN=2$ vector multiplet has charge 0 under $\U(1)_{2R-f}$ so that it becomes a 2d (0,2) adjoint chiral multiplet. Consequently, 4d $\cN=2$ vector multiplet gives rise to 2d $\cN=(2,2)$ vector multiplet.
Under this twist, the half-hypermultiplets $(q,\tilde q)$ transform into 2d (0,2) chiral and Fermi multiplet, forming a (2,2) chiral multiplet. Here, the $\U(1)_f$ flavor symmetry plays an important role. In summary, under this twisted compactification, we obtain the following mapping
\bea \nonumber
\textrm{ 4d $\cN=2$ vector multiplet } &\rightsquigarrow \textrm{ 2d $\cN=(2,2)$ vector multiplet } ~,\cr 
\textrm{ 4d $\cN=2$ hypermultiplet } &\rightsquigarrow \textrm{ 2d $\cN=(2,2)$ chiral multiplet }~.
\eea

\begin{table}[ht]\centering
$$\begin{array}{c|ccccc|cc|cc}
 & \SU(2)_1 & \SU(2)_2 & \SU(2)_R & \U(1)_r  & \U(1)_f & \U(1)_{T^2} & \U(1)_{S^2} & \U(1)^{(0,2)} & \U(1)^{(2,2)}  \\
\hline \hline 
Q_{-}^1 & -\frac{1}{2} & 0 & \frac{1}{2}& 1&0& -1& -1& 0 & 0 \\
Q_{+}^1 & \frac{1}{2} & 0 & \frac{1}{2}& 1&0& 1& 1& 2 & 2 \\
Q_{-}^2 & -\frac{1}{2} & 0 & -\frac{1}{2}& 1&0& -1& -1& -1 & -2  \\
Q_{+}^2 & \frac{1}{2} & 0 & -\frac{1}{2}& 1&0& 1& 1& 1 & 0  \\
\tilde{Q}_{\dot-}^1 & 0 & -\frac{1}{2} & \frac{1}{2}& -1&0& -1& 1& 1 & 2 \\
\tilde{Q}_{\dot+}^1 & 0 & \frac{1}{2} & \frac{1}{2}& -1&0& 1& -1& -1 & 0 \\
\tilde{Q}_{\dot-}^2 & 0 & -\frac{1}{2} & -\frac{1}{2}& -1&0& -1& 1& 0 & 0 \\
\tilde{Q}_{\dot+}^2 & 0 & \frac{1}{2} & -\frac{1}{2}& -1&0& 1& -1& -2 & -2 \\
\hline 
q  & 0  & 0 & \frac{1}{2} &0&1&0&0&0&0\\ 
\tilde q  & 0  & 0 &\frac{1}{2}&0&-1&0&0&1&2\\ 
\Phi  & 0  & 0 &0&2&0&0&0&1&0\\ 
\end{array}$$
\caption{Symmetries of 4d $\mathcal{N}=2$ supercharges and fields. The 4d $\mathcal{N}=1$ chiral fields $(q, \tilde{q})$ form an $\mathcal{N}=2$ hypermultiplet, while $\Phi$ represents the $\mathcal{N}=1$ adjoint chiral in an $\mathcal{N}=2$ vector multiplet. The fifth column indicates the $\U(1)_f$ flavor symmetry that differentiates between $q$ and $\tilde{q}$. The topological twist of $\U(1)_{S^2}$ with $\U(1)_{R + \frac{1}{2}(r - f)}$ gives rise to $\mathcal{N}=(0,2)$ supersymmetry, whereas twisting with $\U(1)_{2R-f}$ results in $\mathcal{N}=(2,2)$ supersymmetry.}
\label{tab:top-twist}
\end{table}

\begin{figure}[htb]
\centering
\includegraphics[width=4cm]{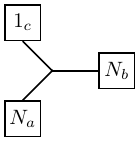}
\caption{Basic building block $U_N^{(2,2)}$ for SU($N$) theory}
\label{fig:SUN-trinion}
\end{figure}

In class $\cS$ theories of type $A_{N-1}$, punctures are labeled by partitions of $N$, and these theories generally lack a Lagrangian description. To perform the (2,2) reduction, we focus on class $\cS$ theories with Lagrangian descriptions. The fundamental building block, in this case, is a sphere with two maximal punctures and one minimal puncture, corresponding to $N^2$ hypermultiplets with a flavor symmetry group $\U(N^2)$. This group contains a subgroup $\SU(N)_a \times \SU(N)_b \times \U(1)_c$. 
In the (2,2) reduction, $\U(1)_c$ plays the role of the $\U(1)_f$ introduced earlier, used in the topological twist. Under this twisted compactification, it reduces to $N^2$ $\cN=(2,2)$ chiral multiplets. This reduced (2,2) theory, denoted $U_N^{(2,2)}$, serves as the basic building block for the $\cN=(2,2)$ theories we consider, and its quiver is depicted in Fig.~\ref{fig:SUN-trinion}.

Using elliptic genera, we will study $\cN=(2,2)$ quiver gauge theories constructed from the building block $U_N^{(2,2)}$. The minimal review of the elliptic genera is given in Appendix \ref{sec:fundamental}.
The contribution to the elliptic genus from the basic building block $U_N^{(2,2)}$, which consists of $N^2$ chiral multiplets, is 
\begin{align}
   \cI_{U_N}^{(2,2)} (\boldsymbol{a},\boldsymbol{b},c)=\prod_{i,j=1}^{N}\frac{\vartheta_1(ya_{i} b_{j} c)}{\vartheta_1(a_{i} b_{j} c)}~,
\end{align}
where we impose the condition $\prod_{i=1}^N a_i=1=\prod_{i=1}^N b_i$ for the $\SU(N)$ fugacities.

We construct $\cN=(2,2)$ quiver gauge theories by gauging the flavor symmetries of the building blocks $U_N^{(2,2)}$. To gauge the $\SU(N)$ flavor symmetries, we include the contribution from the $\SU(N)$ vector multiplet in the Jeffrey-Kirwan integral of the elliptic genus
\begin{align}
   \cI_{\SU(N)}^{(2,2),\mathrm{V}}(\boldsymbol{\zeta})=
\frac{1}{N!}\left(\frac{\eta(q)^3}{\vartheta_1(y)}\right)^{N-1}\prod_{\substack{i,j=1 
\\ i\ne j }}^N\frac{\vartheta_1(\zeta_{i}/\zeta_{j})}{\vartheta_1(y\zeta_{i}/\zeta_{j})}~,
\end{align}
where the condition $\prod_{i=1}^N \zeta_i$ is imposed on the gauge fugacities.
We do not introduce a superpotential in a quiver theory.

As in the case of $\mathcal{N}=(0,4)$ theories \cite{Nawata:2023aoq}, the $\mathcal{N}=(2,2)$ elliptic genera for quiver types of genus greater than zero also turn out to be remarkably simple. First, consider the theory of genus one with one minimal puncture, which consists of the free chiral plus the $\SU(N)$ gauge theory with adjoint chiral. The $\SU(N)$ gauge theory with adjoint chiral is indeed the $\cN=(4,4)$ vector multiplet, where the potential for the scalar fields is
\be 
V= \frac{1}{2 e^2} \Tr[\sigma, \sigma^{\dagger}]^2+\frac{e^2}{2} \Tr[\phi, \phi^{\dagger}]^2 + \frac{1}{2}\Tr[\phi^{\dagger}, \sigma^{\dagger}][\sigma, \phi]+ \frac{1}{2}\Tr[\phi^{\dagger}, \sigma][\sigma^{\dagger}, \phi]~.
\ee 
Here, $\phi$ is the lowest component of the $\cN=(2,2)$  adjoint chiral while $\sigma$ is the scalar in the $\cN=(2,2)$ vector multiplet. Therefore, the moduli space is spanned by the mutually commuting eigenvalues of $\phi$ and $\sigma$ up to the action of the Weyl group, which leaves the maximal torus $\U(1)^{N-1}$ of the gauge group unbroken. Consequently, the vacuum moduli space of the theory is 
\be 
\bC\times \frac{\frakt_\bC\times \frakt_\bC}{S_N}
\ee 
where $\frakt_\bC$ is the Cartan subalgebra of $\SL(N,\bC)$. The IR CFT is the orbifold CFT with this target, and therefore the central charge is
\be \label{cc-genus1-1}
c_L= c_R=3(N^2-(N^2-1))+6(N-1)= 3+6(N-1)~.
\ee 
The first term accounts for the contributions from $N^2$ chiral multiplets and $N^2 - 1$ gauginos, while the second term arises from the unbroken gauge group $\U(1)^{N-1}$.
The elliptic genus of the theory is then expressed as\footnote{Disclaimer: In this paper, we present numerous conjectural identities for elliptic genera, such as \eqref{EG-genus1-puncture1}, expressed in terms of JK residue integrals. However, these identities are not rigorously proven. Instead, we verify them by explicitly computing JK residue integrals up to rank five and performing $q$-expansions up to $\cO(q^2)$. Establishing formal proof of these identities remains an intriguing open problem.}
\begin{align}\label{EG-genus1-puncture1}
\begin{split}
   \cI_{g=1,n=1}^{(2,2),N}&=
\oint_{\JK} \frac{d\boldsymbol{a} }{2\pi  i \boldsymbol{a}}\mathcal{I}_{U_N}^{(2,2)} (\boldsymbol{a},\boldsymbol{a}^{-1},c)\cI_{\SU(N)}^{(2,2),\mathrm{V}}(\boldsymbol{a})\cr 
&=\frac{\vartheta_1(y^Nc^N)\vartheta_1(y)}{\vartheta_1(c^N )\vartheta_1(y^N)}
=\frac{\vartheta_{1}(yc)}{\vartheta_{1}(c)}\cdot \frac{\vartheta_1(y^Nc^N)\vartheta_1(c )\vartheta_1(y)}{\vartheta_1(yc ) \vartheta_1(c^N )\vartheta_1(y^N)}~.
\end{split}
\end{align}
Similar to \cite[Eqn.~(3.37)]{Nawata:2023aoq}, the first term represents the contribution from a free hypermultiplet, while the second term corresponds to the elliptic genus of an \(\mathcal{N} = (4,4)\) vector multiplet. The chiral ring of the symmetric product \((\mathfrak{t}_\mathbb{C} \times \mathfrak{t}_\mathbb{C}) / S_N\) is generated by the operators \(\Tr (\phi^i \sigma^j)\) for \(i + j > 0\). 
The elliptic genus of the \(\mathcal{N} = (4,4)\) vector multiplet is expected to be expressed in terms of contributions from these chiral operators, along with Fermi fields that impose their relations. Remarkably, these contributions cancel out, resulting in the compact form of the elliptic genus given above. 
Currently, the authors do not have an explicit formulation of the relations among the generators or a detailed understanding of the cancellation mechanism. (For the SU(2) gauge group, such a  chiral ring relation is given in \cite[\S2.2.3]{Putrov:2015jpa}.)

\begin{figure}[ht]
\centering
\includegraphics[width=9cm]{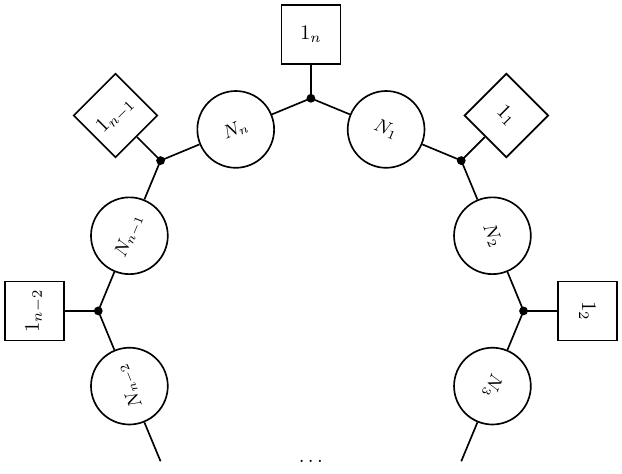}
\caption{Quiver theory of genus one with $n$ punctures. }
\label{fig:SUN-genus1}
\end{figure}

Let us consider the quiver theory of genus one with $n$ punctures as illustrated in Fig.~\ref{fig:SUN-genus1}. The theory can be understood as the (2,2) reduction of the corresponding Lagrangian class $\cS$ theory. At a generic point on the moduli space of chiral multiplets, the diagonal Cartan subgroup $\U(1)^{N-1}$ is unbroken at the infra-red. Therefore, the corresponding twisted chiral field can take the vacuum expectation value. Therefore, the complex dimension of the vacuum moduli space is $2(N-1)+n$, and the central charge of the IR CFT is
\be 
c_L=c_R=3n(N^2-(N^2-1))+6(N-1)=3n+6(N-1)~,
\ee 
where the first term accounts for the contributions from $n$ sets of $N^2$ chiral multiplets and $n$ sets of $N^2 - 1$ gauginos, while the second term arises from the unbroken gauge group $\U(1)^{N-1}$.
Performing the JK residue integral, we verify the equality of the elliptic genus indeed takes the form of a product of \eqref{EG-genus1-puncture1}:
\begin{align}\label{EG-genus1}
\cI_{g=1,n}^{(2,2),N}(c_1,\ldots,c_n) =\prod_{s=1}^n\frac{\vartheta_1(y^Nc^N_s)\vartheta_1(y)}{\vartheta_1(c_s^N)\vartheta_1(y^N)}
\end{align}
where $c_i$ are the U(1) flavor symmetries associated to the $n$ puncture. Thus, this indicates that the elliptic genera receive local contributions from minimal punctures for this class of $\cN=(2,2)$ theories. 

\begin{figure}[ht]
\centering
\includegraphics[width=13cm]{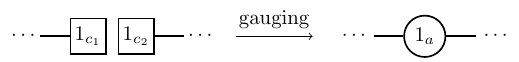}
\caption{U(1) gauging }
\label{fig:U1-gauging}
\end{figure}

\bigskip
Unlike class $\cS$ theories, we consider U(1) gauging in 2d $\cN=(2,2)$ theories when gluing the building blocks $U_N$. However, gauging a U(1) node results in a theory that is no longer a (2,2) reduction of a class $\cS$ theory. Nonetheless, as in Fig.~\ref{fig:U1-gauging}, we can gauge the anti-diagonal part of U(1) while preserving the diagonal U(1) as a global symmetry. 

\noindent At the level of the elliptic genus, the U(1) gauging procedure is given by
\begin{equation}
\cI_{\cT_1}(c_1,\ldots) \cI_{\cT_2}(c_2,\ldots)  \quad \rightarrow \quad \frac{\eta(q)^3}{\vartheta_1(y)} \oint_{\mathrm{JK}} \frac{d a}{2 \pi i a} \cI_{\cT_1}( ad,\ldots )\cI_{\cT_2}(a^{-1}d,\ldots   ) 
\end{equation}
where $d$ is the diagonal U(1) flavor fugacity.

\begin{figure}[ht]
\centering
{\raisebox{.7cm}{\includegraphics[width=7cm]{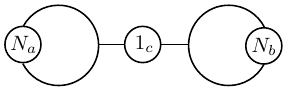}}} \qquad \includegraphics[width=3.8cm]{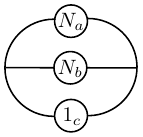}
\caption{Genus two theories. }
\label{fig:SUN-genus2}
\end{figure}

Let us consider the quiver theories of genus two, which involves the U(1) gauging. There are two types of quivers, as shown in Fig.~\ref{fig:SUN-genus2}. In the theory of genus one with one puncture, the Cartan subgroup $\U(1)^{N-1}$ is unbroken. For the genus-two theory, the Cartan subgroup $\U(1)^{2(N-1)}$ of the gauge group $\SU(N)\times\SU(N)$ is unbroken. Consequently, the corresponding twisted chiral fields can take the expectation value, and the central charge of the genus-two theory is
\be 
c_L= c_R =3 (2N^2-2(N^2-1)-1)+ 12(N-1) =3+ 12(N-1)~.
\ee 
Here, the first term accounts for the contributions from two sets of $N^2$ chiral multiplets and two sets of $N^2 - 1$ gauginos, along with an additional single U(1) gaugino. The second term originates from the unbroken gauge group $\U(1)^{2(N-1)}$.
Since the two quiver descriptions have different Lagrangians, they have different expressions by JK residue integrals for the elliptic genera
\bea
\cI^{(2,2),N}_{\includegraphics[width=1cm]{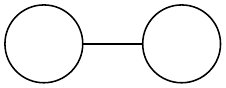}}=&\frac{\eta(q)^3}{\vartheta_1(y)}\int_{\JK}\frac{d\boldsymbol{a}
}{2\pi i\boldsymbol{a}}\frac{d\boldsymbol{b}}{2\pi i\textbf{b}}\frac{dc}{2\pi ic}\mathcal{I}_{U_N}^{(2,2)}(\boldsymbol{a},\boldsymbol{a}^{-1};dc)\mathcal{I}_{U_N}^{(2,2)}(\boldsymbol{b},\boldsymbol{b}^{-1};dc^{-1})  \mathcal{I}_{\SU(N)}^{(2,2),\textrm{V}} (\boldsymbol{a})\mathcal{I}_{\SU(N)}^{(2,2),\textrm{V}} (\boldsymbol{b})~,\cr 
\cI^{(2,2),N}_{\includegraphics[width=.5cm]{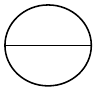}}=&\frac{\eta(q)^3}{\vartheta_1(y)} \int_{\JK}\frac{d\boldsymbol{a}}{2\pi i\boldsymbol{a}}\frac{d\boldsymbol{b}}{2\pi i\boldsymbol{b}}\frac{dc}{2\pi ic}\mathcal{I}_{U_N}^{(2,2)}(\boldsymbol{a},\boldsymbol{b};dc) \mathcal{I}_{U_N}^{(2,2)}(\boldsymbol{a}^{-1},\boldsymbol{b}^{-1};dc^{-1})   \mathcal{I}_{\SU(N)}^{(2,2),\textrm{V}} (\boldsymbol{a})\mathcal{I}_{\SU(N)}^{(2,2),\textrm{V}} (\boldsymbol{b}) ~.
\eea 
However, the explicit evaluation of the JK residue integrals verifies their agreement:
\bea\label{EG-genus2}
   \cI^{(2,2),N}_{\includegraphics[width=1cm]{figures/genus2-2.pdf}}=\cI^{(2,2),N}_{\includegraphics[width=.5cm]{figures/genus2-1.pdf}}&= \frac{\eta(q)^3\vartheta_1(y)}{\vartheta_1(y^N)^2} \oint_{\mathrm{JK}} \frac{d a}{2 \pi i a}  \frac{\vartheta_1(y^Na^Nd^N)}{\vartheta_1(a^Nd^N)} \frac{\vartheta_1(y^Na^{-N}d^N)}{\vartheta_1(a^{-N}d^N)}\cr &=N\frac{\vartheta_1(y)\vartheta_1(y^Nd^{2N})}{\vartheta_1(y^N)\vartheta_1(d^{2N})}~,
\eea
indicating that they are dual to each other. This suggests that the IR theory is independent of specific descriptions of genus-two theories, implying an underlying TQFT structure in this class of theories.

For a $(2,2)$ quiver theory of genus $g>0$ constructed from the $U_2^{(2,2)}$ theory, the minimum number of $\mathrm{U}(1)$ gauge groups required is $g-1$. Hence, extending the previous results, we can consider a $(2,2)$ quiver theory of genus $g>0$ with $n$ punctures, where the numbers of $\mathrm{SU}(N)$ and $\mathrm{U}(1)$ gauge groups are $2(g-1)+n$ and $g-1$, respectively. At a generic point in the moduli space of the chiral multiplets, we conjecture that the $\mathrm{U}(1)^{g(N-1)}$ gauge group remains unbroken. Consequently, the central charge of the theory is:
\be 
c_L = 6g(N-1) + 3(n + g - 1) = c_R~.
\ee

For this class of theories, the elliptic genus depends only on $(g,n)$, and is independent of the quiver descriptions (or frames), which is expressed as
\be 
\cI_{g>0,n}^{(2,2),N}(c_1,\ldots,c_n) = \cI_{g=1,n}^{(2,2),N}(c_1,\ldots,c_n)  \prod_{i=1}^{g-1} \frac{N \vartheta_1(y) \vartheta_1(y^N d_i^{2N})}{\vartheta_1(y^N) \vartheta_1(d_i^{2N})}~,
\ee
where $\cI_{g=1,n}^{(2,2),N}$ is given by \eqref{EG-genus1}. Thus, theories with different quiver descriptions are all dual to each other. 
Moreover, the integral formula \eqref{EG-genus2} guarantees that the above form of the elliptic genera is consistent with the TQFT structure as 
\bea
\mathcal{I}_{g=g_1+g_2, n_1+n_2-2}^{(2,2),N}=&\frac{\eta(q)^3}{\vartheta_1(y)}\int_{\mathrm{JK}} \frac{d a}{2 \pi i a} \mathcal{I}^{(2,2),N}_{g_1, n_1}(\ldots,d_{g-1}a) \mathcal{I}^{(2,2),N}_{g_2, n_2}(d_{g-1}a^{-1},\ldots) ~,\cr 
    \mathcal{I}_{g + 1, n - 2}^{(2,2),N} = &\frac{\eta(q)^3}{\vartheta_1(y)}\int_\text{JK} \frac{da}{2\pi i a} 
    \mathcal{I}_{g, n}^{(2,2),N}(\ldots,d_{g} a, d_{g}a^{-1})~.
\eea

As a result of JK residue integrals, these elliptic genera are all expressed as simple products of theta functions, suggesting that they are dual to $\cN=(2,2)$ Landau-Ginzburg models. The challenge lies in identifying the correct superpotential. The method for this is given by using $S^2$ partition functions \cite{Gomis:2012wy} based on the techniques of \cite{Hori:2000kt,Benini:2012ui,Doroud:2012xw}. We leave this problem for future investigation.

Furthermore, one can introduce FI parameters for the U(1) gauge groups and discrete theta angles for $\SU(N)$ gauge groups. It is also desirable to study the phases of the $\cN=(2,2)$ quiver theories considered in this paper with respect to these parameters, following the approach of \cite{Witten:1993yc}.

\section{\texorpdfstring{$\cN=(0,2)$}{N=(0,2)} dualities}\label{sec:(0,2)-dual}
In this section, we explore deeper into the rich landscape of 2d $\mathcal{N}=(0,2)$ dualities. 
2d $\cN=(0,2)$ supersymmetric theories are an important class of quantum field theories with chiral supersymmetry. They appear in contexts ranging from heterotic string worldsheet models to effective theories on branes. Despite having only two supercharges, techniques developed in recent years allow us to study their rich dynamics.  In particular, understanding their infrared behavior and uncovering possible dualities remains a central topic of interest.

A foundational breakthrough in this direction was the discovery of the triality of 2d $\mathcal{N}=(0,2)$ U($N$) gauge theories \cite{Gadde:2013lxa}, which demonstrated that three distinct unitary gauge theories become identical in the infra-red. This phenomenon closely parallels 4d $\cN=1$ Seiberg duality.  Inspired by this example, more examples of 2d (0,2) dualities are found by reducing known 4d dual pairs on $S^2$ with topological twist \cite{Putrov:2015jpa,Gadde:2015wta,Dedushenko:2017osi,Sacchi:2020pet,Nawata:2023aoq}. In particular,  Seiberg-like dualities for 2d $\mathcal{N}=(0,2)$ SU and Sp gauge theories were uncovered by compactifying their 4d $\mathcal{N}=1$ counterparts on $S^2$ \cite{Gadde:2015wta,Sacchi:2020pet}. Similarly, 2d $\mathcal{N}=(0,2)$ gauge/Landau-Ginzburg (LG) dualities emerged from the twisted compactification of 4d $\mathcal{N}=2$ SCFTs \cite{Nawata:2023aoq}. 

Motivated by these developments, we extend 2d $\mathcal{N}=(0,2)$ Seiberg-like dualities to novel trialities for Sp and SO gauge groups. Moreover, using these dualities (or trialities) as building blocks, we derive a broader class of 2d $\mathcal{N}=(0,2)$ dualities.

 \subsection{Comments on 2d (0,2) Seiberg-like dualities with SU gauge groups}\label{sec:SU-Seiberg}

Let us first review the duality of 2d $\cN=(0,2)$ SU gauge theories obtained by the twisted compactification of the original 4d $\mathcal{N}=1$ Seiberg duality \cite{Seiberg:1994pq} on $S^2$. This duality is illustrated in Fig.~\ref{fig:seiberg-SU-N1-N2-N3}.\footnote{This duality was originally discovered in \cite{Gadde:2015wta}, which we regrettably overlooked in the first version of our paper. In that version, we presented a special case of this duality in a manner that might have implied it was our original derivation. We would like to clarify that full credit for this discovery belongs to \cite{Gadde:2015wta}. The purpose of this subsection is to elucidate key subtleties and provide additional insights into the duality. }

\begin{figure}[htb]
\centering
  \includestandalone[width=0.75\textwidth]{figures/SUNc-N1-N2-N3-1}
  \figuretag{SU-dual}
\caption{2d (0,2) duality for SQCDs with SU gauge groups, where $n_3=\frac{N_1+N_2-N_3}{2}$ and $n_2=\frac{N_1+N_3-N_2}{2}$ ($N_1\ge N_2+N_3$). The solid line represents a chiral multiplet while the dashed line represents a Fermi multiplet. }
\label{fig:seiberg-SU-N1-N2-N3}
\end{figure}

The two dual theories are described as follows:
\begin{itemize}
\item The first theory is an $\SU(n_3)$ gauge theory with:
\begin{itemize}
\item $N_1$ fundamental chiral multiplets $X_1$,
\item $N_2$ anti-fundamental chiral multiplets $X_2$,
\item $N_3$ anti-fundamental Fermi multiplets $\Psi$,
\end{itemize}
where $\bZ \ni n_3=\frac{N_1+N_2-N_3}{2}$. 
Additionally, the theory contains a gauge-singlet meson $M$, which couples via the superpotential:
\be 
\cW= \Tr (\Psi M X_1).
\ee 

\item The second theory is an $\SU(n_2)$ gauge theory with:
\begin{itemize}
\item $N_1$ anti-fundamental chiral multiplets $\widetilde{X}_1$,
\item $N_2$ fundamental Fermi multiplets $\widetilde{\Psi}$,
\item $N_3$ fundamental chiral multiplets $\widetilde{X}_2$,
\end{itemize}
where $\bZ \ni  n_2=\frac{N_1+N_3-N_2}{2}$. 
This theory also contains a gauge-singlet meson $\widetilde{M}$, which couples via the superpotential:
\be 
\cW= \Tr (\widetilde{\Psi} \widetilde{X}_1 \widetilde{M}).
\ee 
\end{itemize}
For the duality to hold, we impose the condition:
\be\label{rank-condition}
N_1 \geq N_2 + N_3.
\ee
The charges of each multiplet under the global symmetry are summarized in the following table:
\begin{table}[ht]\centering
\renewcommand{\arraystretch}{1.2}
\begin{tabular}{c|cccc|cccc}
   & $X_{1}$ & $X_{2}$& $\Psi$& $M$& $\widetilde{X}_{1}$  & $\widetilde{X}_{2}$  & $\widetilde{\Psi}$ & $\widetilde{M}$\\ \hline
$\mathrm{SU}(N_1)$ & $\overline{\Box}$ & $\mathbf{1}$ & $\mathbf{1}$ & $\Box$& $\Box$ & $\mathbf{1}$  & $\mathbf{1}$& $\overline{\Box}$ \\
$\mathrm{SU}(N_2)$ & $\mathbf{1}$ & $\Box$  & $\mathbf{1}$ & $\mathbf{1}$ & $\mathbf{1}$   & $\mathbf{1}$   & $\overline{\Box}$  & $\mathbf{1}$  \\
$\mathrm{SU}(N_3)$ & $\mathbf{1}$ & $\mathbf{1}$ & $\Box$  & $\overline{\Box}$ & $\mathbf{1}$   & $\overline{\Box}$   & $\mathbf{1}$  & $\Box$\\
$\mathrm{U}(1)_{1}$ & $-1$& $0$  & $0$  & $1$  & $1-\frac{N_1}{n_2}$&$\frac{N_1}{n_2}$& $\frac{N_1}{n_2}$  & $-1$ \\
$\mathrm{U}(1)_{2}$ & $0$ & $1$  & $0$  & $0$  & $0$& $0$   & $-1$& $0$ \\
$\mathrm{U}(1)_{3}$ & $0$ & $0$  & $1$  & $-1$ &$0$ & $-1$   & $0$& $1$ 
\end{tabular}
\end{table}

\bigskip

\paragraph{Why the triality fails with SU gauge groups:} 
However, unlike in the case of the unitary gauge group, this duality does not extend to a triality.  The failure of triality when transitioning from U($N_c$) to SU($N_c$) can be attributed to the absence of a Fayet-Iliopoulos (FI) term, as explained below.

The 2d $\cN=(0,2)$ triality \cite{Gadde:2013lxa} is an equivalence of three unitary gauge theories in the infrared. This triality can be understood through the relationship among the target geometry of the non-linear sigma models in the infrared \cite{Jia:2014ffa,Gadde:2014ppa}. The target geometry involves the tautological bundle $S$ of rank $k$ and the quotient bundle $Q$ of rank $(n-k)$ on a Grassmannian $\operatorname{Gr}(k, n)$, related by the short exact sequence:
$$
0 \longrightarrow S \longrightarrow \mathcal{O}^n \longrightarrow Q \longrightarrow 0 ~.
$$
For the 2d $\mathcal{N}=(0,2)$ triality \cite{Gadde:2013lxa}, an essential bundle isomorphism plays a key role:
\begin{equation}\label{bundle-isom} 
\begin{gathered}
S \\
\downarrow \\
\operatorname{Gr}(k,n)
\end{gathered} \quad \cong \quad \begin{gathered}
Q^{*} \\
\downarrow \\
\operatorname{Gr}(n-k,n)~.
\end{gathered}
\end{equation}
This isomorphism reflects the interchange between the tautological bundle $S$ and the dual of the quotient bundle $Q^*$ under the duality transformation 
$ \operatorname{Gr}(k,n) \longleftrightarrow  \operatorname{Gr}(n-k,n)$.

For unitary gauge groups, one can introduce the Fayet-Iliopoulos (FI) term $\zeta$. For a positive FI term ($\zeta>0$), one of the triality theories flows to the non-linear sigma model with the target space
\begin{equation}\label{geometry1} 
\begin{gathered}
S^{\oplus N_3} \oplus Q^{\oplus N_2} \\
\downarrow \\
\operatorname{Gr}(n_3, N_1)
\end{gathered} \ \cong \  \begin{gathered}
S^{* \oplus N_2} \oplus Q^{* \oplus N_3} \\
\downarrow \\
\operatorname{Gr}(n_2, N_1)
\end{gathered}
\end{equation}
where $n_i=\frac{-N_i+N_{i+1}+N_{i+2}}{2}$ (indices are considered mod 3).
Conversely, for a negative FI term ($\zeta<0$), the target space configuration is:
\begin{equation}\label{geometry2}
\begin{gathered}
S^{* \oplus N_3} \oplus Q^{* \oplus N_1} \\
\downarrow \\
\operatorname{Gr}(n_3, N_2) 
\end{gathered} \ \cong\ 
\begin{gathered}
S^{\oplus N_1} \oplus Q^{\oplus N_3} \\
\downarrow \\
\operatorname{Gr}(n_1, N_2)
\end{gathered}
\end{equation}
Permuting the indices ($N_1, N_2, N_3$), we see that $(0,2)$ triality manifests as equivalence among these target spaces. We refer to \cite[Fig.~5]{Gadde:2014ppa} for a detailed illustration.

However, the triality does \emph{not} hold, namely, the three theories are \emph{inequivalent} in the infra-red if we use a special unitary gauge group $\SU(N_c)$ instead of $\U(N_c)$, removing Fermi fields in the determinant representation of the unitary gauge group. In fact, elliptic genera distinguish the three theories if the gauge groups are special unitary groups $\SU(N_c)$.

One key obstruction is that the condition \eqref{rank-condition} for the duality in Fig. \ref{fig:seiberg-SU-N1-N2-N3} is incompatible with the triality condition $N_i+N_{i+1} \geq N_{i+2}$ \cite{Gadde:2013lxa}.
Furthermore, the FI term cannot be introduced for an SU($N_c$) gauge group. Hence, the two geometric configurations described by \eqref{geometry1} and \eqref{geometry2} cannot be connected through an $\SU(N_c)$ gauge theory. In fact, this underpins why the triality fails when naively changing from U($N_c$) to SU($N_c$).

\bigskip

\paragraph{Non-linear sigma model:} With the SU gauge groups, the dual pair in Fig.~\ref{fig:seiberg-SU-N1-N2-N3} flows to the non-linear sigma model with the same target manifold
\begin{equation}\label{geometry3} 
\begin{gathered}
\det(S) \oplus S^{\oplus N_2} \oplus Q^{\oplus N_3} \\
\downarrow \\
\operatorname{Gr}(n_3, N_1)
\end{gathered} \ \cong \  \begin{gathered}
\det(Q^{*}) \oplus Q^{* \oplus N_2} \oplus  S^{* \oplus N_3}  \\
\downarrow \\
\operatorname{Gr}(n_2, N_1)
\end{gathered}
\end{equation}
In fact, the condition \eqref{rank-condition} ensures that the base Grassmannian is defined as the moduli space of subspaces in $\bC^{N_1}$. 
Note that, due to the SU gauge group, the $N_1$ (anti-)fundamental chirals parametrize 
\begin{equation}\label{det-bundle} 
\begin{gathered}
\det(S) \\
\downarrow \\
\operatorname{Gr}(n_3,N_1)
\end{gathered} \ \cong \  \begin{gathered}
 \det (Q^{*} )\\
\downarrow \\
\operatorname{Gr}(n_2,N_1)~.
\end{gathered}
\end{equation}
where $\det(S)$ and $\det(Q^*)$ denote the determinant line bundles of $S$ and $Q^*$, respectively.
Thus, the proposed duality for $\SU$ gauge groups can be understood as the equivalence of the target geometries in \eqref{geometry3}.

\bigskip 

\paragraph{Central charges:}

Since the target space is non-compact, one of the fundamental assumptions of $c$-extremization in \cite{Benini:2012cz,Benini:2013cda} is violated. Specifically, the issue arises because a non-holomorphic current may exist for the flavor symmetry associated with the non-compact direction. This current cannot mix with the $R$-symmetry current, which invalidates the naive application of $c$-extremization \cite{Benini:2013cda,Sacchi:2020pet}. 

To address this issue, it is necessary to identify the chiral operators that parametrize these non-compact directions and assign them an $R$-charge of 0. In our example, while the Grassmannian itself is compact, the total space \eqref{det-bundle} of the determinant line bundle, parametrized by $N_1$ (anti-)fundamental chirals, is non-compact. Since these $N_1$ (anti-)fundamental chirals transform with the same charge under the $\U(1)_1$ flavor symmetry, their $R$-charges must all be set to zero.

Consequently, in both dual theories, the $R$-charges of all chiral fields must be uniformly set to 0. Additionally, the $R$-charge of the Fermi field is fixed at 1 to ensure that the superpotential retains an $R$-charge of 1 \cite{Benini:2013cda,Sacchi:2020pet}.

With this assignment, we verify the agreement of the central charges for the dual theories:
\begin{equation}
c_{L}=2(n_3^2+N_{1}N_{3}+1)~, \qquad c_{R}=3(n_3^{2}+N_{1}N_{3}+1)~.
\end{equation}
Note that the right-moving central charge is three times the dimension of the target space \eqref{geometry3}.

Let us compare this with the case of $\cN=(0,2)$ triality studied in \cite{Gadde:2013lxa}. For unitary gauge groups, the (anti-)fundamental chiral multiplets in the theory simply parameterize a compact Grassmannian manifold (without the determinant line bundle in this case). In addition, the presence of a superpotential imposes further constraints on the $R$-charges of the supermultiplets. Consequently, the $c$-extremization must be performed to determine the $R$-symmetry. 
Hence, although the vacuum moduli space is non-compact as seen in \eqref{geometry1} and \eqref{geometry2},  the application of the $c$-extremization is still required.

\bigskip

\paragraph{Other checks for the duality:}

It is also straightforward to check that 't Hooft anomaly matches under the duality for each global symmetry. 
In addition, the elliptic genera can be computed using the JK residue integral:
\begin{align}
\begin{split}
\mathcal{I}_{A}&=\frac{1}{n_3!}\oint_{\mathrm{JK}}\frac{d\boldsymbol{a}}{2\pi i \boldsymbol{a}}\prod_{i=1}^{n_3}\frac{ \prod_{j\ne i}\vartheta_{1}(a_{i}a_{j}^{-1})\cdot \prod_{j=1}^{N_{3}}\vartheta_{1}(a_{i}^{-1}d_{j}x_3)}{\prod_{{j=1}}^{N_1}\vartheta_{1}(a_{i}b_{j}^{-1}x_1^{-1})\cdot \prod_{{j=1}}^{N_{2}}\vartheta_{1}(a_{i}^{-1}c_{j}x_2)}\frac{\eta(q)^{n_{3}^2+3n_{3}+N_{1}N_{3}-2}}{\prod_{{i=1}}^{N_{1}}\prod_{{j=1}}^{N_{3}}\vartheta_{1}(x_1x_3^{-1}b_{i}d_{j}^{-1})}~, \\
\mathcal{I}_{B}&=\frac{1}{n_{2}!}\oint_{\mathrm{JK}}\frac{d\boldsymbol{a}}{2\pi i \boldsymbol{a}}\prod_{i=1}^{n_{2}}\frac{\prod_{j\ne i}^{}\vartheta_{1}(a_{i}a_{j}^{-1})\cdot \prod_{j=1}^{N_{2}}\vartheta_{1}(a_{i}c_{j}^{-1}x_{2}^{-1}x_{1}^{\frac{N_{1}}{n_{2}}})}{\prod_{j=1}^{N_{1}}\vartheta_{1}(a_{i}^{-1}b_{j}x_{1}^{-\frac{n_{3}}{n_{2}}})\cdot \prod_{j=1}^{N_{3}}\vartheta_{1}(a_{i}d_{j}^{-1}x_{3}^{-1}x_{1}^{\frac{N_{1}}{n_{2}}})}
\frac{\eta(q)^{{n_{2}^{2}+3n_{2}+N_{1}N_{2}-2}}}{\prod_{{i=1}}^{N_{1}}\prod_{j=1}^{N_{2}}\vartheta_{1}(x_{2}x_{1}^{-1}c_{j}b_{i}^{-1})}~.
\end{split}
\end{align}
We verify the identity $\mathcal{I}_A=\mathcal{I}_B$ up to the rank-five integral. Notably, when $N_1 = N_2 + N_3$, an additional condition is required: 
\be\label{additional-cond}
x_1^{N_1}= x_2^{N_2}x_3^{N_3}~.
\ee

\subsection{From 2d (0,2) Seiberg-like dualities to new trialities}\label{sec:triality}

In addition to SU gauge groups, the twisted compactification of 4d $\mathcal{N}=1$ dualities with Sp gauge groups \cite{Intriligator_1995} on $S^2$ was first explored in \cite{Gadde:2015wta}, and later refined in \cite{Sacchi:2020pet}. This leads to the 2d (0,2) duality between Sp SQCD and a Landau-Ginzburg theory. Following their approach, we will find a new 2d (0,2) duality from the twisted compactification of the Intriligator-Seiberg duality \cite{Intriligator:1995id} for the $\mathrm{SO}$ gauge group on $S^2$. Furthermore, in this subsection, we extend these 2d (0,2) Seiberg-like dualities to new trialities.

\paragraph{Sp gauge group:} The twisted compactification on $S^2$  is applied to 4d $\mathcal{N}=1$ dualities with $\Sp(2N)$ gauge groups \cite{Intriligator_1995}, leading to the identification of a 2d $\mathcal{N}=(0,2)$ duality between $\Sp(2N)$ SQCD (AS-1) and an LG model (AS-3) \cite{Gadde:2015wta,Sacchi:2020pet}. In this work, we further generalize this result and propose a new triality, which we will outline below.

Before proceeding, let us remark on an important distinction between the 2d $\mathcal{N}=(0,2)$ Seiberg-like dualities involving the SU and Sp gauge groups. A key difference is that Sp gauge theories only have chiral multiplets in the fundamental representation, whereas SU gauge theories involve both fundamental and anti-fundamental chirals in their duality. Consequently, upon the compactification on $S^2$, we obtain a duality between a 2d $\mathcal{N}=(0,2)$ $\Sp(2N)$ gauge theory and a Landau-Ginzburg (LG) model. However, to our knowledge, a duality between two 2d $\mathcal{N}=(0,2)$ $\Sp(2N)$ gauge theories has not yet been discovered.

\begin{figure}[htb]
\centering
\includestandalone[width=0.8\textwidth]{figures/Sp-SU-AS-AS-even}
  \figuretag{AS-dual}
\caption{A triality among three distinct theories: (AS-1) an $\mathrm{Sp}(2N)$ gauge theory with $2N+2$ fundamental chirals, (AS-2) an $\mathrm{SU}(2N)$ gauge theory with one anti-symmetric and $2N+2$ fundamental chirals, coupled to a neutral Fermi multiplet $\Psi$, and (AS-3) a Landau-Ginzburg model consisting of $(N+1)(2N+1)$ chiral multiplets coupled to a Fermi multiplet $\Psi$.}
\label{fig:seiberg-Sp}
\end{figure}

\begin{enumerate}
\item[AS-1.] $\Sp(2 N)$ gauge theory with $2 N+2$ fundamental chirals $Z$ and no superpotential.
\item[AS-2.] $\SU(2N)$ gauge theory with one anti-symmetric chiral $X$ and $2N+2$ fundamental chirals $Y$. Additionally, there is a neutral Fermi multiplet $\Psi$, forming a superpotential
\be \label{AS_2}
\cW=\Psi\Pf X~.
\ee 

 \item[AS-3.]  LG model of one Fermi $\Psi$ and $(N+1)(2 N+1)$ chirals, forming an anti-symmetric $(2 N+2) \times(2 N+2)$ matrix $A$ with a superpotential
\be 
\cW=\Psi \Pf A~.
\ee 
\end{enumerate}

The $\U(1)_x$ charges of the fields are given as follows:
\be 
\renewcommand{\arraystretch}{1.2}
\begin{tabular}{c|ccccc}
   & $\Psi$ & $A$& $Z$& $X$& $Y$\\ \hline
$\mathrm{U}(1)_x$ & $-(2N+2)$ & $2$ & $1$ & $\frac{2N+2}{N}$&  $\frac{1}{N}$ 
\end{tabular}
\ee
As a simple check, their central charges agree as
\be
c_L=2N(2N+3) ,\qquad c_R=3N(2N+3)~.
\ee
The elliptic genera also agree
\begin{align}\label{eq:Sp-duality}
\cI&=-\frac{\eta(q)^{2 N(N+3)}}{2^{N}N!}\oint_{\textrm{JK}}\frac{d\boldsymbol{a}}{2\pi i \boldsymbol{a}}\prod_{i=1}^{N} \frac{ \vartheta_1(a_i^{ \pm 2}) \prod_{j>i} \vartheta_1(a_j^{ \pm} a_i^{ \pm})}{\prod_{j=1}^{2N+2} \vartheta_1(x a_i^\pm b_j)}\cr 
&=\frac{\eta(q)^{(2N^2+9N-3)}\vartheta_1(x^{-(2N+2)})}{(2N)!}\oint_{\mathrm{JK}}\frac{d\boldsymbol{a}}{2\pi i \boldsymbol{a}}\prod_{i=1}^{2N}\frac{{\prod_{i\ne j}}\vartheta_{1}(a_{i}/a_{j})}{\prod_{i<j}\vartheta_{1}(x^{\frac{2N+2}{N}}a_{i}a_{j})\prod_{j=1}^{2N+2}\vartheta_{1}(x^{\frac{1}{N}}a_{i}b_{j})}\cr 
&=\frac{\eta(q)^{N(2N+3)}\vartheta_1(x^{-(2N+2)})}{\prod_{i< j} \vartheta_1(x^2b_ib_j)}~.
\end{align}

We find this triality by considering the (0,2) reduction of 4d $\mathcal{N}=2$ SCFTs \cite{Gadde:2015wta,Nawata:2023aoq} on $S^2$. For a 4d $\mathcal{N}=2$ theory with gauge group $G$, the $\beta$-function is given by
\begin{equation} 
\beta \propto 2T(\mathbf{adj}) - \sum_i T(R_i)~,
\end{equation} 
where $T(R)$ denotes the Dynkin indices of a representation $R$ for a half-hypermultiplet. On the other hand, the gauge anomaly of the corresponding (0,2) theory is expressed as
\begin{equation} 
\mathrm{Tr}(\gamma_3 G^2) = \sum_i T(R_i) - T(\mathbf{adj}).
\end{equation} 
Since only half of the 4d half-hypermultiplets become (0,2) chiral multiplets upon reduction (see Tab.~\ref{tab:top-twist}), the condition for the vanishing $\beta$-function in 4d $\cN=2$ theory is equivalent to the condition for the vanishing 2d gauge anomaly. Therefore, the (0,2) theory obtained from the Lagrangian 4d $\cN=2$ SCFT is well-defined. In \cite{Nawata:2023aoq}, the (0,2) reduction of 4d $\mathcal{N}=2$ Lagrangian SCFTs of type $A$ class $\mathcal{S}$ is considered. The elliptic genera of these theories can be expressed in a surprisingly simple form, involving products of theta functions and eta functions. In some cases, the LG dual theories are identified.

Here, we consider the 4d $\cN=2$ SCFTs: $\Sp(2N)$ gauge theory with $2N+2$ fundamental half-hypermultiplets, and $\SU(2N)$ gauge theory with one anti-symmetric and $2N+2$ fundamental hypermultiplets. These theories arise from D4-NS5-O6$^-$ brane system (See \cite[\S3.1]{Kim:2024vci}). Their $(0,2)$ reductions yield the AS-1 and AS-2 theories shown in Fig.~\ref{fig:seiberg-Sp}, and they turn out to be dual to a $(0,2)$ LG model (AS-3). Notably, the integrand of the elliptic genus \eqref{eq:Sp-duality} coincides with that of the corresponding 4d $\mathcal{N}=2$ Schur indices (up to the modification of the $\U(1)_x$-charges due to the superpotential \eqref{AS_2}) \cite{Gadde:2015wta,Nawata:2023aoq}.

\bigskip

\begin{figure}[htb]
\centering
  \includestandalone[width=0.5\textwidth]{figures/Sp-ASodd}
\caption{A duality between a free theory with $N(N-1)/2$ chirals and Sp($N-3$) gauge theory with $N$ fundamental chirals, one fundamental Fermi multiplet and $N$ neutral mesons.}
\label{fig:Sp-ASodd}
\end{figure}

Additionally, we find a similar triality with the flavor group $\SU(N)$ with odd $N$, which is given in Appendix \ref{app:odd}. As another generalization, we also propose a duality between the following theories (Fig.~\ref{fig:Sp-ASodd}).
\begin{itemize}
\item Sp($N-3$) gauge theory with $N$ fundamental chirals $X$, one fundamental Fermi multiplet $\Psi$ and $N$ neutral mesons $M$ with a superpotential 
\be 
\cW= \omega^{\a\b}\Psi_\a X_{\b,i} M^i 
\ee 
where $\omega^{\a\b}$ is the symplectic form that projects the tensor product $\Box \otimes \Box$ to the trivial representation for representations of the gauge group Sp($N-3$).
\item a free theory with $N(N-1)/2$ chirals 
\end{itemize}
This duality is used frequently in what follows. The elliptic genera are given by
\begin{equation}
\begin{aligned}
\mathcal{I}&=\frac{\eta(q)^{(N^{2}+2N-9)/2}}{2^{(N-3)/2}\frac{N-3}{2}!}\oint_{\mathrm{JK}}\frac{d\boldsymbol{a}}{2\pi i \boldsymbol{a}} \prod_{i=1}^{\frac{N-3}{2}}\frac{\vartheta_{1}(a_{i}^{\pm}x^{-N})\vartheta_{1}(a_{i}^{\pm2})\prod_{j<i}\vartheta_{1}(a_{i}^{\pm} a_{j}^{\pm})}{\prod_{j=1}^{N}\vartheta_{1}(xa_{i}^{\pm}b_{j})}\cdot \frac{1}{\prod_{i=1}^{N}\vartheta_{1}(x^{N-1}b_{i}^{-1})}
\\
&=\prod_{i<j}\frac{\eta(q)}{\vartheta_{1}(x^{2}b_{i}b_{j})}~.
\end{aligned}
\end{equation}
This duality is used to derive other dualities in what follows.

\paragraph{SO gauge group:}
We also consider the twisted compactification of the 4d $\cN=1$ Intriligator-Seiberg duality \cite{Intriligator:1995id} on $S^2$, which results in a duality between a 2d $\cN=(0,2)$ SO SQCD (Sym-1) and an LG model (Sym-3). This procedure is detailed in Appendix \ref{app:SO-compactification}. Like in the case of the Sp gauge group, the compactification of the 4d $\cN=1$ duality on $S^2$ does not produce a duality between two SO gauge theories, as it involves only fundamental chiral multiplets. 
Moreover, we extend this duality to a triality, as illustrated in Fig.~\ref{fig:seiberg-SO}. 
\begin{figure}[htb]
\centering
  \includestandalone[width=0.8\textwidth]{figures/SO-SU-sym}
  \figuretag{Sym-dual}
\caption{A triality among three distinct theories: (Sym-1) an $\mathrm{SO}(N)$ gauge theory of SQCD type with $N-2$ fundamental chirals coupled to a neutral chiral $X$ and Fermi $\Psi$, (Sym-2) an $\mathrm{SU}(N)$ gauge theory with one symmetric and $N-2$ fundamental chirals, coupled to a neutral Fermi multiplet $\Psi$, and (Sym-3) a Landau-Ginzburg model consisting of $(N-2)(N-1)/2$ chiral multiplets coupled to a Fermi multiplet $\Psi$.}
\label{fig:seiberg-SO}
\end{figure}

\begin{enumerate}
\item[Sym-1.] $\SO(N)$ gauge theory with $N-2$ fundamental chirals $Y$. Additionally, there is a gauge neutral chiral $X$ and Fermi $\Psi$, forming a superpotential
\be 
\cW=\Psi(X^2+ \det A)~,
\ee 
where $A$ is defined by the projector $\eta^{\a\b}$ of the tensor product $\Box\otimes \Box$ to the trivial representation for SO($N$) gauge group as
\be
A_{ij} := \eta^{\a\b} Y_{\a,i}Y_{\b,j}  ~.
\ee 
\item[Sym-2.] $\SU(N)$ gauge theory with one symmetric chiral $Z$ and $N-2$ fundamental chirals $W$. Furthermore, there is a neutral Fermi multiplet $\Psi$, which forms a superpotential
\be 
\cW=\Psi \det A~,
\ee 
where 
\be 
A_{ij}= Z_{\a\b}W^{\a}_iW^{\b}_j~.
\ee
\item[Sym-3.]  LG model of one Fermi $\Psi$ and $(N-2)(N-1)/2$ chirals, forming a symmetric $(N-2) \times(N-2)$ matrix $A$ with a superpotential
\be 
\cW=\Psi \det A~.
\ee 
\end{enumerate}

The $\U(1)_x$ charges of the fields are given as follows:
\be 
\renewcommand{\arraystretch}{1.2}
\begin{tabular}{c|cccccc}
   & $\Psi$ & $A$& $X$& $Y$& $Z$ & $W$\\ \hline
$\mathrm{U}(1)_x$ & $-2(N-2)$ & $2$ & $N-2$ & $1$&  $\frac{2(N-2)}{N}$ & $\frac{2}{N}$
\end{tabular}
\ee 
 These theories have the same central charges
\be 
c_L=N(N-3) ~,\qquad c_R=\frac{3N(N-3)}{2}
\ee 
Furthermore, the elliptic genera agree
\begin{align}\label{eq:SO-duality}
   \cI&=\frac{2\eta(q)^{(N^2-3\chi)/2}\vartheta_{1}(x^{-2(N-2)})}{2^{\lfloor{(N-1)/2}\rfloor}\lfloor{N/2}\rfloor!\vartheta_{1}(x^{N-2})}
\oint_{\textrm{JK}}\frac{d\boldsymbol{a} }{2\pi  i \boldsymbol{a}} 
\frac{
\prod_{i<j}^{\lfloor{N/2}\rfloor}\vartheta_{1}(a_{i}^{\pm}a_{j}^{\pm})
}{
\prod_{i=1}^{\lfloor{N/2}\rfloor}\prod_{j=1}^{N-2}\vartheta_{1}(xa_i^{\pm} b_{j})
}\Bigg(\frac{\prod_{i=1}^{\lfloor{N/2}\rfloor}\vartheta_{1}(a_{i}^{\pm})}{\prod_{j=1}^{N-2}\vartheta_{1}(xb_{j})}\Bigg)^{\chi}
\nonumber
\\
&=- \frac{\eta(q)^{\frac{1}{2}(N^2+3 N-6)}\vartheta_{1}(x^{-2(N-2)})}{N!} \oint_{\textrm{JK}}\frac{d\boldsymbol{a} }{2\pi  i \boldsymbol{a}}  \prod_{i=1}^N \frac{\prod_{i \neq j} \vartheta_1(a_i / a_j)}{\prod_{i \leq j} \vartheta_1(x^{\frac{2(N-2)}{N}} a_i a_j) \prod_{j=1}^{N-2} \vartheta_1(x^{\frac{2}{N}} b_j / a_i)} \cr 
&=\frac{\eta(q)^{N(N-3)/2}\vartheta_{1}(x^{-2(N-2)})}{\prod_{i\leq j}\vartheta_{1}(x^2 b_i b_{j})}
\end{align}
where $\chi\equiv N \mod 2$.

In fact, we arrive at this triality by considering the (0,2) reduction of the 4d $\cN=2$ SCFTs: $\SO(2N)$ gauge theory with $2N-2$ fundamental half-hypermultiplets, and $\SU(2N)$ gauge theory with one symmetric and $2N-2$ fundamental hypermultiplets. These theories arise from D4-NS5-O6$^+$ brane system (See \cite[\S3.1]{Kim:2024vci}). Their $(0,2)$ reductions yield the Sym-1 and Sym-2 theories shown in Fig.~\ref{fig:seiberg-SO}, and they turn out to be dual to a $(0,2)$ LG model (Sym-3).

\begin{figure}[htb]
\centering
\includestandalone[width=0.75\textwidth]{figures/SU2-SU2k-N1-N2-derivation}
\caption{A chain of dualities. Starting with the first theory (top), we apply  \ref{fig:seiberg-Sp} between AS-2 and AS-3 repeatedly to obtain the quiver gauge theory shown in the second line. Next, we apply  \ref{fig:seiberg-SU-N1-N2-N3} to the SU$(N_1 - N_2 - 2)$ gauge group in the second theory, resulting in the third theory. Finally, successive applications of \ref{fig:seiberg-SU-N1-N2-N3} to the remaining SU gauge groups, proceeding from right to left, yield the fourth theory at the bottom. The additional Fermi and chiral fields $\Psi$ and $X$ are included as indicated throughout the chain, and they form a superpotential term $\cW=\Psi X$.}
\label{fig:chain}
\end{figure}

\subsection{Duality between SU and Sp gauge theory}\label{sec:SU-Sp}

It is natural to ask whether the duality between the Sp and SU gauge theories (i.e. AS-1 and AS-2) in Fig.~\ref{fig:seiberg-Sp} can be generalized to include the fundamental Fermi multiplets. To explore this, adding anti-fundamental Fermi multiplets to the AS-2 theory, we consider the following theory:
\begin{itemize}
\item SU$(N_1-N_2-2)$ gauge theory with one anti-symmetric chiral $X$ ($\U(1)_x$-charge 2), $N_1$ fundamental chirals $Y$ and $N_2$ anti-fundamental Fermi's $\Gamma$ where $N_1-N_2$ is even. In addition, there is one free Fermi multiplet $\Psi$ ($\U(1)_x$-charge $N=N_1-N_2-2$), forming a superpotential 
\be 
\cW=\Psi~ \Pf X~.
\ee 
The elliptic genus of the theory is
\begin{align}\label{I_SU}
\mathcal{I}_{\textrm{SU}}=\frac{\vartheta_{1}(x^{N})\eta(q)^{(N^{2}+9N-6)/2}}{N!}\oint_{\mathrm{JK}}\frac{d\boldsymbol{a}}{2\pi i \boldsymbol{a}}\prod_{i=1}^{N}\frac{\prod_{j\ne i}^{}\vartheta_{1}(a_{i}/a_{j})\cdot \prod_{j=1}^{N_{2}}\vartheta_{1}(a_{i}^{-1}c_{j}y_{2})}{\prod_{j<i}^{}\vartheta_{1}(x^{2}a_{i}a_{j})\prod_{j=1}^{N_{1}}\vartheta_{1}(a_{i}b_{j}y_{1}^{-1})}~.
\end{align}
\end{itemize}

We aim to establish a duality with an Sp gauge theory by adding anti-fundamental Fermi multiplets to the AS-1 theory. (See the right panel of Fig. \ref{fig:Sp-SU-linear}.) However, as we shall see, the resulting Sp gauge theory lacks the $\U(1)_x$ symmetry. Therefore, we perform a sequence of duality transformations starting from the theory described above (illustrated in Fig.~\ref{fig:chain}) and then apply a Higgsing procedure to eliminate the $\U(1)_x$ symmetry. Let us explain this process step by step.

We begin by recursively applying the duality between the AS-2 and AS-3 theories in Fig.~\ref{fig:seiberg-Sp}. Specifically, we apply this duality to the $\SU$ gauge node that contains a single anti-symmetric chiral multiplet, as demonstrated in  \cite{Sacchi:2020pet,Amariti:2024usp}. This transformation yields a linear quiver theory with SU gauge groups, as shown in the second line of Fig.~\ref{fig:chain}. It is important to note that for $\SU(2)$, the anti-symmetric representation is trivial. Consequently, the dual description includes an additional chiral multiplet $X$ and a superpotential term $\cW = \Psi X$, where $\Psi$ is a Fermi multiplet. In this intermediate theory,  the bifundamental chiral multiplets between gauge groups, as well as $\Psi$ and $X$,  are charged under the $\U(1)_x$ symmetry.

Next, we recursively apply the SU Seiberg-like duality from Fig.~\ref{fig:seiberg-SU-N1-N2-N3} to each gauge node, proceeding from right to left. This sequence of dualities ultimately produces the theory shown at the bottom of Fig.~\ref{fig:chain}, where all the theories in Fig.~\ref{fig:chain} are dual to each other.

The final theory is characterized by the superpotential
\begin{align}
\cW = \Psi X + \Tr \left[\Gamma \, Y^{(1)} \cdots Y^{\left(\frac{N_1 - N_2}{2}\right)}\right]~,
\end{align}
where $\Gamma$ denotes the Fermi meson, and each $Y^{(i)}$ is a bifundamental chiral multiplet transforming under the $\SU(N_2 + 2i) \times \SU(N_2 + 2i - 2)$ gauge or flavor groups for $i = 1, \ldots, \frac{N_1 - N_2}{2}$. These fields are charged under a linear combination of the $\U(1)_{y_1}$ and $\U(1)_{y_2}$ global symmetries.
At this point, the $\U(1)_x$ symmetry can be removed by giving the vacuum expectation value $\langle X \rangle$, effectively decoupling the chiral multiplet $X$ and the Fermi multiplet $\Psi$.  This procedure completes the elimination of the $\U(1)_x$ symmetry and ensures that the global symmetry structure of the resulting theory matches that of the target $\Sp$ gauge theory.

\begin{figure}[htb]
\includestandalone[width=0.9\textwidth]{figures/Sp-SU-linear}
\caption{A duality between the quiver gauge theory with SU gauge groups and Sp($N_1-N_2-2$) gauge theory. The labels with a linear combination of $\mathsf{y}_1$ and $\mathsf{y}_2$ represent U(1) flavor chemical potentials for the corresponding bifundamental fields.}
\label{fig:Sp-SU-linear}
\end{figure}

As a result, we propose a duality between the $\Sp$ gauge theory and the quiver gauge theory with $\SU$ gauge nodes obtained through this sequence of dualities and Higgsing, as illustrated in Fig.~\ref{fig:Sp-SU-linear}.  The specifics of these dual theories are detailed below:
\begin{itemize}
    \item 
The quiver theory with SU gauge groups features a superpotential:  
\be
\cW=  \Tr \Gamma Y^{(1)} \cdots Y^{(\frac{N_1 - N_2}{2})}~,
\ee  
where $\Gamma$ is the Fermi meson, and $Y^{(i)}$ represents the bifundamental chiral fields between the $\SU(N_2 + 2i)$ and $\SU(N_2 + 2i - 2)$ gauge/flavor groups ($i=1,\ldots,\frac{N_1-N_2}{2}$). The elliptic genus of the theory is
\begin{multline}\label{I_quiver}
   \mathcal{I_{\mathrm{quiver}}}=\prod_{k=1}^{\frac{N}{2}}\frac{1}{(N_1-2k)!}~\int_{\mathrm{JK}}^{}\frac{\mathrm{d}\boldsymbol{a}}{2\pi \mathrm{i} \boldsymbol{a}} \prod_{k=1}^{\frac{N}{2}}\frac{\prod_{i\ne j}^{N_{1}-2k}\vartheta_{1}\Big(a_{i}^{(k)}/a_{j}^{(k)}\Big)}{\prod_{i=1}^{N_{1}-2k+2 }\prod_{j=1}^{N_{1}-2k}\vartheta_{1}\Big(y_{1}^{\frac{-2N_{1}}{(N_1-2k+2)(N_1-2k)}}a_{i}^{(k-1)}/a^{(k)}_{j}\Big)}
        \\
        \qquad\frac{\eta(q)^{\frac{1}{4}(3N_1-N_2)N}\prod_{i=1}^{N_{1}}\prod_{j=1}^{N_{2}}\vartheta_{1}\Big(y_{2}y_{1}^{-1}b_{i}^{-1}c_{j}\Big)}{ \prod_{i=1}^{N_{2}+2}\prod_{j=1}^{N_{2}}\vartheta_{1}\Big(a^{(\frac{N}{2})}_{i}c_{j}^{-1}y_{2}^{-1}y_{1}^{\frac{N_{1}}{N_{2}+2}}\Big)}
\end{multline}
where the fugacity $a^{(0)}=b$.
\item Sp$(N_1-N_2-2)$ gauge theory with $N_1$ fundamental chirals and $N_2$ fundamental Fermi's. The elliptic genus of the theory is
\begin{align}\label{I_Sp}
\mathcal{I}_{\mathrm{Sp}}=\frac{\eta(q)^{\frac{N^{2}}{2}+3N}}{2^{\frac{N}{2}}\frac{N}{2}!}\oint_{\mathrm{JK}}\frac{d\boldsymbol{a}}{2\pi i \boldsymbol{a}}\prod_{i=1}^{\frac{N}{2}} \frac{\vartheta_{1}(a_{i}^{\pm2})\prod_{j<i}^{}\vartheta_{1}(a_{i}^{\pm}a_{j}^{\pm})\cdot \prod_{j=1}^{N_{2}}\vartheta_{1}(a_{i}^{\pm}c_{j}y_{2})}{\prod_{j=1}^{N_{1}}\vartheta_{1}(a_{i}^{\pm}b_{j}^{-1}y_{1}^{-1})}~.
\end{align}
\end{itemize}
Note that, throughout this subsection, we assume that $N \equiv N_1 - N_2 - 2$ is even.

In fact, the central charges of both theories match: \be c_L = (N_1 - N_2 - 2)(N_1 - N_2 + 1), \qquad c_R = \frac{3}{2}(N_1 - N_2 - 2)(N_1 - N_2 + 1)~. \ee
Furthermore, we have verified that the elliptic genera agree, $\mathcal{I}_{\mathrm{quiver}} = \mathcal{I}_{\mathrm{Sp}}$, up to the rank-six JK integrals ($N_1=7$ and $N_2=1$ case).

\bigskip

\begin{figure}[htb]
\centering
  \includestandalone[width=0.9\textwidth]{figures/SU2-SU2k-Sp2k}
\caption{The quiver gauge theory on the left features an Sp$(2m)$ gauge group at the most left, followed by multiple SU gauge nodes in the middle, where the ranks increase by two at each step from left to right, with $2m < N$. The quiver theory on the right represents an Sp$(N)$ gauge theory. The two theories are related by Higgsing at the chemical potential $\sfx$ is set to be zero.  The elliptic genus of the left quiver gauge theory matches that of the right Sp$(N)$ gauge theory when the $\U(1)_x$ fugacities associated with the bifundamental chiral fields are turned off.}
\label{fig:SU-Sp}
\end{figure}

At the level of elliptic genera, the Higgsing procedure can be understood in the following way. The elliptic genus \eqref{I_SU} reduces precisely to \eqref{I_Sp} when the $\U(1)_x$ flavor fugacity is turned off:
\be\label{identity}
\lim_{x \to 1} \mathcal{I}_{\mathrm{SU}} = \mathcal{I}_{\mathrm{Sp}}~.
\ee
Since all theories shown in Fig.~\ref{fig:chain} are dual to each other, their elliptic genera must coincide with \eqref{I_SU}. Consequently, the identity \eqref{identity} holds equally between the second theory in Fig.~\ref{fig:chain} and the Sp gauge theory. Furthermore, reversing the direction of arrows in the quiver diagrams on both sides establishes a similar Higgsing relation between the two theories depicted in Fig.~\ref{fig:SU-Sp}. Specifically, once the $\U(1)_x$ fugacity associated to the bifundamental chials is turned off, the elliptic genus of the quiver theory on the left-hand side  in Fig.~\ref{fig:SU-Sp} coincides with that of the Sp gauge theory on the right-hand side. 
This relation will play a role in deriving additional dualities in subsequent subsections. (See Fig. \ref{figure: Sp-adj-der} and Fig. \ref{fig:SU-Sym-AS-even}.)

\subsection{SU gauge theories with anti-symmetric chiral}\label{sec:AS}

Using the dualities above, we can derive another duality. In this subsection, we will show a duality between SU gauge theories with one anti-symmetric chiral, (anti-)fundamental chirals, and potentially fundamental Fermi multiplets. 

We consider an $\SU(N_A)$ gauge theory with the following matter content:
\begin{itemize}\setlength\itemsep{.05em}
\item One anti-symmetric chirals $Y$
\item $n_1$ anti-fundamental chirals $X_1$
\item $N$ fundamental chirals $X_2$ 
\item $m_1$ anti-fundamental Fermi $\Psi$ where $m_1$ is either zero or one
\item A Fermi meson $\Gamma$, transforming as  $(\overline\Box,\Box)$ under the flavor symmetries $\SU(n_1)\times \SU(N)$
\item A chiral meson $M_1$, transforming as  $(-1,\Box)$ under the flavor symmetries $\U(m_1)\times \SU(N)$
\item 
A chiral meson $M_2$, transforming as  $(1,\Box)$ under the flavor symmetries $\U(m_1)\times \SU(n_1)$
\item A Fermi meson $\Lambda$, transforming as $\overline{\protect\yng(1,1)}=\wedge^{n_1-2}\Box$ under the flavor symmetry $\SU(n_1)$
\item $1-m_1$ neutral chiral $P$ under the flavor symmetries $\U(m_1)$ 
\end{itemize}
The gauge group is determined by anomaly-free condition, which is given by
\begin{equation}
N_A = N + n_1 - m_1 - 2~.
\end{equation}
The superpotential takes the form
\begin{equation}
\cW= \Gamma X_2 X_1 + \Psi M_1 X_2 + \Lambda^{ij} X_{1,i}^\alpha X_{1,j}^\beta Y_{\alpha\beta}~.
\end{equation}

The dual theory has a similar structure: it consists of an $\SU(N_B)$ gauge theory, where the matter content is modified by replacing $n_1$ and $m_1$ in the original theory with $n_2$ and $m_2$, respectively, where $m_2 \in \{0,1\}$. The gauge anomaly-free condition in the dual theory is given by
\begin{equation}
N_B = N + n_2 - m_2 - 2~.
\end{equation}
Note that the duality holds only under the condition
\begin{equation}\label{N-cond}
N \geq m_1 + m_2 + 2~.
\end{equation}

\begin{table}[ht]\centering
\renewcommand{\arraystretch}{1.1}
\begin{tabular}{c|ccccccccc}
& $X_{1}$   & $X_{2}$   & $\Psi$  & $M_{1}$& $M_{2}$& $\Gamma$  & $Y$& $\Lambda$ & $P$   \\ \hline
$\mathrm{SU}(N_{A})$& $\overline{\Box}$ & $\Box$& $\overline{\Box}$& $\mathbf{1}$   & $\mathbf{1}$& $\mathbf{1}$   & $\yng(1,1)$& $\mathbf{1}$   & $\mathbf{1}$\\
$\mathrm{SU}(N)$& $\mathbf{1}$ & $\overline{\Box}$ & $\mathbf{1}$ & $\Box$ & $\mathbf{1}$& $\Box$ & $\mathbf{1}$   & $\mathbf{1}$   & $\mathbf{1}$  \\
$\mathrm{SU}(n_1)$ & $\Box$& $\mathbf{1}$ & $\mathbf{1}$ & $\mathbf{1}$   & $\Box$ & $\overline{\Box}$ & $\mathbf{1}$   & $\overline{\protect\yng(1,1)}$ & $\mathbf{1}$\\
$\mathrm{U}(1)_{0}$& $-1$ & $1$  & $-N_{A}-1$   & $N_{A}$& $-N_{A}$& $0$& $2$& $0$& $-N_{A}$\\
$\mathrm{U}(1)_{1}$& $0$  & $-1$ & $N$  & $1-N$  & $N$ & $1$& $0$& $0$& $N$\\
$\mathrm{U}(1)_{2}$& $1$  & $0$  & $-n_{1}$& $n_{1}$& $1-n_{1}$   &$-1$& $0$& $-2$   & $-n_{1}$\\ 
\end{tabular}
\vspace{1pt}
\begin{tabular}{c|ccccccccc}
& $\widetilde{X}_{1}$   & $\widetilde{X}_{2}$   & $\widetilde\Psi$  & $\widetilde{M}_{1}$   & $\widetilde{M}_{2}$& $\widetilde{\Gamma}$  & $\widetilde{Y}$& $\widetilde{\Lambda}$   & $\widetilde{P}$   \\ \hline
$\mathrm{SU}(N_{B})$& $\overline{\Box}$& $\Box$& $\overline{\Box}$ & $\mathbf{1}$  & $\mathbf{1}$  & $\mathbf{1}$  & $\yng(1,1)$& $\mathbf{1}$& $\mathbf{1}$   \\
$\mathrm{SU}(N)$& $\mathbf{1}$  & $\overline{\Box}$& $\mathbf{1}$   & $\Box$& $\mathbf{1}$  & $\Box$& $\mathbf{1}$  & $\mathbf{1}$& $\mathbf{1}$ \\
$\mathrm{SU}(n_2)$ & $\Box$& $\mathbf{1}$  & $\mathbf{1}$   & $\mathbf{1}$  & $\Box$& $\overline{\Box}$  & $\mathbf{1}$  & $\overline{\protect\yng(1,1)}$  & $\mathbf{1}$\\
$\mathrm{U}(1)_{0}$& $\frac{N_{A}}{N_{B}}$ & $-\frac{N_{A}}{N_{B}}$& $N_{A}+\frac{N_{A}}{N_{B}}$& $-N_{A}$ & $N_{A}$& $0$& $-2\frac{N_{A}}{N_{B}}$& $0$& $N_{A}$\\
$\mathrm{U}(1)_{1}$& $-\frac{N}{N_{B}}$& $\frac{N}{N_{B}}-1$   & $-\frac{N}{N_{B}}$& $1$   & $0$& $1$& $2\frac{N}{N_{B}}$ & $0$& $0$\\
$\mathrm{U}(1)_{3}$& $1$   & $0$   & $-n_{2}$  & $n_{2}$  & $1-n_{2}$  &$-1$& $0$& $-2$& $-n_{2}$   \\ 
\end{tabular} \caption{Summary of symmetries for Theory A (upper one) and Theory B (lower one) in Fig. \ref{fig:SU-AS-N-n-m}.}\label{tab:SU-AS-N-n-m}
\end{table} 
\begin{figure}[ht]
\centering
\includestandalone[width=0.7\textwidth]{figures/SU-AS-N-n-m}
\caption{Duality between $\SU(N_A)$ and $\SU(N_B)$ gauge theories with 1AS where $N_A=N+n_1-m_1-2$ and $N_B=N+n_2-m_2-2$. Note that $m_1$ and $m_2$ are either zero or one. }
\label{fig:SU-AS-N-n-m}
\end{figure}

As illustrated in Fig.~\ref{fig:SU-AS-N-n-m}, this duality can be derived through a sequence of duality transformations.
Roughly speaking, an SU anti-symmetric chiral multiplet is replaced by an Sp gauge group with a bifundamental chiral multiplet via the duality in either Fig.~\ref{fig:seiberg-Sp} or Fig.~\ref{fig:Sp-ASodd}. Thus, the process begins by applying one of these dualities—selecting Fig.~\ref{fig:seiberg-Sp} or Fig.~\ref{fig:Sp-ASodd} depending on whether $N_A$ is even or odd. Following this initial step, the duality shown in Fig.~\ref{fig:seiberg-SU-N1-N2-N3} is applied. To obtain the final dual theory, a second application of either Fig.~\ref{fig:seiberg-Sp} or Fig.~\ref{fig:Sp-ASodd} is required, depending on whether the intermediate gauge group rank is even or odd. This systematic procedure ultimately leads to the dual theory.

The symmetries of the theory are summarized in Tab.~\ref{tab:SU-AS-N-n-m}. It is straightforward to verify that the 't Hooft anomalies of each flavor symmetry match under the duality, providing a consistency check for this duality.

Moreover, the elliptic genera for these theories are expressed as
\bea \label{eq:SU-AS-N-n-m}
\mathcal{I}_{A}&=\frac{\eta(q)^{\frac{N_{A}^2+9N_{A}-4}{2}}}{N_{A}!}\oint_{\mathrm{JK}}\frac{d\boldsymbol{a}}{2\pi i \boldsymbol{a}}
\prod_{i=1}^{N_{A}}\frac{\prod_{j\ne i}^{}\vartheta_{1}(a_{i}a_{j}^{-1})\cdot \prod_{j=1}^{m_{1}}\vartheta_{1}(a_{i}^{-1}e_{j}x_{0}^{-1})}{\prod_{j<i}^{}\vartheta_{1}(x_{0}^{2}a_{i}a_{j})\cdot \prod_{j=1}^{N}\vartheta_{1}(a_{i}b_{j}^{-1}x_0x_{1}^{-1})\cdot \prod_{j=1}^{n_{1}}\vartheta_{1}(a_{i}^{-1}c_{j}x_0^{-1}x_{2})}
\cr &\quad\times 
\frac{\prod_{i=1}^{n_{1}}\prod_{j<i}^{}\vartheta_{1}(x_{2}^{-2}c_{i}^{-1}c_{j}^{-1})}{\eta(q)^{\frac{n_{1}(n_{1}-1)}{2}-N(m_{1}-n_{1})-m_{1}n_{1}-1+m_{1}}} \prod_{i=1}^{N}\frac{\prod_{j=1}^{n_{1}}\vartheta_{1}(b_{i}c_{j}^{-1}x_{1}x_{2}^{-1})}{\prod_{j=1}^{m_{1}}\vartheta_{1}(b_{i}e_{j}^{-1}x_{1})}\cdot \frac{\vartheta_{1}(e_{1})^{m_{1}-1}}{\prod_{i=1}^{m_{1}}\prod_{j=1}^{n_{1}}\vartheta_{1}(e_{i}c_{j}x_{2})}
\cr 
\mathcal{I}_{B}&=\frac{\eta(q)^{\frac{N_{B}^2+9 N_{B}-4}{2}}}{N_{B}!}\oint_{\mathrm{JK}}\frac{d\boldsymbol{a}}{2\pi i \boldsymbol{a}}
\prod_{i=1}^{N_{B}}\resizebox{0.63\textwidth}{!}{$\frac{\prod_{j\ne i}^{}\vartheta_{1}(a_{i}a_{j}^{-1})\cdot \prod_{j=1}^{m_{2}}\vartheta_{1}(a_{i}^{-1}f_{j}x_{0}^{-1}y^{-1})}{\prod_{j<i}^{}\vartheta_{1}(x_{0}^{2}y^{2}a_{i}a_{j})\cdot \prod_{j=1}^{N}\vartheta_{1}(a_{i}b_{j}^{-1}x_0 y x_{1}^{-1})\cdot \prod_{j=1}^{n_{2}}\vartheta_{1}(a_{i}^{-1}d_{j}x_0^{-1}y^{-1}x_{3})}$}
\cr &\quad\times 
\frac{\prod_{i=1}^{n_{2}}\prod_{j<i}^{}\vartheta_{1}(x_{3}^{-2}d_{i}^{-1}d_{j}^{-1})}{\eta(q)^{\frac{n_{2}(n_{2}-1)}{2}-N(m_{2}-n_{2})-m_{2}n_{2}-1+m_{2}}}\prod_{i=1}^{N}\frac{\prod_{j=1}^{n_{2}}\vartheta_{1}(b_{i}d_{j}^{-1}x_{1}x_{3}^{-1})}{\prod_{j=1}^{m_{2}}\vartheta_{1}(b_{i}f_{j}^{-1}x_{1})}\cdot \frac{\vartheta_{1}(f_{1})^{m_{2}-1}}{\prod_{i=1}^{m_{2}}\prod_{j=1}^{n_{2}}\vartheta_{1}(f_{i}c_{j}x_{3})}
\eea
where $n_1+n_2+m_1+m_2$ is even. 
 The fugacities $f_1$, $e_1$ and $y$ are defined as
 \begin{equation}
 f_1=x_0^{N_A}x_3^{-n_2},\quad e_1=x_0^{-N_A}x_1^{N}x_2^{-n_1},\quad y^{N_B}=x_0^{-N_A-N_B}x_1^{N}~.
 \end{equation}
For the special case where $N = m_1 + m_2 + 2$, an additional condition must be satisfied:
\begin{align}
x_{1}^{(1-m_{1})N}x_{2}^{-(1-m_{1})n_{1}}x_{3}^{-(1-m_2)n_{2}}=x_{0}^{(m_{2}-m_{1}) (n_{1}+m_{2})}~.
\end{align}
This condition originates from \eqref{additional-cond}, which is based on the relation $N_1 = N_2 + N_3$ in Fig.~\ref{fig:seiberg-SU-N1-N2-N3}.

The elliptic genus $\cI_A$ is independent of the $\SU(n_1)$ fugacity and the $\U(1)_2$ fugacity $x_2$. Likewise, $\cI_B$ is independent of the $\SU(n_2)$ fugacity and the $\U(1)_3$ fugacity $x_3$. Furthermore, the agreement between $\cI_A$ and $\cI_B$ has been verified up to the rank-five JK integrals through an expansion in $q$.

When $m_1 = 0$ and $N = 2,3,4$ in Theory A, Theory B becomes an LG model, as shown in \cite{Amariti:2024usp}. To obtain the LG dual found in \cite{Amariti:2024usp}, we first transfer the Fermi mesons $\Gamma$ and $\Lambda$ from Theory A to Theory B, where they are reinterpreted as chiral mesons. By appropriately choosing $n_2$ and $m_2$, as described below, the LG duality can be established.

\begin{itemize}
\item \textbf{$N = 2$ and even $n_1$:} Set $n_2 = 2$ and $m_2 = 0$ in Theory B. This results in a gauge group rank $N_B = N + n_2 - m_2 - 2 = 2$. Using the duality in Fig.~\ref{fig:seiberg-Sp}, where $\mathrm{SU}(2)$ is identified with $\mathrm{Sp}(2)$, the corresponding LG dual theory \cite[\S3.1]{Amariti:2024usp} is obtained.

\item \textbf{$N = 2$ and odd $n_1$:} Assign $n_2 = 1$ and $m_2 = 0$, reducing the gauge group rank to $N_B = 1$. This corresponds to an LG theory \cite[\S3.2]{Amariti:2024usp}.

\item \textbf{$N = 3$ and even $n_1$:} Choose $n_2 = m_2 = 0$, yielding $N_B = 1$, which directly describes an LG theory \cite[\S3.3]{Amariti:2024usp}.

\item \textbf{$N = 3$ and odd $n_1$:} Set $n_2 = 1$ and $m_2 = 0$, giving $N_B = 2$. Applying the duality in Fig.~\ref{fig:seiberg-Sp}, we obtain the LG dual \cite[\S3.4]{Amariti:2024usp}.

\item \textbf{$N = 4$ and even $n_1$:} Select $n_2 = m_2 = 0$, resulting in $N_B = 2$. The LG dual \cite[\S3.5]{Amariti:2024usp} follows from the duality in Fig.~\ref{fig:seiberg-Sp}.

\item \textbf{$N = 4$ and odd $n_1$:} Assign $n_2 = 0$ and $m_2 = 1$, leading to $N_B = 1$. This implies Theory B is an LG theory \cite[\S3.6]{Amariti:2024usp}.
\end{itemize}

The duality holds under the condition in \eqref{N-cond}, which explicitly excludes the cases where the number of fundamental chirals is $N = 1$ or $N = 0$. Alternative dualities for these excluded cases are discussed in Appendices~\ref{app:N=1} and \ref{app:N=0}.

\subsection{SU gauge theories  with symmetric chiral}\label{sec:Sym}
In a similar manner, we establish a duality between SU gauge theories containing one symmetric chiral multiplet, as depicted in Fig.~\ref{fig:SU-Sym-chiral}. The gauge groups of the dual theories are given by  
\begin{equation}
N_A = N + n_1 - m_1 + 2, \qquad N_B = N + n_2 - m_2 + 2~.
\end{equation}
The matter content and superpotential closely resemble those of the previous duality, with the key difference being the replacement of anti-symmetric representations by symmetric representations. To keep our discussion concise, we do not repeat the detailed discussion of matter content and symmetries.  

Essentially, through the duality depicted in Fig.~\ref{fig:seiberg-SO}, an SU symmetric chiral can be replaced by an SO gauge group with a bifundamental chiral. Consequently, as in Fig.~\ref{fig:SU-AS-N-n-m}, this duality is derived by sequentially applying the known dualities illustrated in Fig.~\ref{fig:seiberg-SO} and \ref{fig:seiberg-SU-N1-N2-N3}.

\begin{figure}[ht]
\centering
\includestandalone[width=0.9\textwidth]{figures/SU-Sym-N-n-m}
\caption{Duality for $\SU(N_A)$ and $\SU(N_B)$ gauge theories with 1Sym where $N_A=N+n_1-m_1+2$ and $N_B=N+n_2-m_2+2$. Note that $m_1$ and $m_2$ are either zero or one.}
\label{fig:SU-Sym-chiral}
\end{figure}

The expressions for the elliptic genera of these theories are provided below:
\bea
\mathcal{I}_{A}&=\frac{\eta(q)^{\frac{N_{A}^2+3N_{A}-4}{2}}}{N_{A}!}\oint_{\mathrm{JK}}\frac{d\boldsymbol{a}}{2\pi i \boldsymbol{a}}
\prod_{i=1}^{N_{A}}\frac{\prod_{j\ne i}^{}\vartheta_{1}(a_{i}a_{j}^{-1})\cdot \prod_{j=1}^{m_{1}}\vartheta_{1}(a_{i}^{-1}e_{j}x_{0}^{-1})}
{\prod_{j\leq i}^{}\vartheta_{1}(x_{0}^{2}a_{i}a_{j})\cdot \prod_{j=1}^{N}\vartheta_{1}(a_{i}b_{j}^{-1}x_{0}x_{1}^{-1})\cdot \prod_{j=1}^{n_{1}}\vartheta_{1}(a_{i}^{-1}c_{j}x_{0}^{-1}x_{2})}
\\&\quad\times 
\frac{\prod_{i=1}^{n_{1}}\prod_{j\leq i}^{}\vartheta_{1}(x_{2}^{2}c_{i}c_{j})}{\eta(q)^{\frac{n_{1}(n_{1}+1)}{2}-N(m_{1}-n_{1})-m_{1}n_{1}-1}}\cdot \prod_{i=1}^{N}\frac{\prod_{j=1}^{n_{1}}\vartheta_{1}(b_{i}c_{j}^{-1}x_{1}x_{2}^{-1})}{\prod_{j=1}^{m_{1}}\vartheta_{1}(b_{i}e_{j}^{-1}x_{1})}\cdot \frac{\vartheta_{1}(e_{1}^{2})^{-m_{1}} \vartheta_{1}(e_{1})^{m_{1}-1}}{\prod_{i=1}^{m_{1}}\prod_{j=1}^{n_{1}}\vartheta_{1}(e_{i}c_{j}x_{2})}
\cr 
\mathcal{I}_{B}&=\frac{\eta(q)^{\frac{N_{B}^2+3N_{B}-4}{2}}}{N_{B}!}\oint_{\mathrm{JK}}\frac{d\boldsymbol{a}}{2\pi i \boldsymbol{a}}
\prod_{i=1}^{N_{B}}\resizebox{0.63\textwidth}{!}{$\frac{\prod_{j\ne i}^{}\vartheta_{1}(a_{i}a_{j}^{-1})\cdot \prod_{j=1}^{m_{2}}\vartheta_{1}(a_{i}^{-1}f_{j}x_{0}^{-1}y^{-1})}{\prod_{j\leq i}^{}\vartheta_{1}(x_{0}^{2}y^{2}a_{i}a_{j})\cdot \prod_{j=1}^{N}\vartheta_{1}(a_{i}b_{j}^{-1}x_0 y x_{1}^{-1})\cdot \prod_{j=1}^{n_{2}}\vartheta_{1}(a_{i}^{-1}d_{j}x_0^{-1}y^{-1}x_{3})}$}
\\&\quad\times 
\frac{\prod_{i=1}^{n_{2}}\prod_{j\leq i}^{}\vartheta_{1}(x_{3}^{2}d_{i}d_{j})}{\eta(q)^{\frac{n_{2}(n_{2}+1)}{2}-N(m_{2}-n_{2})-m_{2}n_{2}-1}}\cdot \prod_{i=1}^{N}\frac{\prod_{j=1}^{n_{2}}\vartheta_{1}(b_{i}d_{j}^{-1}x_{1}x_{3}^{-1})}{\prod_{j=1}^{m_{2}}\vartheta_{1}(b_{i}f_{j}^{-1}x_{1})}\cdot \frac{\vartheta_{1}(f_{1}^{2})^{-m_{2}}\vartheta_{1}(f_{1})^{m_{2}-1}}{\prod_{i=1}^{m_{2}}\prod_{j=1}^{n_{2}}\vartheta_{1}(f_{i}d_{j}x_{3})}
\eea
where the fugacities $f_1$, $e_1$ and $y$ are defined as
 \begin{equation}
 f_1=x_0^{N_A}x_3^{-n_2},\quad e_1=x_0^{-N_A}x_1^{N}x_2^{-n_1},\quad y^{N_B}=x_0^{-N_A-N_B}x_1^{N}~.
 \end{equation}

The elliptic genus $\cI_A$ does not depend on the $\SU(n_1)$ fugacity or the $\U(1)_2$ fugacity $x_2$, while $\cI_B$ is independent of the $\SU(n_2)$ fugacity and the $\U(1)_3$ fugacity $x_3$. Moreover, the agreement between $\cI_A$ and $\cI_B$ has been confirmed, via an expansion in $q$, up to the rank-five JK integrals.

\subsection{Dualities with adjoint chiral}\label{sec:adj}

Consider the 4d $\cN=4$ SCFT. Its (0,2) reduction results in the (0,2) gauge theory only with a (0,2) adjoint chiral multiplet $\phi$, which is indeed a (2,2) vector multiplet. The IR central charge of the $\cN=(2,2)$ pure Yang-Mills theory with gauge group $G$ is given by
\be 
c_L=c_R=3 \operatorname{rank} G~.
\ee 
As we will demonstrate, the theory is indeed dual to $\operatorname{rank} G$ free $\cN=(2,2)$ chiral multiplets.

\paragraph{Sp($2N$)+1{Adj}:} Since the duality for the $\SU(N)$ gauge theory with a single adjoint matter field is well-studied in \cite{Aharony:2016jki,Nawata:2023aoq}, we now extend our analysis to other classical gauge groups. We begin by considering the $\Sp(2N)$ gauge theory with an adjoint chiral field.

A useful observation is that $\textrm{Adj}\cong \textrm{Sym}$ for the $\Sp(2N)$ gauge group. This allows us to apply the duality depicted in Fig.~\ref{fig:seiberg-SO}. As illustrated in Fig.~\ref{figure: Sp-adj-der}, a sequence of duality transformations relates $\Sp(2N) + 1\mathrm{Adj}$ to $\Sp(2N-2) + 1\mathrm{Adj}$ and $\SU(2) + 1\mathrm{Adj}$, with an extra chiral multiplet $X$ and a Fermi multiplet $\Psi$.

Furthermore, it is known \cite{Aharony:2016jki,Nawata:2023aoq} that the $\SU(2) + 1\mathrm{Adj}$ theory is dual to an $\mathcal{N}=(2,2)$ twisted chiral multiplet consisting of a free $\mathcal{N}=(0,2)$ chiral multiplet $\Tr Y^2$ and a Fermi multiplet $\Gamma$. We claim that they form a superpotential 
\be 
\cW= \Gamma X~, \label{superpotential}
\ee 
so that eliminates both the chiral multiplet $X$ and the Fermi multiplet $\Gamma$. As a result, the duality transformation establishes an equivalence between $\Sp(2N) + 1\mathrm{Adj}$ and $\Sp(2N-2) + 1\mathrm{Adj}$ with the additional free chiral multiplet $\Tr Y^2$ and the Fermi multiplet $\Psi$.

This duality recursively reduces the rank of the $\Sp(2N)$ gauge group, and at the end, we obtain the following duality:
\begin{equation} \label{Sp-adj}
\Sp(2N) + 1\mathrm{Adj} \ \Longleftrightarrow \ N \textrm{ free chirals } + N \textrm{ free Fermi's } ~.
\end{equation} 
Note that $N$ free chirals correspond to the gauge-invariant operators made out of the adjoint chiral $Z$:
\begin{equation}
\Tr(Z^{2k}) ~, \qquad k=1,\ldots, N~,
\end{equation}
and $N$ free Fermi's are necessary to form $\cN=(2,2)$ twisted chiral multiplets. 
The duality can be verified by the JK residue computation
\begin{align}\label{Sp}
\begin{split}
  \cI^\Adj_{\mathrm{Sp}(2N)}&= \frac{\eta(q)^{3N}\vartheta_{1}(y)^N}{2^N N!}
\oint_{\textrm{JK}}\frac{d\boldsymbol{a} }{2\pi  i \boldsymbol{a}}
\frac{
\prod_{i< j}^{N}\vartheta_{1}(a_{i}^{\pm}a_{j}^{\pm})
\cdot 
\prod_{i=1}^{N}\vartheta_{1}(a_{i}^{2\pm})
}{\prod_{i\le j}^{N}\vartheta_{1}(y a_{i}^{\pm}a_{j}^{\pm})}
\\
&=(-1)^N\prod_{k=1}^N\frac{\vartheta_{1}(y^{2k-1})}{\vartheta_{1}(y^{2k})}~.
\end{split}
\end{align}

\begin{figure}[htb]
\centering
  \includestandalone[width=\textwidth]{figures/Sp-adj}
\caption{The recursion relation of $\Sp(2N) + 1\text{Adj}$. The leftmost diagram represents the original $\Sp(2N)$ gauge theory with one adjoint chiral. Using a sequence of duality
transformations, this theory is mapped to a different description. In particular, from the third to the fourth diagram, we first apply the duality in Fig.~\ref{fig:seiberg-SU-N1-N2-N3} to the $\SU(2N)$ gauge node, followed by the duality in Fig. \ref{fig:seiberg-SO} applied to the $\Sp(2N-2)$ gauge node. As a result, the theory is ultimately mapped to $\Sp(2N-2) + 1\mathrm{Adj}$ and $\SU(2)+1\mathrm{Adj}$, along with additional chiral $X$ and Fermi $\Psi$, forming the superpotential \eqref{superpotential}. This establishes the recursion pattern, illustrating how $\Sp(2N)$ gauge group transforms into  $\Sp(2N-2)$ gauge group, and iterative applications of this procedure lead to lower-rank cases.}
\label{figure: Sp-adj-der}
\end{figure}

\paragraph{SO($n$)+1{Adj}:}
Now, let us consider the $\SO(n)$ gauge group. Note that $\textrm{Adj}\cong \textrm{AS}$ for the $\SO(n)$ gauge group. Depending on whether $n$ is even or odd, we can apply the dualities presented in Fig.~\ref{fig:seiberg-Sp} and Fig.~\ref{fig:Sp-ASodd}.

As illustrated in Fig.~\ref{fig:adj}, these dualities, when followed by the duality in Fig.~\ref{fig:seiberg-SO}, establish an equivalence between the $\SO(n) + 1\textrm{Adj}$ theory and the $\Sp$ gauge theory with an adjoint matter field, supplemented by additional chiral and Fermi multiplets. This connects the $\SO(n)$ case to the previously discussed duality for $\Sp(2N) + 1\text{Adj}$ in Eq.~\eqref{Sp-adj}.

In the upper part of Fig.~\ref{fig:adj}, the $\SO(2N) + 1\text{Adj}$ theory is dual to the $\Sp(2N-2) + 1\text{Adj}$ theory, along with an additional free chiral multiplet $X$ and a Fermi multiplet $\Psi$. Examining the U(1) charge, we identify that the chiral multiplet $X$ corresponds to the gauge-invariant operator $\Pf Z$, where $Z$ represents the adjoint chiral superfield of the $\SO(2N)$ gauge theory. Consequently, this duality leads to the relation:
\begin{align} \label{SO-even-adj}
\SO(2N) + 1\mathrm{Adj} \ \Longleftrightarrow \ N \textrm{ free chirals } + N \textrm{ free Fermi's } ~,
\end{align} 
where the $N$ free chiral multiplets correspond to gauge-invariant operators constructed from the adjoint chiral superfield $Z$:
\begin{equation} 
\Pf(Z), \quad \Tr(Z^{2k}), \qquad k=1,\ldots, N-1~.
\end{equation} 

In the lower part of Fig.~\ref{fig:adj}, a similar sequence of dualities establishes that the $\SO(2N+1) + 1\text{Adj}$ theory is dual to the $\Sp(2N-2) + 1\text{Adj}$ theory, again with an additional free chiral multiplet $X$ and Fermi multiplet $\Psi$. By examining the U(1) charge, we recognize that the chiral multiplet $X$ corresponds to the gauge-invariant operator $\Tr Z^{2N}$, where $Z$ is the adjoint chiral multiplet of the $\SO(2N+1)$ gauge theory. Incorporating Eq.~\eqref{Sp-adj}, we arrive at the following triality:
\begin{align} \label{SO-odd-adj}
\begin{tikzpicture}
\node () at (0,0)  {$\SO(2N+1) + 1\mathrm{Adj}$};
 \node () at (2.5,0)  {$\Longleftrightarrow$};
 \node () at (6,0)  {$N$ free chirals  + $N$ free Fermi's };
 \node[rotate=40] () at (0.8,-0.75) {$\Big\Updownarrow$};
\node[rotate=-40] () at (4,-0.75) {$\Big\Updownarrow$};
\node () at (2.5,-1.5)  {$\Sp(2N) + 1\mathrm{Adj}$};
\end{tikzpicture}
\end{align} 
where the $N$ free chiral multiplets correspond to gauge-invariant operators constructed from the adjoint chiral multiplet $Z$:
\begin{equation} 
\Tr(Z^{2k}) ~, \qquad k=1,\ldots, N~.
\end{equation}
In both even and odd cases, $N$ free Fermi multiplets are necessary to form $N$ (2,2) free twisted chiral multiplets. 

In fact, the explicit computations of the elliptic genera verify the results
\begin{align}
\begin{split}
  \cI^\Adj_{\mathrm{SO}(2N)}&=
\frac{2\eta(q)^{3N}}{2^{N-1}N! \cdot \vartheta_{1}(y)^N}
\oint_{\textrm{JK}}\frac{d\boldsymbol{a} }{2\pi  i \boldsymbol{a}}
\prod_{i<j}^{N}\frac{
\vartheta_{1}(a_{i}^{\pm}a_{j}^{\pm})
}{
\vartheta_{1}(ya_{i}^{\pm}a_{j}^{\pm})
}
\\
&=(-1)^N\frac{\vartheta_1(y^{N-1})}{\vartheta_{1}(y^N)}\cdot\prod_{k=1}^{N-1} \frac{\vartheta_{1}(y^{2k-1})}{\vartheta_{1}(y^{2k})}~,\label{SO-even}
\end{split}
\\
\begin{split}
  \cI^\Adj_{\mathrm{SO}(2N+1)}&=
\frac{2\eta(q)^{3N}}{2^{N}N! \cdot \vartheta_{1}(y)^N}
\oint_{\textrm{JK}}\frac{d\boldsymbol{a} }{2\pi  i \boldsymbol{a}}
\prod_{i<j}^{N}
\frac{
\vartheta_{1}(a_{i}^{\pm}a_{j}^{\pm})
}{
\vartheta_{1}(ya_{i}^{\pm}a_{j}^{\pm})
}
\cdot
\prod_{i=1}^{N}\frac{
\vartheta_{1}(a_{i}^{\pm})
}{
\vartheta_{1}(ya_{i}^{\pm})
}
\\
&=(-1)^N\prod_{k=1}^{N} \frac{\vartheta_{1}(y^{2k-1})}{\vartheta_{1}(y^{2k})}~.\label{SO-odd}
\end{split}
\end{align}

\begin{figure}[htb]
\centering
  \includestandalone[width=0.7\textwidth]{figures/SO-adj-even}
  \hspace{1cm}
\includestandalone[width=0.8\textwidth]{figures/SO-adj-odd}
\caption{A chain of dualities provides the relation between $\SO + 1\text{Adj}$ and $\Sp + 1\text{Adj}$. In the upper part of the diagram, the $\SO(2N) + 1\text{Adj}$ theory is dual to the $\Sp(2N-2) + 1\text{Adj}$ theory, with additional free chiral $X$ and Fermi  $\Psi$. The chiral  $X$ corresponds to the gauge-invariant operator $\Pf Z$, where $Z$ is the adjoint chiral of the SO($2N$) gauge theory.\\
In the lower part of the diagram, a similar chain of dualities relates the $\SO(2N+1)+1\textrm{Adj}$ theory to $\Sp(2N-2) + 1\text{Adj}$, again with additional free chiral $X$ and Fermi $\Psi$. Here, the free chiral $X$ corresponds to the gauge-invariant operator $\Tr Z^{2N}$, where $Z$ is the adjoint chiral of the SO($2N+1$) gauge theory. Consequently, $\SO(2N+1)+1\textrm{Adj}$ is dual to $\Sp(2N)+1\textrm{Adj}$.}
\label{fig:adj}
\end{figure}

It is easy to verify the identities of the elliptic genera coming from the isomorphisms between gauge algebras:
\be 
\cI^{\Adj}_{\SO(4)}= (\cI^{\Adj}_{\SU(2)})^2~, \qquad \cI^{\Adj}_{\SO(5)}= \cI^{\Adj}_{\Sp(2)}~,\qquad \cI^{\Adj}_{\SO(6)}= \cI^{\Adj}_{\SU(4)}~.
\ee 
The chiral algebra of the 4d $\cN=4$ theory with gauge group $G$ admits the free-field realization by the $bc\b\g$ system \cite{Bonetti:2018fqz}. The elliptic genus (\ref{SO-even}, \ref{SO-odd}, \ref{Sp}) is
the vacuum character of the corresponding $bc\b\g$ system up to a factor \cite{Pan:2021ulr}. Consequently, the elliptic genera are reducible module characters of the chiral algebra of the 4d $\cN=4$ theory with the corresponding gauge groups.

\subsection{SU gauge theories with symmetric and anti-symmetric chirals}\label{sec:AS-Sym}
We now consider the $\SU(N)$ gauge theory that includes both symmetric and anti-symmetric chiral multiplets. As discussed earlier, an SU symmetric chiral multiplet can be replaced by an SO gauge group with bifundamental fields, while an SU anti-symmetric chiral multiplet can be replaced by an Sp gauge group with bifundamental fields. We then apply a sequence of duality transformations to obtain an LG dual model. This procedure closely parallels the approach in the previous subsection, where we derived a recursion relation to systematically reduce the ranks of the gauge groups.

The replacement of the SU anti-symmetric chiral multiplet with an Sp gauge field depends on whether $N$ is even or odd—specifically, via Fig.~\ref{fig:seiberg-Sp} for even $N$ and Fig.~\ref{fig:Sp-ASodd} for odd $N$. Therefore, we divide the discussion into two cases: even and odd $N$.

Consider the case that $N$ is even. As shown in Fig.~\ref{fig:SU-Sym-AS-even}, a sequence of the known dualities provides a ``recursion relation'' of SU+1Sym+1AS theories in terms of the ranks of gauge groups.  Starting with SU($N$)+1Sym+1AS, supplemented by additional 1 chiral multiplets $Z^{(0)}$ and 3 Fermi multiplets $\Psi^{(0)}_{1,2,3}$, with a superpotential 
\bea
\label{SUN1AS1Sym-superpotential}
\cW=&\Psi^{(0)}_1\det X^{(0)}+\Psi^{(0)}_2\Pf Y^{(0)}+\Psi^{(0)}_3 \left((Z^{(0)})^2+B^{(0)}(2)\right)
\eea
As shown in Fig.~\ref{fig:SU-Sym-AS-even},
a series of duality transformations systematically map the theory to a similar theory: SU$(N-2$)+1Sym+1AS, along with additional chiral multiplets $Z^{(1)}, \widetilde{Z}^{(1)}$ and Fermi multiplets $\Psi^{(1)}_{1,2,3,4}$. (Here, $\Psi^{(0)}_2$ forms a superpotential with a newly generated chiral field $P^{(1)}$, which then decouples from the remaining theory. Since $\Psi^{(0)}_2$ and $P^{(1)}$ do not contribute to either the central charge or the elliptic genus, we will ignore it in the subsequent discussion for convenience.)
The superpotential of the dual side takes a similar form:
\bea
\label{SUN-1-1AS1Sym-superpotential}
\cW=&\Psi^{(1)}_1(\det X^{(1)}+\widetilde{Z}^{(1)})+\Psi^{(1)}_2\Pf Y^{(1)}+\Psi^{(1)}_3 \left((Z^{(1)})^2+B^{(1)}(2)\right)+\Psi^{(0)}_3\left((Z^{(0)})^2+\Tr[(A^{(0)})^2]\right)
\eea
where $B^{(i)}(n)$ is a neutral baryon with $\U(1)_{x^{(i)}}\times \U(1)_{y^{(i)}}$ charge $(2n,2(N-2i-n))$ that defined as:
\be
B^{(i)}(n)\equiv\e^{\alpha_1\ldots\alpha_{N-2i}}\e^{\beta_1\ldots\beta_{N-2i}}Y^{(i)}_{\alpha_1\beta_1}\ldots Y^{(i)}_{\alpha_{N-n-2i}\beta_{N-n-2i}}X^{(i)}_{\alpha_{N-n-2i+1}\beta_{N-n-2i+1}}\ldots X^{(i)}_{\alpha_{N-2i}\beta_{N-2i}}
\ee

Moreover, the dual side also has an SU(2)+1Sym theory with the symmetric tensor \(A^{(0)}\), and couples, via \eqref{SUN-1-1AS1Sym-superpotential}, with the chiral $Z^{(0)}$ and Fermi $\Psi_3^{(0)}$ from the original theory, as shown in the last quiver of Fig.~\ref{fig:SU-Sym-AS-even}. Again, the SU(2)+1Sym theory is dual to one chiral $\Tr (A^{(0)})^2$ and Fermi with opposite charges with the chiral $Z^{(0)}$ and Fermi $\Psi_3^{(0)}$. As a result, their combined contribution to the elliptic genus cancels out:
\be 
\underbrace{\frac{\vartheta_1(x^{-2}y^{-N+2})}{\vartheta_1(x^4y^{2N-4})}}_{\textrm{SU(2)+1Sym}}\cdot  \frac{\overbrace{\vartheta_1(x^{-4}y^{-2N+4})}^{\Psi_3^{(0)}} }{\underbrace{\vartheta_1(x^2y^{N-2})}_{Z^{(0)}}}=1~.
\ee 
Although their contributions to the elliptic genus cancel out, these fields are not entirely negligible, as their combined central charge contribution from SU(2)+1Sym, $\Psi^{(0)}_3$, and $Z^{(0)}$ is $3$. 
This establishes a recursive pattern of SU+1Sym+1AS demonstrating how an $\SU(N)$ gauge group systematically reduces to $\SU(N-2)$ through a series of duality transformations, as emphasized in Fig.~\ref{fig:SU-Sym-AS-even}.  
We summarize the charges of the fields in the SU($N$)+1AS+1Sym theory in Tab.~\ref{tab:SU(N)-1AS-1Sym-and-SU(N-2)-even}, as well as the charges of the fields in the SU($N-2$)+1AS+1Sym theory obtained after performing this series of duality transformations.

By repeatedly applying this process, the gauge group rank is systematically reduced, allowing the theory to be expressed in terms of lower-rank cases. The duality recursion reaches its lowest order in the case of SU(2)+1Sym+1AS along with additional Fermi and chiral fields.

\begin{figure}[ht]
\centering
\includestandalone[width=\textwidth]{figures/SU-Sym-AS-even}
\caption{The first quiver on the left corresponds to the theory SU($N$)+1AS+1Sym, along with three Fermi's $\Psi^{(0)}_{1,2,3}$ and one chiral field $Z^{(0)}$. By applying the Fig.~\ref{fig:seiberg-SO} and Fig.~\ref{fig:seiberg-Sp} transformations, we obtain the second quiver, which describes a new theory. This new theory generates two neutral chirals $P^{(1)}$, $Z^{(1)}$: $P^{(1)}$ couples with $\Psi^{(0)}_2$ and decouples from the other fields in the theory. From the third to the fourth diagram, we first apply the duality in Fig.~\ref{fig:seiberg-SU-N1-N2-N3} to the $\SU(N-2)$ gauge node, followed by the same duality applied to the $\SU(N)$ gauge node. As a result, the theory is ultimately mapped to $\SU(N-2) + 1\mathrm{Sym} + 1\mathrm{AS}$ and $\SU(2)+1\mathrm{Sym}$, along with additional chirals $Z^{(0)},Z^{(1)},\widetilde{Z}^{(1)}$ and Fermi's $\Psi^{(0)}_{3},\Psi^{(1)}_{1,2,3,4}$. This establishes the recursion pattern, illustrating how $\SU(N)$ gauge group transforms into $\SU(N-2)$ gauge group and iterative applications of this procedure lead to lower-rank cases.}
\label{fig:SU-Sym-AS-even}
\end{figure}

\begin{table}[ht]
\centering
\begin{tabular}{c|ccc}
&$\SU(N)$&$\U(1)_{x^{(0)}}$&$\U(1)_{y^{(0)}}$  \\ \hline $X^{(0)}_{\alpha\beta}$
& $\yng(2)$ &$2$ &$0$\\ $Y^{(0)}_{\alpha\beta}$ &$\yng(1,1)$&$0$&$2$\\$\Psi^{(0)}_1$&$\mathbf{1}$&$-2N$&$0$\\$\Psi^{(0)}_2$&$\mathbf{1}$&$0$&$-N$\\$\Psi^{(0)}_3$&$\mathbf{1}$&$-4$&$-2N+4$\\$Z^{(0)}$&$\mathbf{1}$&$2$&$N-2$
\end{tabular}
\qquad
\begin{tabular}{c|ccc}
&$\SU(N-2)$&$\U(1)_{x^{(1)}}$&$\U(1)_{y^{(1)}}$  \\ \hline $X^{(1)}_{\alpha\beta}$
& $\yng(2)$ &$2$ &$0$\\ $Y^{(1)}_{\alpha\beta}$ &$\yng(1,1)$&$0$&$2$\\$\Psi^{(1)}_1$&$\mathbf{1}$&$-2(N-2)$&$0$\\$\Psi^{(1)}_2$&$\mathbf{1}$&$0$&$-(N-2)$\\$\Psi^{(1)}_3$&$\mathbf{1}$&$-4$&$-2(N-2)+4$\\$\Psi^{(1)}_4$&$\mathbf{1}$&$-2N+2$&$2$\\$Z^{(1)}$&$\mathbf{1}$&$2$&$(N-2)-2$\\$\widetilde{Z}^{(1)}$&$\mathbf{1}$&$2N-4$&$0$
\end{tabular}
\caption{We start from the theory in the left table, which includes the Sym chiral $X^{(0)}$ and the AS chiral $Y^{(0)}$ under $\SU(N)$, as well as three Fermi fields $\Psi^{(0)}_{1,2,3}$ and a chiral field $Z^{(0)}$. After performing the duality depicted in Fig.~\ref{fig:SU-Sym-AS-even}, we obtain the fields in the right table, which include the Sym chiral $X^{(1)}$ and the AS chiral $Y^{(1)}$ under $\SU(N-2)$, four Fermi fields $\Psi^{(1)}_{1,2,3,4}$, and two chiral fields $Z^{(1)}$ and $\widetilde{Z}^{(1)}$. 
The \(\U(1)_{x(i)}\) and \(\U(1)_{y(i)}\) charges are introduced for convenience, but they can also be rewritten in terms of \(\U(1)_{x(0)}\) and \(\U(1)_{y(0)}\) using the recursion relations:
$x^{(i+1)} = \frac{4}{N-2i-2} x^{(i)} + \frac{N-2i-4}{N-2i-2} y^{(i)}$ and
$y^{(i+1)} = \frac{N-2i+2}{N-2i-2} x^{(i)} - \frac{2}{N-2i-2} y^{(i)}$.}
\label{tab:SU(N)-1AS-1Sym-and-SU(N-2)-even}
\end{table}

With this definition of $B^{(i)}(n)$, it is straightforward to observe that 
\be
B^{(i)}(0) \propto \det Y^{(i)}, \quad B^{(i)}(N-2i) \propto \det X^{(i)}.
\ee
This allows us to identify part of the free field relations between the two dual theories:
\bea
\Psi^{(i)}_3 &= \Psi^{(i-1)}_1, \quad Z^{(i)} = \Tr[(X^{(i-1)})^{N/2}], \quad \widetilde{Z}^{(i)} = B^{(i-1)}(4).
\eea

At the final stage of the duality sequence, we obtain $\frac{N-2}{2}$ copies of $\SU(2) + 1\text{Sym}$, along with $N + 1$ Fermi multiplets, $N - 1$ chiral multiplets, and an $\SU(2) + 1\text{AS} + 1\text{Sym}$ theory, as illustrated in Fig.~\ref{fig:SU-Sym-AS-EG-even}. The corresponding superpotential takes the form:
\bea
\label{superpotentialSU21AS1Sym}
\mathcal{W} &= \Psi_1^{(\frac{N-2}{2})} \left( \det X^{(\frac{N-2}{2})} + \widetilde{Z}^{(\frac{N-2}{2})} \right) + \Psi^{(\frac{N-2}{2})}_2 \Pf Y^{(\frac{N-2}{2})} + \Psi^{(\frac{N-2}{2})}_3 \det X^{(\frac{N-2}{2})} \cr
&\quad + \sum^{1}_{i=0} \Psi^{(i)}_3 \left( (Z^{(i)})^2 + \Tr[(A^{(i)})^2] \right) + \sum^{\frac{N-2}{2}}_{i=2} \Psi^{(i)}_3 \left( \widetilde{Z}^{(i-1)} + (Z^{(i)})^2 + \Tr[(A^{(i)})^2] \right).
\eea

For $\SU(2)$, the antisymmetric representation is trivial. As discussed earlier, $\SU(2) + 1\text{Sym}$ is dual to a pair of chiral and Fermi fields. Therefore, the recursive duality takes the following form
\begin{align} \label{SU-even-Sym-AS}
\begin{tikzpicture}
\node () at (0,0)  {$\SU(N) + 1\mathrm{Sym} + 1\mathrm{AS}+ 3\textrm{ Fermi's}+ 1\textrm{ chiral}$};
\node () at (0,-0.5)  {with superpotential \eqref{SUN1AS1Sym-superpotential}};
 \node () at (7.5,0)  {$\Big(\frac{3N}{2}\Big) \textrm{ chirals } + \Big(\frac{3N}{2}+1\Big) \textrm{ Fermi's }$};
 \node () at (7.5,-0.5)  {with superpotential \eqref{superpotentialSU21AS1Sym}};
 \node[rotate=40] () at (1.8,-1.25) {$\Big\Downarrow$};
\node[rotate=-40] () at (6,-1.25) {$\Big\Uparrow$};
\node () at (4,-2.0)  {$\SU(N-2) + 1\mathrm{Sym} + 1\mathrm{AS}+ 6\textrm{ Fermi's}+ 4\textrm{ chirals}$};
\node () at (4,-2.5)  {with superpotential \eqref{SUN-1-1AS1Sym-superpotential}};
\end{tikzpicture}
\end{align} 

At each step of this iterative procedure, we neglect the fields \( \Psi^{(i)}_2 \) and \( P^{(i+1)} \), as they decouple from the theory due to the superpotential term \( \mathcal{W} = \Psi^{(i)}_2 P^{(i+1)} \). The matter content on the right side of the first row is obtained by applying this duality iteration to the final stage, as illustrated in Fig.~\ref{fig:SU-Sym-AS-EG-even}. 

For free Fermi and free chiral fields, we assign an R-charge of \( 0 \). For interacting Fermi and chiral fields, we assign R-charges of \( 1 \) and \( 0 \), respectively. With these assignments, we compute the central charge of the theory as:
\bea
c_L=2N-4,\qquad c_R=3N-6~.
\eea

The elliptic genus for even $N$ is given by:
\bea
\mathcal{I}=&\frac{\eta(q)^{3N-4}}{N!}\frac{\vartheta_{1}(x^{-2N})\vartheta_{1}(y^{-N})\vartheta_1(x^{-4}y^{4-2N})}{\vartheta_1(x^2y^{N-2})}\oint_{\mathrm{JK}}\frac{d\boldsymbol{a}}{2\pi i \boldsymbol{a}}\prod_{i=1}^{N}\frac{\prod_{j\ne i}^{}\vartheta_{1}(a_{i}/a_{j})}{\prod_{j\leq i}^{}\vartheta_{1}(x^{2}a_{i}a_{j})\cdot \prod_{j<i}^{}\vartheta_{1}(y^{2}a_{i}a_{j})}
\cr=&(-1)^{\frac{N}{2}+1}\frac{\vartheta_1(x^{2N})}{\eta(q)}\prod_{i=1}^{\frac{N-2}{2}}\frac{\vartheta_1(x^{4i+2}y^{2N-4i-2})}{\vartheta_1(x^{4i+4}y^{2N-4i-4})}
\eea

\begin{figure}[ht]
\centering
\includestandalone[width=0.9\textwidth]{figures/SU-Sym-AS-EG-even}
\caption{In the final stage of the iterative procedure, we have $\frac{N-2}{2}$ copies of SU(2)+1Sym, where $ A^{(i)} $ is the corresponding symmetric tensor with charges $(2, N-2i-2)$ under $ \U(1)_{x^{(i)}} \times \U(1)_{y^{(i)}} $. Since each SU(2)+1Sym is dual to a pair of Fermi and chiral fields, so these SU(2)+1Sym theories are dual to $\frac{N-2}{2}$ pairs of chiral and Fermi fields. Additionally, we have a total of $ N + 1 $ Fermi fields, denoted as $ \Psi_{3,4}^{(i)} $, and a total of $ N - 1 $ chiral fields, denoted as $ Z^{(i)} $ and $ \widetilde{Z}^{(i)} $. Finally, there is the last-stage SU(2)+1AS+1Sym, where $ Y^{(\frac{N-2}{2})} $ is the AS tensor and $ X^{(\frac{N-2}{2})} $ is the Sym tensor. As previously mentioned, the AS representation of $ \SU(2) $ is trivial, so this SU(2)+1AS+1Sym theory is dual to two chiral fields and one Fermi field. The corresponding final superpotential is given by \eqref{superpotentialSU21AS1Sym}}
\label{fig:SU-Sym-AS-EG-even}
\end{figure}

When $N$ is odd, we can start with the theory SU($N$)+1AS+1Sym, along with three Fermi's $\Psi_{1,2,3}$ and a free chiral $Z$. By applying a series of dualities as illustrated in Fig.~\ref{fig:SU-Sym-AS-odd}, we can transform it into the even case $(N-1)$, which corresponds to the theory SU($N-1$)+1AS+1Sym, along with three Fermi's $\Psi^{(0)}_{1,2,3}$ and one chiral $Z^{(0)}$. This allows us to use the aforementioned formulas and dualities for iterative derivation. The corresponding charges for these two theories are provided in Tab~\ref{tab:SUodd1AS1Sym-SUeven1AS1Sym}.   Through this duality, we can apply the iterative dualities illustrated in Fig.~\ref{fig:SU-Sym-AS-even}, ultimately resulting in $\frac{N-1}{2} + 3$ free Fermi's and $\frac{N-1}{2} +2$ free chirals.
\begin{figure}[ht]
\centering
\includestandalone[width=\textwidth]{figures/SU-Sym-AS-odd}
\caption{A chain of duality transformations relate SU($N$)+1Sym+1AS to SU($N-1$)+1Sym+1AS where $N$ is odd. 
In the final quiver, $Z^{(0)}$ and $P^{(0)}$ are chiral fields generated during the duality process, while $\Psi^{(0)}_2$ and $\Gamma^{(0)}$ are newly generated Fermions. The boxed components do not contribute to the elliptic genus. Furthermore, $\Psi_2$ couples to $P^{(0)}$, $Z$ is a free chiral field, and $\Gamma^{(0)}$ is a free Fermion.
}
\label{fig:SU-Sym-AS-odd}
\end{figure}

\begin{table}[ht]
\centering
\begin{tabular}{c|ccc}
&$\SU(N)$&$\U(1)_x$&$\U(1)_y$  \\ \hline $X_{\alpha\beta}$
& $\yng(2)$ &$2$ &$0$\\ $Y_{\alpha\beta}$ &$\yng(1,1)$&$0$&$2$\\$\Psi_1$&$\mathbf{1}$&$-2N$&$0$\\$\Psi_2$&$\mathbf{1}$&$-2$&$-2N+2$\\$\Psi_3$&$\mathbf{1}$&$-6$&$-2N+6$\\$Z$&$\mathbf{1}$&$4$&$2N-4$
\end{tabular}
\qquad
\begin{tabular}{c|ccc}
&$\SU(N-1)$&$\U(1)_{x^{(0)}}$&$\U(1)_{y^{(0)}}$  \\ \hline $X^{(0)}_{\alpha\beta}$
& $\yng(2)$ &$2$ &$0$\\ $Y^{(0)}_{\alpha\beta}$ &$\yng(1,1)$&$0$&$2$\\$\Psi^{(0)}_1$&$\mathbf{1}$&$-2(N-1)$&$0$\\$\Psi^{(0)}_2$&$\mathbf{1}$&$0$&$-(N-1)$\\$\Psi^{(0)}_3$&$\mathbf{1}$&$-4$&$-2(N-1)+4$\\$Z^{(0)}$&$\mathbf{1}$&$2$&$(N-1)-2$
\end{tabular}
\caption{The table on the left corresponds to the charge table of the theory SU($N$)+1AS+1Sym with three Fermi's $\Psi_{1,2,3}$ and one free chiral $Z$ field when $N$ is odd. The table on the right represents the theory obtained after applying a series of transformations as depicted in Fig.~\ref{fig:SU-Sym-AS-odd}, namely SU($N-1$)+1AS+1Sym with three Fermi's $\Psi^{(0)}_{1,2,3}$ and one chiral $Z^{(0)}$. Here, $\U(1)_{x^{(0)}}$ and $\U(1)_{y^{(0)}}$ are introduced for convenience, but they can be rewritten in terms of $\U(1)_x$ and $\U(1)_y$, with the charge relations given by $\mathsf x^{(0)} = \frac{3}{N-1}\mathsf x +\frac{N-3}{N-1}\mathsf y$ and $\mathsf y^{(0)} = \frac{N+2}{N-1}\mathsf x - \frac{2}{N-1}\mathsf y$.}
\label{tab:SUodd1AS1Sym-SUeven1AS1Sym}
\end{table}
The superpotential for the odd $N$ case is given by:
\bea
\cW=\Psi_1\det X+\Psi_2 B(2)+\Psi_3 B(3)
\eea
We identify some of the relations of the two sides as:
\bea
\Psi^{(0)}_1=\Psi_3,\quad\Psi^{(0)}_3=\Psi_1,\quad Z^{(0)}=\Tr[X^{N/2}]~.
\eea
The corresponding elliptic genus when $N$ is odd is:  
\bea
\mathcal{I}&=\frac{\eta(q)^{3N-4}}{N!}\frac{\vartheta(x^{-2N})\vartheta(x^{-2}y^{-2N+2})\vartheta(x^{-6}y^{-2N+6})}{\vartheta(x^4y^{2N-4})}\oint_{\mathrm{JK}}\frac{d\boldsymbol{a}}{2\pi i \boldsymbol{a}}\prod_{i=1}^{N}\frac{\prod_{j\ne i}^{}\vartheta_{1}(a_{i}/a_{j})}{\prod_{j\leq i}^{}\vartheta_{1}(x^{2}a_{i}a_{j})\cdot \prod_{j<i}^{}\vartheta_{1}(y^{2}a_{i}a_{j})}
\cr&=(-1)^{\frac{N+1}{2}}\frac{\vartheta(x^6y^{2N-6})}{\eta(q)}\prod_{i=1}^{\frac{N-3}{2}}\frac{\vartheta(x^{4i+4}y^{2N-4i-4})}{\vartheta(x^{4i+2}y^{2N-4i-2})}~.
\eea

\section{Conclusions}

This paper has uncovered new dualities in 2d $\mathcal{N}=(2,2)$ and $\mathcal{N}=(0,2)$ supersymmetric theories, expanding the known landscape of 2d dualities and showing their connections to 4d physics. Table~\ref{tab:summary} summarizes the 4d $\mathcal{N}=1$ and $\mathcal{N}=2$ theories considered in this work, along with the corresponding 2d dualities obtained via twisted compactification on $S^2$. Furthermore, we derive additional 2d dualities from these constructions.

\begin{table}[h]
\centering
\renewcommand{\arraystretch}{1.4}
\begin{tabular}{|c|c|}
\hline
\textbf{4d theories} & \textbf{2d dualities} \\
\hline
$\mathcal{N}=2$ Lagrangian class $\mathcal{S}$ theories & $\mathcal{N}=(2,2)$ dualities in \S\ref{sec:(2,2)} \\
\hline
$\mathcal{N}=1$ Seiberg dual theories & $\mathcal{N}=(0,2)$ duality in Fig.~\ref{fig:seiberg-SU-N1-N2-N3} \\
\hline
$\mathcal{N}=1$ Intriligator–Pouliot dual theories &\\
$\mathcal{N}=2$ Sp$(2N) + (2N+2)\mathbf{F}$ & $\mathcal{N}=(0,2)$ triality in Fig.~\ref{fig:seiberg-Sp} \\
$\mathcal{N}=2$ SU$(2N) + 1\mathbf{AS} + (2N+2)\mathbf{F}$ &\\
\hline
$\mathcal{N}=1$ Intriligator–Seiberg dual theories & \\
$\mathcal{N}=2$ SO$(N) + (N-2)\mathbf{F}$&$\mathcal{N}=(0,2)$ triality in Fig.~\ref{fig:seiberg-SO} \\
$\mathcal{N}=2$ SU$(N) + 1\mathbf{Sym} + (N-2)\mathbf{F}$ & \\
\hline
$\mathcal{N}=4$ SYM theory & $\mathcal{N}=(2,2)$ pure-YM/LG duality in \S\ref{sec:adj} \\
\hline
\end{tabular}
\caption{Summary of 4d $\mathcal{N}=1$ and $\mathcal{N}=2$ theories and their corresponding 2d $\mathcal{N}=(2,2)$ and $\mathcal{N}=(0,2)$ dualities obtained via twisted compactification on $S^2$. Each entry in the right column points to the relevant section or figure where the associated 2d duality is discussed.}
\label{tab:summary}
\end{table}

Specifically, we derive new 2d $\mathcal{N}=(0,2)$ dualities by the twisted compactification of 4d $\mathcal{N}=1$ dualities on $S^2$. We also show that the $(0,2)$ reduction of a 4d $\mathcal{N}=2$ SCFT yields 2d $\mathcal{N}=(0,2)$ gauge theory that is dual to a Landau-Ginzburg model—even when the parent theory does not belong to class $\mathcal{S}$. This indicates that Gauge/LG dualities with \(\mathcal{N}=(0,2)\) supersymmetry arising from \(\mathcal{N}=2\) SCFTs in 4d are more general than previously understood. As shown in \cite{Beem:2013sza}, every 4d $\mathcal{N}=2$ SCFT admits a chiral algebra structure derived from its Schur BPS sector. Building on the results of \cite{Dedushenko:2017osi,Eager:2019zrc}, it was proposed in \cite{Nawata:2023aoq} that the elliptic genus of the $(0,2)$ theory obtained from a Lagrangian class $\mathcal{S}$ theory of type $A$ can be expressed as a linear combination of characters of the corresponding chiral algebra. It would be intriguing to investigate whether such a correspondence between elliptic genera and chiral algebra characters extends more broadly to the $(0,2)$ reductions studied in this work.

While we have found various novel 2d dualities, this study certainly represents only a modest step into what is undoubtedly a vast and intricate domain, with many open questions remaining to be explored.  
One natural avenue for future investigation is to seek brane constructions that realize the new dualities and trialities presented in this work. Brane setups have proven to be a powerful tool in understanding 2d $(0,2)$ gauge theories, as demonstrated in \cite{Mohri:1997ef,Garcia-Compean:1998sla,Franco:2015tna,Franco:2015tya}.  Identifying brane realizations of these newly discovered dualities could provide a more systematic framework for understanding their structure and potential generalizations. 

Another promising direction is to explore connections between these new dualities and integrable structures. The interplay between 2d supersymmetric gauge theories and integrable systems has been studied in \cite{Benini:2014mia,Yagi:2015lha,de-la-Cruz-Moreno:2020xop}, and it would be interesting to investigate whether the dualities we have found correspond to known integrable models or lead to new integrable structures.

2d supersymmetric theories represent exceptionally fertile ground in theoretical physics.  Although intensive study has been devoted to 2d supersymmetric theories since the foundational work \cite{Witten:1993yc}, our understanding of both $\mathcal{N}=(2,2)$ and $\mathcal{N}=(0,2)$ theories remains limited. Despite their rich theoretical landscape, a vast ``\emph{Terra incognita}'' remains largely unexplored.  We believe that the following topics need further study:
\begin{itemize}  \setlength\itemsep{.05em}
\item Further discovery and analysis of $\mathcal{N}=(2,2)$ and $\mathcal{N}=(0,2)$ non-Abelian dualities
\item Supersymmetric enhancement from $\mathcal{N}=(0,2)$ to $\mathcal{N}=(2,2)$ theories
\item Construction and investigation of $\mathcal{N}=(2,2)$ and $\mathcal{N}=(0,2)$ non-Lagrangian theories
\item Supersymmetric boundary conditions in $\mathcal{N}=(2,2)$ and $\mathcal{N}=(0,2)$ theories
\end{itemize}
We conclude this short paper with the hope that motivated readers will further develop this field.

\acknowledgments
First and foremost, we would like to express our sincere gratitude to SciPost referees for their detailed and insightful comments, as well as for their patience in carefully reviewing our manuscript. Their invaluable feedback has greatly improved the quality of our paper.

We are grateful to Yiwen Pan for his collaboration on our previous work \cite{Nawata:2023aoq}, which laid the foundation for this paper. Our thanks extend to Runkai Tao for his contributions in the early stages of this project, and to Yiwen Pan and Yuji Tachikawa for their valuable comments on the draft. In addition, we would like to thank Wei Cui, Junkang Huang, Zixiao Huang, Du Pei, and Shutong Zhuang for their insightful discussions. In particular, we would like to extend special thanks to Mauricio Romo for his substantial contributions, illuminating discussions, and generous assistance throughout this work. 

This work is supported by the Shanghai Municipal Science and Technology Major Project (No. 24ZR1403900).

\appendix

\section{Conventions}\label{sec:conventions}
For symplectic groups, we use the notation $\operatorname{Sp}(N)$, where $N$ is twice the rank, ensuring that $N$ is always even. For various $\mathrm{SU}$, $\mathrm{Sp}$, and $\mathrm{SO}$ gauge theories, we frequently adopt the following concise notations for a representation of a chiral multiplet under a gauge group:
\begin{itemize}\setlength{\itemsep}{.5pt}
\item Fundamental representation: $\mathbf{F}$
\item Anti-fundamental representation: $\mathbf{AF}$
\item Adjoint representation: $\Adj$
\item anti-symmetric representation: $\AS$
\item Symmetric representation: $\Sym$
\end{itemize}

\bigskip

\noindent Additionally, all quiver diagrams for $\mathcal{N}=(0,2)$ theories follow the conventions outlined below,
\begin{itemize}
\item Gauge node for a gauge group $G$:  
$\begin{array}{l}
\includegraphics[width=1.3cm]{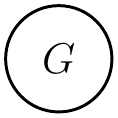}
\end{array}$

\item Flavor node corresponding to a flavor group $F$ with chemical potential $\mathsf{f}$:
$\begin{array}{l}
\includegraphics[width=1.2cm]{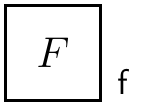}
\end{array}$

\item Chiral multiplet:  
\includegraphics[width=4cm]{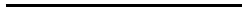}

\item Fermi multiplet:  
\includegraphics[width=4cm]{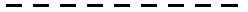}

\item In the case of the SU($N$) group, we  simply write $N$ in both flavor and gauge nodes.
\item For representations of $\SU(N)$ groups, an inward-pointing arrow indicates a multiplet in the fundamental representation $\mathbf{F}$, while an outward-pointing arrow denotes a multiplet in the anti-fundamental representation $\mathbf{AF}$. For example, $N_1\times N_2$ chiral mesons in the representation $\mathbf{\overline{N_1}}\otimes\mathbf{N_2}$ under the flavor group $\SU(N_1)\times \SU(N_2)$ are represented as:
$$ 
\includegraphics[width=5cm]{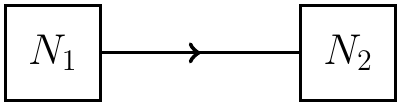}
$$
\item For clarity, the charges and chemical potentials of the U(1) flavor symmetries associated with various multiplets are sometimes indicated in the quiver diagram. For instance, the chiral multiplet carries a charge of 2 under the $\mathrm{U}(1)_{x}$ flavor symmetry is represented as follows:
$$
\includegraphics[width=4cm]{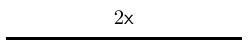}
 $$
Here, the fugacity $x$ is related to the chemical potential $\mathsf{x}$ by $x = e^{2\pi i \mathsf{x}}$.
\end{itemize}
\section{Fundamentals of 2d \texorpdfstring{$\cN=2$}{N=2} theories}\label{sec:fundamental}

In this appendix, we provide a concise review of 2d $\mathcal{N}=2$ supersymmetric theories relevant to this paper. To ensure that our discussion is self-contained, we present the necessary background and clarify the notation used in the main content and analyses of this paper. Specifically, we examine the anomaly coefficients in 2d $\mathcal{N}=(0,2)$ supersymmetric theories and explain the computation of elliptic genera for both $\mathcal{N}=(0,2)$ and $\mathcal{N}=(2,2)$ supersymmetric theories.

Given that $(0,2)$ theories are chiral, it is crucial to pay attention to anomalies. Consider a $(0,2)$ theory with a global symmetry $F$ described by a simple Lie algebra. The 't Hooft anomaly coefficient $k_F$ associated with this symmetry can be determined by
\begin{equation}\label{anomaly1}
\Tr \gamma_3 f^a f^b = k_F \delta^{ab}~,
\end{equation}
where $f^a$ are the generators of $F$, $\gamma_3$ is the gamma matrix that quantifies chirality, and the trace is taken over the Weyl Fermions in the theory.

In particular, the anomaly associated with the $\U(1)_R$-symmetry is related to the right-moving central charge $c_R$ as
\begin{equation}\label{cR}
c_R = 3\Tr(\gamma_3R^2)~.
\end{equation}
To determine the $\U(1)_R$-charges of various fields, $c$-extremization is performed \cite{Benini:2012cz,Benini:2013cda} if the theory meets the following two assumptions:
\begin{enumerate}\setlength{\itemsep}{1pt}
\item The theory is bounded, and the energy spectrum is bounded from below.
\item The vacuum moduli space is compact, and the ground state wavefunction is normalizable.
\end{enumerate}
If chiral multiplets parametrize compact directions, we need to apply the $c$-extremization on them. However, if there is no restriction on $R$-charges by a superpotential, the chiral fields spanning non-compact directions must have zero $R$-charge.

In fact, the vacuum moduli space in every $\cN=(0,2)$ theory considered in this paper is non-compact, and all the chiral multiplets span some non-compact directions, making $c$-extremization inapplicable. In such cases, the right-moving central charge is determined as three times the complex dimension of the moduli space, which serves as the target space of the non-linear sigma model in the infra-red limit \cite{Nawata:2023aoq}.

Once $c_R$ is determined, the left-moving central charge can be obtained from the gravitational anomaly, which is the difference between the number of chiral and Fermi multiplets
\begin{equation}\label{grav}
c_R - c_L = \Tr(\gamma_3)~.
\end{equation}

To study 2d supersymmetric theories, the elliptic genus is a powerful observable that can be evaluated from the UV description \cite{Benini:2013nda,Benini:2013xpa}. In this paper, we consider the elliptic genus in the Ramond sector defined by
\begin{align}
\cI^{(0,2)}(q,z) &= \Tr_{\mathrm{R}}(-1)^F q^{H_L} \overline{q}^{H_R} \prod_a z_a^{f_a},
\end{align}
where the left- and right-moving Hamiltonians are $2H_L = H + iP$ and $2H_R = H - iP$, respectively, in the Euclidean signature. In a superconformal theory, these operators correspond to the zero-mode generators $L_0$, $\overline{L}_0$ of the superconformal algebra. Due to supersymmetry, only the right-moving ground states ($H_R = 0$) contribute to the elliptic genus in the Ramond sector. Consequently, the elliptic genus is a holomorphic function of $q$.

Given a Lagrangian of $\cN=(0,2)$ theory, the computation of the elliptic genus is straightforward \cite{Benini:2013nda,Benini:2013xpa}. Let us review the contributions from different types of multiplets.
The contribution of an $\cN=(0,2)$ chiral multiplet in a representation $\lambda$ of the gauge and flavor group is
\be \label{chiral}
\cI_{\textrm{chi}}^{(0,2)}(q,z)=\prod_{w \in \lambda} i \frac{\eta(q)}{\vartheta_1(z^w)}~ .
\ee
where $w$ is a weight of the representation $\lambda$.
The contribution of an $\cN=(0,2)$ Fermi multiplet in a representation $\lambda$ of the gauge and flavor group is given by
\be \label{Fermi}
\cI_{\textrm{fer}}^{(0,2)}(q,z)=\prod_{w \in \lambda} i \frac{\vartheta_1(z^w)}{\eta(q)} ~ .
\ee
The contribution of an $\cN=(0,2)$ vector multiplet with gauge group $G$ is 
\be \label{Vector}
\cI_{\textrm{vec}}^{(0,2)}(q,z)= \frac{(-i\eta(q))^{2 \,\textrm{rk}G} }{\left|W_G\right|}\prod_{\alpha \in \Delta} i \frac{\vartheta_1(z^\alpha)}{\eta(q)}~.
\ee
where $\Delta$ is the set of roots of the gauge group $G$.
Then, the elliptic genus of an $\cN=(0,2)$ quiver gauge theory can be schematically expressed as the Jeffrey-Kirwan (JK) residue integral \cite{jeffrey1995localization,brion1999arrangement,szenes2003toric,Benini:2013xpa}
\begin{equation}\label{EG}
\cI^{(0,2)}=\oint_{\textrm{JK}} \prod_{\textrm{gauge}} \frac{d z}{2 \pi i z} \cI_{\mathrm{vec}}^{(0,2)}(q,z) \prod_{\textrm{matter}} \cI_{\textrm{chi}}^{(0,2)}(q,z) \cI_{\textrm{fer}}^{(0,2)}(q,z) .
\end{equation}

Note that the notations and conventions are the same as in \cite{Nawata:2023aoq}. The Dedekind eta function is defined as
\be \label{eta}
\eta(q)=q^{\frac{1}{24}} \prod_{n=1}^{\infty}(1-q^n)
\ee 
where $q=e^{2 \pi i \tau}$ and $\Im \tau>0$.
The Jacobi theta functions are defined by
\begin{align}
 \vartheta_1(z|q):&=-i \sum_{r \in \mathbb{Z}+\frac{1}{2}}(-1)^{r-\frac{1}{2}} z^{r} q^{\frac{r^2}{2}},\cr 
&=i q^{1 / 8} z^{1 / 2} \prod_{k=1}^{\infty}(1-q^k)(1-z^{-1} q^k)(1-zq^{k-1}) ~.
\end{align}
For the sake of brevity, the $q$ is often omitted, and we simply write $\vartheta_1(z)$.

\bigskip
For $\mathcal{N}=(2,2)$ theories, the elliptic genus in the Ramond-Ramond sector is given by
\begin{equation}
\mathcal{I}^{(2,2)}(q,y,z) = \Tr_{\mathrm{RR}}  (-1)^F q^{H_L} \bar{q}^{H_R} y^{-J} \prod_a z_a^{f_a} ~,
\end{equation}
where $J$ represents a left-moving $\mathrm{U}(1)_R$ symmetry.

To understand the contributions of $\cN=(2,2)$ multiplets, note that a $\mathcal{N}=(2,2)$ chiral multiplet can be decomposed into $\mathcal{N}=(0,2)$ chiral and Fermi multiplets. Consequently, the contribution of an $\mathcal{N}=(2,2)$ chiral multiplet with $R$-charge $r$ in a representation $\lambda$ of the gauge and flavor group is
\begin{equation}
\mathcal{I}_{\textrm{chi}}^{(2,2)}(q,y,z) = \prod_{w \in \lambda} \frac{\vartheta_1( y^{1-r/2} z^w)}{\vartheta_1( y^{-r/2} z^w)}~.
\end{equation}
In addition, an $\mathcal{N}=(2,2)$ vector multiplet consists of $\mathcal{N}=(0,2)$ vector and adjoint chiral multiplets. Thus, the contribution of an $\mathcal{N}=(2,2)$ vector multiplet with gauge group $G$ is
\begin{equation}
\mathcal{I}_{\textrm{vec}}^{(2,2)}(q,y,y,z) = \frac{1}{|W_G|}\left( \frac{\eta(q)^3}{\vartheta_1( y)} \right)^{\mathrm{rk} G} \prod_{\alpha \in \Delta} \frac{\vartheta_1( z^\alpha)}{\vartheta_1( y z^\alpha)}~,
\end{equation}
where $|W_G|$ is the order of the Weyl group associated with the gauge group $G$.

Similarly, performing the JK residue integral, we obtain the elliptic genus of an $\cN=(2,2)$ quiver gauge theory as 
\begin{equation}\label{EG2}
\cI^{(2,2)}=\oint_{\textrm{JK}} \prod_{\textrm{gauge}} \frac{d z}{2 \pi i z} \cI_{\mathrm{vec}}^{(2,2)}(q,y,z) \prod_{\textrm{matter}} \cI_{\textrm{chi}}^{(2,2)}(q,y,z) ~.
\end{equation}

In this paper, we extensively use elliptic genera to check new dualities and trialities. However, the identities of elliptic genera presented here remain conjectural and are not rigorously proven. Instead, we verify these conjectures by explicitly evaluating elliptic genera using JK integrals up to rank five and performing $q$-expansions up to $\mathcal{O}(q^2)$.

\section{Some details on 2d \texorpdfstring{$\cN=(0,2)$}{N=(0,2)} dualities}\label{app:details}
This appendix presents the detailed and technical aspects of the 2d $\mathcal{N} = (0,2)$ dualities discussed in \S\ref{sec:(0,2)-dual}. While the main text focuses on key results, some derivations, additional dualities, and further supporting evidence are presented here. Although these details are too technical for the main discussion, they are essential for an understanding of the underlying structure of $\mathcal{N} = (0,2)$ dualities.

\subsection{Twisted compactification of 4d Intriligator-Seiberg duality on \texorpdfstring{$S^2$}{S2}}\label{app:SO-compactification}

We consider the twisted compactification of 4d $\cN=1$ Intriligator-Seiberg duality \cite{Intriligator:1995id} on $S^2$, resulting in a 2d $(0,2)$ $\SO$ gauge theory dual to Landau-Ginzburg model. Our approach follows a method similar to that used in \cite{Gadde:2015wta}. Tab.~\ref{tab:4d-SO-seiberg-dual} summarizes the matter contents of the 4d $\cN=1$ Intriligator-Seiberg duality where the theory in the right has a superpotential
\be 
\cW=\hat Q^i \hat Q^j M_{ij}~.
\ee 

\begin{table}[ht]\centering
\bgroup
\def\arraystretch{1.5}
\begin{tabular}{|c|c|c|c|}
\hline & $\mathrm{SO}(N_{c})$& $\mathrm{SU}(N_{f})$   & $\mathrm{U}(1)_{R}$ \\ \hline
$Q$ & $\Box$ & $\Box$   & $\frac{N_f-N_c+2}{N_f}$ \\ \hline
\end{tabular}\qquad \qquad
\begin{tabular}{|c|c|c|c|}
\hline & $\mathrm{SO}(N_{f}-N_{c}+4)$& $\mathrm{SU}(N_{f})$   & $\mathrm{U}(1)_{R}$ \\ \hline
$\hat Q$ & $\Box$ & $\overline{\Box}$   & $\frac{N_c-2}{N_f}$  
\\ \hline
$M$ & $\mathbf{1}$ & $\mathbf{Sym}$   & $\frac{2(N_f-N_c+2)}{N_f}$\\ \hline  
\end{tabular}
\egroup
\caption{Matter contents of 4d $\mathcal{N}=1$ $\SO$ Intriligator-Seiberg duality}
\label{tab:4d-SO-seiberg-dual}
\end{table}

Upon compactification on $S^2$, we reassign non-negative integer $R$-charges $r_a$ to each fundamental chiral multiplet in the left theory and $\hat{r}_a$ to those in the right theory \cite{Closset:2013sxa}. In 4d, the $\U(1)_R\SO(N_c)\SO(N_c)$ mixed gauge anomaly can be calculated as
\begin{align}
\Tr RGG={T}_{\SO}(\Box)\sum_{a=1}^{N_f}(r_a-1)+{T}_{\SO}(\mathrm{adj})=0~.
\end{align}
This anomaly cancellation imposes the following constraint on the $R$-charges of the fundamental chirals:
\be 
\sum_{a=1}^{N_f}r_a=N_f-N_c+2~.
\ee 
Similarly, in the dual theory, the anomaly condition leads to:
\be \sum_{a=1}^{N_f}\hat{r}_a=N_c-2~.\ee

To satisfy these conditions, we assign $r_a = 1$ for $a = 1, \dots, N_f - N_c + 2$ and $r_a = 0$ for the remaining indices. The dual $R$-charges are then given by $\hat{r}_a = 1 - r_a$. Consequently, the $R$-charges of the mesons in the symmetric representation of the dual theory are determined as:
\be 
r(M_{ab}) = 2 - \hat{r}_a - \hat{r}_b~, \qquad a \leq b~.
\ee

Following \cite{Closset:2013sxa}, upon compactification on $S^2$, the 2d $(0,2)$ chiral multiplets arise from 4d fields with integral $R$-charges less than 1, while 2d $(0,2)$ Fermi multiplets originate from 4d fields with integral $R$-charges greater than 1. 4d fields with $R$-charge exactly equal to 1 do not contribute to the 2d theory.

\begin{figure}
\centering
\includegraphics[width=\linewidth]{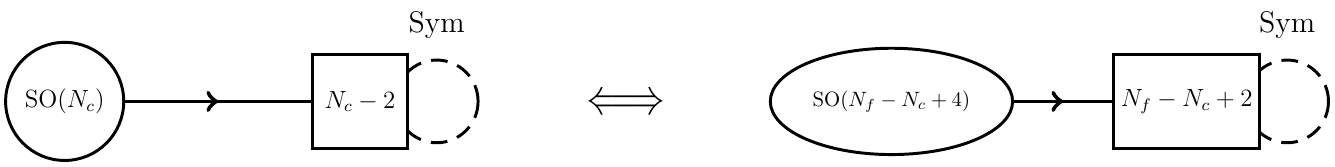}
\caption{A trivial 2d $\cN=(0,2)$ duality obtained from the compactification of 4d $\cN=1$ Intriligator-Seiberg duality on $S^2$.}
\label{fig:SO-reduction}
\end{figure}

Thus, we arrive at the conclusion that the 2d $(0,2)$ SO($N_c$) gauge theory with $N_c - 2$ fundamental chirals is dual to the $\SO(N_f - N_c + 4)$ gauge theory, which includes $N_f - N_c + 2$ fundamental chirals, $(N_c - 2)(N_c - 1)/2$ chiral mesons, and $(N_f - N_c + 2)(N_f - N_c + 3)/2$ Fermi mesons. After transferring the chiral mesons to the other side, we obtain the duality depicted in Fig.~\ref{fig:SO-reduction}.

It is evident from Fig.~\ref{fig:SO-reduction} that the dual theories are independent of the rank of the gauge groups. By evaluating the elliptic genus
   \bea
   \cI&=\frac{2 \eta(q)^{\frac{3N-3\chi-2}{2}}\prod_{i\leq j}\vartheta_{1}(x^2 b_i b_{j})}{2^{\lfloor{(N-1)/2}\rfloor}\lfloor{N/2}\rfloor!}
\oint_{\textrm{JK}}\frac{d\boldsymbol{a} }{2\pi  i \boldsymbol{a}} 
\frac{
\prod_{i<j}^{\lfloor{N/2}\rfloor}\vartheta_{1}(a_{i}^{\pm}a_{j}^{\pm})
}{
\prod_{i=1}^{\lfloor{N/2}\rfloor}\prod_{j=1}^{N-2}\vartheta_{1}(xa_i^{\pm} b_{j})
}\Bigg(\frac{\prod_{i=1}^{\lfloor{N/2}\rfloor}\vartheta_{1}(a_{i}^{\pm})}{\prod_{j=1}^{N-2}\vartheta_{1}(xb_{j})}\Bigg)^{\chi}
\cr
&= \frac{\vartheta_{1}(x^{N-2})}{\eta(q)}~,
\eea 
we conclude that the theory is dual to one free Fermi field. 

By reorganizing the field content, we obtain the 2d $(0,2)$ version of the Intriligator-Seiberg duality between Sym-1 and Sym-3 in Fig.~\ref{fig:seiberg-SO}.

\subsection{Triality with SU(\texorpdfstring{$N$}{N}) flavor symmetry for odd \texorpdfstring{$N$}{N}}\label{app:odd}

We have identified a new triality, as illustrated in Fig. \ref{fig:seiberg-Sp}, in which the flavor symmetry is SU($N$) with $N$ being even. Building on this discovery, we now present a closely related triality that arises when the flavor symmetry is SU($N$) with $N$ being odd. This new case exhibits analogous structures while introducing subtle differences in superpotentials due to the parity of $N$. We propose that the following three theories become equivalent in the infra-red.

\begin{enumerate}
\item[AS-1'.] $\mathrm{Sp}(N-1)$ gauge theory with $N$ fundamental chirals $P$, one fundamental chiral $Q$, and no superpotential.
\item[AS-2'.] $\mathrm{SU}(N-2)$ gauge theory with one anti-symmetric chiral $X$, $N$ fundamental chirals $Y$ and $N$ chiral mesons $Z$. Additionally, there is a neutral Fermi multiplet $\Psi$, forming a superpotential
\be 
\cW=\Psi( \epsilon^{\alpha_1 \cdots \alpha_{N-2}} X_{\alpha_1\alpha_2} \cdots X_{\alpha_{N-4} \alpha_{N-3}} Y_{\alpha_{N-2}}^i Z_i)~.
\ee 
\item[AS-3'.]  LG model of one Fermi $\Psi$, $N$ chiral mesons $Z$ and $\frac12N(N-1)$ chirals, forming an anti-symmetric $N \times N$ matrix $A$ with a superpotential
\be 
\cW=\Psi(\epsilon^{i_1\cdots i_N} A_{i_1 i_2}\cdots A_{i_{N-2} i_{N-1}} Z_{i_N}).
\ee 
\end{enumerate}

\begin{figure}[htb]
\centering
\includestandalone[width=0.8\textwidth]{figures/Sp-SU-AS-AS-odd}
\caption{A triality among three distinct theories: (AS-1') an $\mathrm{Sp}(N-1)$ gauge theory of SQCD type with $N+1$ fundamental chirals, (AS-2') an $\mathrm{SU}(N-2)$ gauge theory with one anti-symmetric, $N$ fundamental chirals and $N$ chiral mesons, coupled to a neutral Fermi multiplet $\Psi$, and (AS-3') a Landau-Ginzburg model consisting of $\frac12N(N-1)$ chiral multiplets and $N$ chiral mesons coupled to a Fermi multiplet $\Psi$.}
\label{fig:seiberg-Sp-2}
\end{figure}
The $\U(1)_x$ and the $\U(1)_y$ charges of the fields are given as follows:
\be 
\renewcommand{\arraystretch}{1.2}
\begin{tabular}{c|cccccccc}
& $P$& $Q$& $\Psi$ & $A$& $X$& $Y$ & $Z$\\ \hline
$\mathrm{U}(1)_x$   & $1$& $0$& $-N$   & $2$& $\frac{2N}{N-2}$   &$\frac{2}{N-2}$   &  $1$ \\
$\mathrm{U}(1)_y$   & $0$& $-1$  & $1$& $0$& $0$&$0$  &  $-1$ 
\end{tabular}
\ee

A straightforward verification shows that the central charges of these theories agree, which are given by
\be
c_{L}=(N+2)(N-1) \qquad c_R=\frac{3}{2}(N+2)(N-1)~.
\ee
Furthermore, we verify that their elliptic genera agree
\begin{equation}
\begin{aligned}
\mathcal{I}&=-\frac{\eta(q)^{(N^{2}+4N-5)/2}}{2^{\frac{N-1}{2}}\frac{N-1}{2}!}\oint_{\mathrm{JK}}\frac{d \boldsymbol{a}}{2\pi i \boldsymbol{a}}\prod_{i=1}^{\frac{N-1}{2}} \frac{\vartheta_{1}(a_{i}^{\pm2})\prod_{j<i}^{}\vartheta_{1}(a_{i}^{\pm }a_{j}^{\pm })}{\prod_{j=1}^{N}\vartheta_{1}(a_{i}^{\pm }b_{j}x)\cdot \vartheta_{1}(a_{i}^{\pm } y^{-1})}~,
\\
&=\frac{\eta(q)^{{\frac{N^{2}+7N-20}{2}}}\vartheta_{1}(x^{-N}y)}{(N-2)!~\prod_{i=1}^{N}\vartheta_{1}(b_{i}y^{-1}x)}\oint_{\mathrm{JK}}\frac{d\boldsymbol{a}}{2\pi i \boldsymbol{a}}\prod_{i=1}^{N-2}\frac{\prod_{j\ne i}^{}\vartheta_{1}(a_{i}/a_{j})}{\prod_{j<i}^{}\vartheta_{1}(a_{i}a_{j}x^{\frac{2N}{N-2}})\cdot \prod_{j=1}^{N}\vartheta_{1}(a_{i}b_{j}^{-1}x^{\frac{2}{N-2}})}~,
\\
&=\frac{\eta(q)^{(N^{2}+N-2)/2}}{\prod_{i<j}^{}\vartheta_{1}(x^{2}b_{i}b_{j})}\frac{\vartheta_{1}(x^{-N}y)}{\prod_{i=1}^{N}\vartheta_{1}(b_{i}y^{-1}x)}~.
\end{aligned}
\end{equation}

\subsection{SU gauge theory with one anti-symmetric and one fundamental chiral}\label{app:N=1} 

In \S\ref{sec:AS}, we studied the dualities of $\SU(n)$ gauge theories with one anti-symmetric chiral multiplet. For the duality in Fig.~\ref{fig:SU-AS-N-n-m} to hold, the number of fundamental chirals must satisfy $N \geq 2$, as required by condition \eqref{N-cond}. Here, we consider the special case where the number of fundamental chirals is reduced to one, i.e., $N = 1$.

As illustrated in Fig.~\ref{fig:SU AS with N=1}, a sequence of dualities leads to an equivalence between an $\SU(n)$ gauge theory and an $\Sp(n+m-3)$ gauge theory, where $m$ is either $0$ or $1$ such that $n + m$ is odd. The matter content and superpotentials for both theories are detailed below.

\begin{figure}[ht]
\centering
\includestandalone[width=\textwidth]{figures/02-SU-as-N1-N2-N2=1}
\caption{A duality between $\SU(n)$ and $\Sp(n+m-3)$ gauge theories, where $m$ is either $0$ or $1$ such that $n+m$ is odd. The transformation between Theory A and Theory B follows \ref{fig:seiberg-Sp} and \ref{fig:seiberg-SO}, as indicated. The chemical potential is given by $\mathsf{y} = n \mathsf{x}_2 - (n - 1) \mathsf{x}_1$.}
\label{fig:SU AS with N=1}
\end{figure}

\begin{itemize}
\item \textbf{Theory A:} an $\SU(n-1)$ gauge theory with the following matters:
\begin{itemize}
\item One anti-symmetric chiral $Y$
\item One fundamental chiral $X_2$
\item $n$ anti-fundamental chirals $X_1$
\item One free Fermi $\Psi$ that transforms as $(1-n,m)$ under the symmetries $\U(1)_0\times\U(1)_1$.
\item A Fermi meson $\Gamma$
\end{itemize}
The superpotential of the theory takes the following form:
\begin{equation}
\cW=\left\{\begin{aligned}
&\Psi\Pf Y+\Tr\Gamma X_1 X_2\quad\text{when $n$ is odd}\cr&\Psi\e^{\alpha_1\ldots\alpha_{n-1}}Y_{\alpha_1\alpha_2}\ldots Y_{\alpha_{n-3}\alpha_{n-2}}X_{2,\alpha_{n-1}}+\Tr\Gamma X_1 X_2\quad\text{when $n$ is even}
\end{aligned} \right.
\end{equation}

\item   \textbf{Theory B:} an $\Sp(n+m-3)$ gauge theory, where $m$ is either zero or one such that  $n+m$ is odd, with the following matters:
\begin{itemize}
\item One fundamental chiral $\widetilde X_1$ with charge $(\Box,\Box,1)$ under the flavor symmetries $\Sp(n+m-3)\times\SU(n)\times\U(1)_2$.
\item One fundamental chiral $\widetilde X_2$ with charge $(\Box,-m)$ under the symmetries $\Sp(n+m-3)\times\U(m)_1$.
\item One chiral meson $Z$ that charges $(\overline\Box,1-n,n-1)$ under $\SU(n)\times \U(1)_0\times\U(1)_2$.
\item One fundamental Fermi $\Lambda$ with charge $(\Box,n-1,-n)$ under $\Sp(n+m-3)\times \U(1)_0\times\U(1)_2$.
\item One Fermi meson $\widetilde \Gamma$ with charge $(\overline{\Box},1,-1)$ under $\SU(n)\times\U(m)_1\times\U(1)_2$
\item One single free Fermi $\widetilde \Psi$ with charge $(n-1,1,-n)$ under $\U(1)_0\times\U(1-m)_1\times\U(1)_2$
\end{itemize}
The superpotential of theory B is given by:
\begin{equation}
\cW=\Tr\widetilde{\Gamma}\widetilde{X}_1\widetilde{X}_2+\Tr\Lambda\widetilde{X}_1Z
\end{equation}
\end{itemize}

The central charges agree as
\begin{equation}
c_L=n^2-n-2, \qquad c_R=\frac{3}{2}(n^2-n-2)~.
\end{equation}
Moreover, the elliptic genera agree and are expressed as follows:
\begin{align}
\begin{split}
\mathcal{I}_A&=\frac{\eta(q)^{\frac{n^2+5n-14}{2}}}{(n-1)!}\oint_{\textrm{JK}}\frac{d\boldsymbol{a}}{2\pi i\boldsymbol{a}}~\prod_{i=1}^{n-1}\frac{\prod_{j\ne i}\vartheta_{1}(a_{i}/a_{j})}{\prod_{j<i}\vartheta_{1}(x_{0}^2a_{i}a_{j})\cdot \prod_{j=1}^{n}\vartheta_{1}(a_{i}^{-1}c_{j}x_2x_{0}^{-1})\cdot \vartheta_{1}(a_{i}x_{1}^{-1}x_{0})}
\\
&\qquad\times \vartheta_{1}(x_{0}^{1-n}x_{1}^{m})\prod_{i=1}^{n}\vartheta_{1}(c_{i}^{-1}x_{1}x_{2}^{-1})~,
\end{split}
\\
\begin{split}
\mathcal{I}_{{B}}&=\frac{\eta(q)^{N_B(N_B+6)/2}}{2^{\frac{N_{B}}{2}}\frac{N_{B}}{2}!}\oint_{\text{JK}}\frac{d\boldsymbol{g}}{2\pi i \boldsymbol{g}}~\prod_{i=1}^{\frac{N_{B}}{2}}\frac{\vartheta_{1}(g_{i}^{\pm2})\prod_{j<i}^{}\vartheta_{1}(g_{i}^{\pm}g_{j}^{\pm})\cdot\vartheta_{1}(g_{i}^{\pm}y^{-1})}{\prod_{j=1}^{n}\vartheta_{1}(x_{2}g_{i}^{\pm}c_{j}) \cdot \vartheta_{1}(g_{i}^{{\pm}}x_{1}^{-1})^{m}}
\\
&\qquad\times
\frac{\vartheta_{1}(y^{-1}x_{1})^{1-m}\prod_{i=1}^{n}\vartheta_{1}(x_{1}x_{2}^{-1}c_{i}^{-1})^{m}}{\eta(q)^{(m-1)(n-1)}\prod_{i=1}^{n}\vartheta_{1}(yc_{i}^{-1}x_{2}^{-1})}~,
\end{split}
\end{align}
where $N_B=n+m-3$ and $y=x_1^{-(n-1)}x_2^{n}$~.

\subsection{SU gauge theory with one anti-symmetric and no fundamental chiral}\label{app:N=0} 

Finally, we examine the case where there is no fundamental chiral multiplet, i.e., $N = 0$. This corresponds to an $\SU(n-2)$ gauge theory with one anti-symmetric chiral multiplet and $n$ anti-fundamental chirals. One approach to studying this case is to directly compute the elliptic genus using the JK residue integrals. Alternatively, we can derive the result through a sequence of known dualities and operations.

In this case, we consider transformations that provide the identities of the elliptic genus, focusing only on the field contents while ignoring detailed aspects of the theory, such as the superpotential. As before, this depends on whether $n$ is even or odd: we apply the duality in Fig.~\ref{fig:seiberg-Sp} or Fig.~\ref{fig:seiberg-Sp-2} to the anti-symmetric chiral multiplet, followed by a sequence of additional dualities and transformations.

\begin{figure}[ht]
\centering
\includestandalone[width=0.95\textwidth]{figures/SU-AS-AF-even}
\caption{A sequence of transformations applied to the $\SU(n-2)$ gauge theory with one anti-symmetric chiral and $n$ anti-fundamental chirals where $n$ is even. In the transition from the third to the fourth diagram, the $\Sp(n-4)$ and $\SU(2)$ gauge nodes are merged into a single $\Sp(n-2)$ node. This merging introduces an additional factor of $(n-2)/2 = |W(\Sp(n-2))| / |W(\Sp(n-4)) \times W(\SU(2))|$, reflecting the ratio of Weyl group orders in the elliptic genus. 
Additionally, at this step, the direction of the arrow between nodes $2$ and $n$ is reversed from $(2 \leftarrow n)$ to $(2 \rightarrow n)$, utilizing the identity $\vartheta_1(z) = -\vartheta_1(1/z)$. 
 For the 4th and 5th quiver diagrams, we consider a disjoint union of $(n-2)/2$ copies of the theory. Through these transformations, we establish an identity between the elliptic genera at each step of the sequence.}
\label{fig:SU AS with N=0-even}
\end{figure}

Let us consider the case of even $n$, and Fig. \ref{fig:SU AS with N=0-even} illustrates a series of transformations that provides the identity of elliptic genera. Based on this identity, we propose the duality between the following two theories:
\begin{itemize}
\item An $\mathrm{SU}(n-2)$ gauge theory with one anti-symmetric chiral $X$, $n$ anti-fundamental chirals $Y$, and a neutral Fermi multiplet $\Psi$. The superpotential is given by:
\begin{equation}
\cW= \Psi\Pf X~.
\end{equation}
\item A disjoint union of $\frac{n-2}{2}$ copies of a theory with $\frac{1}{2}n(n-1)$ chirals, forming an anti-symmetric $n\times n$ matrix $A$, and a Fermi multiplet $\Psi$, with the superpotential:
\begin{equation}
\cW= \Psi \Pf A~.
\end{equation}
\end{itemize}

The $\U(1)_x$ charges of the fields are summarized as follows,
\be 
\renewcommand{\arraystretch}{1.2}
\begin{tabular}{c|cccccc}
& $X$& $Y$   & $\Psi$   & $A$     \\ \hline
$\mathrm{U}(1)_x$   & $\frac{2n}{n-2}$ & $-\frac{2}{n-2}$& $-n$  & $2$  \\
\end{tabular}~.
\ee
The elliptic genera are given by
\begin{equation}
\begin{aligned}
\mathcal{I} &= -\frac{\eta(q)^{\frac{n^{2}+5n-20}{2}}\vartheta_1(x^{-n})}{(n-2)!}
\oint_{\text{JK}} \frac{d\boldsymbol{a}}{2\pi i \boldsymbol{a}} 
\prod_{i=1}^{n-2} \frac{\prod_{j\ne i} \vartheta_{1}(a_{i}/a_{j})}{\prod_{j<i} \vartheta_{1}(x^{\frac{2n}{n-2}}a_{i}a_{j}) \cdot \prod_{j=1}^{n} \vartheta_{1}(a_{i}^{-1}b_{j}x^{-\frac{2}{n-2}})}
\\
&= \frac{n-2}{2}\frac{\vartheta_1(x^{-n})}{\eta(q)}\prod_{i<j} \frac{\eta(q)}{\vartheta_{1}(x^{2}b_{i}b_{j})}~.
\end{aligned}
\end{equation}

\begin{figure}[ht]
\centering
\includestandalone[width=0.95\textwidth]{figures/SU-AS-AF-odd}
\caption{A sequence of transformations applied to the $\SU(n-2)$ gauge theory with one anti-symmetric chiral and $n$ anti-fundamental chirals where $n$ is odd. In the transition from the third to the fourth diagram, the $\Sp(n-3)$ and $\SU(2)$ gauge nodes are merged into a single $\Sp(n-1)$ node. This merging introduces an additional factor of $(n-1)/2 = |W(\Sp(n-1))| / |W(\Sp(n-3)) \times W(\SU(2))|$, reflecting the ratio of Weyl group orders in the elliptic genus. 
Additionally, at this step, the direction of the arrow between nodes $2$ and $n$ is reversed from $(2 \leftarrow n)$ to $(2 \rightarrow n)$, utilizing the identity $\vartheta_1(z) = -\vartheta_1(1/z)$. 
 For the 4th and 5th quiver diagrams, we consider a disjoint union of $(n-1)/2$ copies of the theory. Through these transformations, we establish an identity between the elliptic genera at each step of the sequence.}
\label{fig:SU AS with N=0-odd}
\end{figure}

Now we turn to the case of odd $n$, and Fig.~\ref{fig:SU AS with N=0-odd} illustrates a sequence of transformations that provide the identities of the elliptic genus. Based on this, we propose a duality between the following two theories
\begin{itemize}
\item An $\mathrm{SU}(n-2)$ gauge theory with one anti-symmetric chiral $X$, $n$ anti-fundamental chirals $Y$, and one Fermi multiplet $\Psi$. 
The superpotential is given by:
\begin{equation}
\cW= \Psi  \det X ~.
\end{equation}
\item A disjoint union of $\frac{n-1}{2}$ copies of a theory with $\frac{n(n-1)}{2}$ chiral multiplets $A$, and one Fermi multiplet $\Psi$. The superpotential is given by:
\begin{equation}
\cW= \Psi \det A~.
\end{equation}
\end{itemize}
The $\mathrm{U}(1)_x$ and $\mathrm{U}(1)_d$ charges of the fields are summarized as follows:
\be 
\renewcommand{\arraystretch}{1.2}
\begin{tabular}{c|ccccccccc} 
&$\Psi$ &   $X$ & $Y$ & $A$  \\
\hline $\mathrm{U}(1)_x$ &$-2n$  & $\frac{2 n}{n-2}$ & $-\frac{2}{n-2}$ & $2$  
\end{tabular}
\ee
Therefore, under the duality, there is a correspondence between the operator $\Psi  (\det B)^2$ and $\Psi$.
The elliptic genera are given by:
\begin{equation}
\begin{aligned}
\mathcal{I} &  =-\frac{\eta(q)^{{\frac{n^{2}+5n-20}{2}}}\vartheta_{1}(x^{-2n})}{(n-2)!}\oint_{\mathrm{JK}}\frac{d\boldsymbol{a}}{2\pi i \boldsymbol{a}}\prod_{i=1}^{n-2}\frac{\prod_{j\ne i}^{}\vartheta_{1}(a_{i}/a_{j})}{\prod_{j<i}^{}\vartheta_{1}(a_{i}a_{j}x^{\frac{2n}{n-2}})\cdot \prod_{j=1}^{n}\vartheta_{1}(a_{i}^{-1}b_{j}x^{-\frac{2}{n-2}})}~,
\\
&=\frac{n-1}{2}\frac{\vartheta_{1}(x^{-2n})}{\eta(q)}\prod_{i<j}^{}\frac{\eta(q)^{}}{\vartheta_{1}(x^{2}b_{i}b_{j})}~.
\end{aligned}
\end{equation}

\bibliographystyle{JHEP}
\bibliography{references}

\end{document}

%% file: figures/SUNc-N1-N2-N3-1.tex
\begin{tikzpicture}[scale=1]
    \tikzset{
        cir/.style={circle, draw, minimum size=1.9cm, text centered, font=\Large,line width=0.5mm},
        point/.style={circle, fill, inner sep=1.5pt},
        box/.style={draw, minimum width=1.6cm, minimum height=1.6cm, text centered,line width=0.5mm,font=\Large},
        ->-/.style={decoration={
        markings,mark=at position #1 with {\arrow[scale=1.5]{>}}},postaction={decorate},line width=0.5mm},
        -<-/.style={decoration={
        markings,
        mark=at position #1 with {\arrow[scale=1.5]{<}}},postaction={decorate},line width=0.5mm},
        elli/.style={ellipse,draw, minimum width=2.5cm, minimum height=2cm,line width=0.5mm,font=\Large},
        dashed/.style={dash pattern=on 8pt off 5pt,line width=0.5mm}    
    }

\node[box] (N3) at (0,4)  {$N_{3}$};
\node[cir] (SUN) at (0,0) {$n_3$};
\node[box] (N2)   at (3,-3.5)  {$N_{2}$};
\node[box] (N1)   at (-3,-3.5)  {$N_1$};
%AF
\draw[-<-=.5]  (N2.north) -- (SUN.-35) node[midway, xshift=-20pt, yshift=-5pt]{\Large$X_{2}$};
%Fund
\draw[-<-=.5]  (SUN.-145) -- (N1.north) node[midway, xshift=20pt, yshift=-5pt]{\Large$X_{1}$};
%Fermi
\draw[-<-=.6,dashed] (N3.south) -- (SUN.north) node[midway, xshift=20pt, yshift=-5pt]{\Large$\Psi$};
%meson
\draw[->-=.5] (N3.-115) -- (N1.115) node[midway,xshift=-20pt,yshift=5pt]{\Large$M$};

\node[rotate=90] (arr) at (5,0) {\scalebox{2}{$\Big\Updownarrow$}};

\begin{scope}[xshift=10cm]
\node[box] (N3) at (0,4)  {$N_{3}$};
\node[cir] (SUN) at (0,0) {$n_2$};
\node[box] (N2)   at (3,-3.5)  {$N_{2}$};
\node[box] (N1)   at (-3,-3.5)  {$N_1$};
%F
\draw[->-=.6] (N3.south) -- (SUN.north) node[midway, xshift=20pt, yshift=-5pt]{\Large$\widetilde{X}_{2}$};
%AF
\draw[->-=.5]  (SUN.-145) -- (N1.north) node[midway, xshift=20pt, yshift=-5pt]{\Large$\widetilde{X}_{1}$};
%Fermi
\draw[->-=.5,dashed] (N2.north) -- (SUN.-35) node[midway, xshift=-20pt, yshift=-5pt]{\Large$\widetilde{\Psi}$};
%Meson
\draw[->-=.5] (N1.east) -- (N2.west) node[midway,yshift=-20pt]{\Large$\widetilde{M}$};

\end{scope}

%\draw[->, line width=0.5mm] (4.5,0) -- (5.5,0) node[midway,above] 

\end{tikzpicture}

%% file: figures/Sp-SU-AS-AS-even.tex
\begin{tikzpicture}[scale=1]
    \tikzset{
        %>=latex
        cir/.style={circle, draw, minimum size=1.8cm, text centered, font=\Large,line width=0.5mm},
        point/.style={circle, fill, inner sep=1.5pt},
        box/.style={draw, minimum width=1.6cm, minimum height=1.6cm, text centered,line width=0.5mm,font=\Large},
        ->-/.style={decoration={
        markings,mark=at position #1 with {\arrow[scale=1.5]{>}}},postaction={decorate},line width=0.5mm},
        -<-/.style={decoration={
        markings,
        mark=at position #1 with {\arrow[scale=1.5]{<}}},postaction={decorate},line width=0.5mm},
        dashed/.style={dash pattern=on 8pt off 5pt,line width=0.5mm},
        elli/.style={ellipse,draw, minimum width=2.5cm, minimum height=2cm,line width=0.5mm,font=\Large
    }}

% Sp with F
    \node[fill=blue!10!,scale=1.7]  at (2.7,2) {AS-1};
    \node[cir] (gauge) at (0,0) {\normalsize$\mathrm{Sp}(2N)$};
    \node[box] (flavor) at (5,0) {$2N$+$2$};
    \draw[line width=0.5mm] (flavor.west) -- (gauge.east);
    \node[rotate=90] () at (7.5,0) {\scalebox{2}{$\Big\Updownarrow$}};

% AS
\begin{scope}[xshift=10.5cm]
    \node[fill=blue!10!,scale=1.7]  at (2.7,2) {AS-3};
        \node[box] (UN) at (0,0) {};
        \draw[line width=0.5mm]   ($(UN.north)+(0,0.4)$) circle(0.7cm); 
        \node[box,fill=white] () at (UN) {$2N$+$2$} node[yshift=2.4cm]{\Large AS};
        \node[font=\Large,scale=1.3] (Gamma) at ($(UN)+(2,0)$)  {$+\Psi$};
\end{scope}
% SU with AS and F
\begin{scope}[yshift=-4cm,xshift=4cm]
    \node[fill=blue!10!,scale=1.7]  at (2.5,-1.5) {AS-2};
\node (gauge) at (0,0) {};
\node[box] (flavor) at (5,0) {$2N$+$2$};
\draw[->-=.5] (flavor.west) -- (gauge.east);
\draw[line width=0.5mm] ($(gauge)-(1.3,0)$) circle (0.7);
\node[cir,fill=white,scale=0.9] () at (gauge) {$2N$};
\node[font=\Large] (AS) at ($(gauge)-(2.6,0)$) {AS};
\node[font=\Large,scale=1.3] (Gamma) at ($(gauge)+(7,0)$) {$+\Psi$};
\node[rotate=30] () at (-0.8,2) {\scalebox{2}{$\Big\Updownarrow$}};
\node[rotate=-30] () at (5.5,2) {\scalebox{2}{$\Big\Updownarrow$}};
\end{scope}
\end{tikzpicture}

%% file: figures/Sp-ASodd.tex
\begin{tikzpicture}[scale=1]
    \tikzset{
        cir/.style={circle, draw, minimum size=2.2cm, text centered, font=\large,line width=0.5mm},
        point/.style={circle, fill, inner sep=1.5pt},
        box/.style={draw, minimum width=1.6cm, minimum height=1.6cm, text centered,line width=0.5mm,font=\large},
        ->-/.style={decoration={
        markings,mark=at position #1 with {\arrow[scale=1.5]{>}}},postaction={decorate},line width=0.5mm},
        -<-/.style={decoration={
        markings,
        mark=at position #1 with {\arrow[scale=1.5]{<}}},postaction={decorate},line width=0.5mm},
        dashed/.style={dash pattern=on 8pt off 5pt,line width=0.5mm},
        elli/.style={ellipse,draw, minimum width=2.5cm, minimum height=2cm,line width=0.5mm,font=\Large}
    }
    \node[cir,scale=0.9] (SpNm3) at (0,0) {$\mathrm{Sp}(N-3)$};
    \node[box]  (UN)    at (-1.5,-3) {\Large $N$};
    \node[box]  (U1)    at (1.5,-3) {\Large $\mathrm{U}(1)$};
        \node[scale=1.3]    at  (2.8,-3.6)   {$N\mathsf{x}$};
    \draw[->-=0.5] (SpNm3.-135) -- (UN.north) node [midway,left=2pt,scale=1.5] {$\mathsf{x}$};
    \draw[-<-=0.5,dashed] (SpNm3.-45) -- (U1.north);
    \draw[->-=0.5] (UN.east) -- (U1.west) node[below,midway,scale=1.5] {$-\mathsf{x}$};
    \node[rotate=90] () at (4.3,-2.5) {\scalebox{2}{$\Big\Updownarrow$}};

    \begin{scope}[xshift=7cm,yshift=-2cm]
 
        \node[box] (UN) at (0,-1) {\Large $N$};
        \node[font=\Large] at (1.1,0.5) {$2\mathsf{x}$};
        \draw[line width=0.5mm] (UN.120) arc[start angle=135+90,end angle=45-90, radius=0.7cm] node[yshift=0.5cm,midway]{\Large AS};
    \end{scope}
\end{tikzpicture}

%% file: figures/SO-SU-sym.tex
\begin{tikzpicture}[scale=1]
    \tikzset{
        %>=latex
        cir/.style={circle, draw, minimum size=1.8cm, text centered, font=\Large,line width=0.5mm},
        point/.style={circle, fill, inner sep=1.5pt},
        box/.style={draw, minimum width=1.6cm, minimum height=1.6cm, text centered,line width=0.5mm,font=\Large},
        ->-/.style={decoration={
        markings,mark=at position #1 with {\arrow[scale=1.5]{>}}},postaction={decorate},line width=0.5mm},
        -<-/.style={decoration={
        markings,
        mark=at position #1 with {\arrow[scale=1.5]{<}}},postaction={decorate},line width=0.5mm},
        dashed/.style={dash pattern=on 8pt off 5pt,line width=0.5mm},
        elli/.style={ellipse,draw, minimum width=2.5cm, minimum height=2cm,line width=0.5mm,font=\Large
    }}

% SO with F
    \node[fill=blue!10!,scale=1.7]  at (2,2) {Sym-1};
    \node[cir] (gauge) at (0,0) {\large$\mathrm{SO}(N)$};
    \node[box] (flavor) at (5,0) {$N-2$};
    \draw[line width=0.5mm] (flavor.west) -- (gauge.east);
    \node[rotate=90] () at (7.5,0) {\scalebox{2}{$\Big\Updownarrow$}};
    \node[font=\Large,scale=1.3] (Gamma) at ($(gauge)-(2.5,0)$)  {$X+\Psi+$};
% Sym
\begin{scope}[xshift=10.5cm]
    \node[fill=blue!10!,scale=1.7]  at (2.5,2) {Sym-3};
        \node[box] (UN) at (0,0) {};
        \draw[line width=0.5mm]   ($(UN.north)+(0,0.4)$) circle(0.7cm); 
        \node[box,fill=white] () at (UN) {$N-2$} node[yshift=2.4cm]{\Large Sym};
        \node[font=\Large,scale=1.3] (Gamma) at ($(UN)+(2,0)$)  {$+\Psi$};
\end{scope}
% SU with Sym and AF
\begin{scope}[yshift=-4cm,xshift=4cm]
    \node[fill=blue!10!,scale=1.7]  at (2.5,-1.5) {Sym-2};
\node[cir] (gauge) at (0,0) {};
\node[box] (flavor) at (5,0) {$N-2$};
\draw[-<-=.5] (flavor.west) -- (gauge.east);
\draw[line width=0.5mm] ($(gauge)-(1.3,0)$) circle (0.8);
\node[cir,fill=white] () at (gauge) {$N$};
\node[font=\Large] (AS) at ($(gauge)-(2.8,0)$) {Sym};
\node[font=\Large,scale=1.3] (Gamma) at ($(gauge)+(7,0)$) {$+\Psi$};
\node[rotate=30] () at (-0.8,2) {\scalebox{2}{$\Big\Updownarrow$}};
\node[rotate=-30] () at (5.5,2) {\scalebox{2}{$\Big\Updownarrow$}};
\end{scope}
\end{tikzpicture}

%% file: figures/SU2-SU2k-N1-N2-derivation.tex
\begin{tikzpicture}[scale=1]
    \tikzset{
        cir/.style={circle, draw, minimum size=2cm, text centered, font=\Large,line width=0.5mm},
        point/.style={circle, fill, inner sep=1.5pt},
        box/.style={draw, minimum width=1.6cm, minimum height=1.6cm, text centered,line width=0.5mm,font=\Large},
        ->-/.style={decoration={
        markings,mark=at position #1 with {\arrow[scale=1.5]{>}}},postaction={decorate},line width=0.5mm},
        -<-/.style={decoration={
        markings,
        mark=at position #1 with {\arrow[scale=1.5]{<}}},postaction={decorate},line width=0.5mm},
        dashed/.style={dash pattern=on 8pt off 5pt,line width=0.5mm},
        elli/.style={ellipse,draw, minimum width=2.5cm, minimum height=2cm,line width=0.5mm,font=\Large
    }}

    \begin{scope}[yshift=-1.5cm,xshift=1.55cm]
    \node[cir] (SUNm2) at (0,0) {};
    \node[box] (N1) at ($(SUNm2.east)+(4,-1.7)$) {$N_{1}$};
    \node[box] (N2) at ($(SUNm2.east)+(4,1.7)$) {$N_{2}$};
        \draw[line width=0.5mm] ($(SUNm2.west)+(-0.3cm,0)$) circle(0.7cm) node[xshift=-1.2cm]{\Large AS} node[yshift=1.2cm]{};
    \node[cir,fill=white] () at (SUNm2) {$N_{1}$-$N_{2}$-$2$};
    \node[] (arr) at  (-3,-3) {\scalebox{2}{$\Big\Updownarrow$}};
    \node[xshift=-1cm] () at (arr) { \ref{fig:seiberg-Sp}};
    \draw[->-=0.5] (N1.west) -- (SUNm2) ;
    \draw[-<-=0.5,dashed] (N2.west) -- (SUNm2) ;
    \node[scale=2] (Psi) at (8,0) {$+\Psi$};
    \end{scope}

    \begin{scope}[yshift=-8cm]
    \coordinate (size) at (4,0);
    \node[cir] (SU2) at (0,0) {};
    \node[cir] (SU4) at (size) {$4$} node[yshift=-1.3cm]{};%4
    \node[cir,fill=white] () at (SU2) {$2$} node[yshift=-1.3cm]{};%5
    \draw[-<-=.5] (SU2.east) -- (SU4.west);
    \coordinate (lend) at ($(SU4.east)+0.5*(size)-(1,0)$);
    \draw[-<-=.5] (SU4.east) -- (lend) ;
    \node[] (dots) at ($(lend)+(0.3,0)$) {$\ldots $};
    \node[cir] (SUm4) at ($(dots.east)+0.5*(size)$) {$N_{1}$-$N_{2}$-$4$} node[yshift=-1.3cm]{};%6
    \draw[->-=.5] (SUm4.west) -- (dots.east);
    \node[cir] (SUm2) at ($(SUm4)+(size)$) {$N_{1}$-$N_{2}$-$2$} node[yshift=-1.3cm]{};%7
    \node[box] (N1) at ($(SUm2)+(size)+(0,-1.7)$) {$N_{1}$};
    \node[box] (N2) at ($(SUm2)+(size)+(0,1.7)$) {$N_{2}$};
    \node[scale=2] (Psi) at (19.5,0) {$+\Psi+X$};

    \draw[->-=0.5] (SUm2.west) -- (SUm4.east) node[midway,yshift=0.8cm]{};
    \draw[->-=0.5] (N1.west) -- (SUm2) ;
    \draw[-<-=0.5,dashed] (N2.west) -- (SUm2) ;
    \node[] (arr) at  (-1.5,-2.4) {\scalebox{2}{$\Big\Updownarrow$}};
    \node[xshift=-1cm] () at (arr) { \ref{fig:seiberg-SU-N1-N2-N3}};
    \end{scope}

    %figure2
    \begin{scope}[yshift=-13cm]
    \coordinate (size) at (4,0);
    \node[cir] (SU2) at (0,0) {};
    \node[cir] (SU4) at (size) {$4$} node[yshift=-1.3cm]{}; 
    \node[cir,fill=white] () at (SU2) {$2$} node[yshift=-1.3cm]{};
    \draw[-<-=.5] (SU2.east) -- (SU4.west);
    \coordinate (lend) at ($(SU4.east)+0.5*(size)-(1,0)$);
    \draw[-<-=.5] (SU4.east) -- (lend) ;
    \node[] (dots) at ($(lend)+(0.3,0)$) {$\ldots $};
    \node[cir] (SUm4) at ($(dots.east)+0.5*(size)$) {$N_{1}$-$N_{2}$-$4$} ;
    \draw[->-=.5] (SUm4.west) -- (dots.east);
    \node[cir] (SUp2) at ($(SUm4)+(size)$) {$N_{2}+2$} ;
    \node[box] (N2) at ($(SUp2)+(size)$) {$N_{2}$};
    \node[box] (N1) at ($(SUp2)+(0,-3.5)$) {$N_{1}$};
    \draw[-<-=0.5,dashed] (SUp2) -- (SUm4) ;
    \draw[-<-=0.5] (N1) -- (SUp2);
    \draw[-<-=0.5] (SUm4) to (N1.west);
    \draw[-<-=0.5] (SUp2) to (N2);
    \draw[->-=0.5,dashed] (N1.east) -- (N2);
    \node[] (arr) at  (-1.5,-3.4) {\scalebox{2}{$\Big\Updownarrow$}};
    \node[xshift=-1cm] () at (arr) { \ref{fig:seiberg-SU-N1-N2-N3}};
    \node[scale=2] (Psi) at (19.5,-1.75) {$+\Psi+X$};
   \end{scope}

    %figure3
    \begin{scope}[yshift=-20cm]
    \node[box] (N1) at (0,-3.5) {$N_{1}$};
    \node[cir] (N1m2) at (0,0) {$N_{1}-2$};
    \node[cir] (N1m4) at (3.5,0) {$N_{1}-4$};
    \node[] (dots) at (6,0) {$\ldots $};
    \node[cir] (N2p4) at (8.5,0) {$N_{2}+4$};
    \node[cir] (N2p2) at (12,0) {$N_{2}+2$};
    \node[box] (N2) at (12,-3.5) {$N_{2}$};
    \draw[-<-=0.5] (N1)  --  (N1m2) node[midway,xshift=1.3cm] {};
    \draw[-<-=0.5] (N1m2) -- (N1m4) node[midway,yshift=1.5cm] {};
    \draw[-<-=0.5] (N1m4) -- (dots);
    \draw[-<-=0.5] (dots) -- (N2p4);
    \draw[-<-=0.5] (N2p4) -- (N2p2) node[midway,yshift=1.5cm] {};
    \draw[-<-=0.5] (N2p2) -- (N2) node[midway,xshift=-1.2cm] {};
    \draw[-<-=0.5,dashed] (N2) -- (N1) node[midway,yshift=-0.7cm] {};
    \node[scale=2] (Psi) at (15.5,-1.75) {$+\Psi+X$};
    \end{scope}
\end{tikzpicture}

%% file: figures/Sp-SU-linear.tex
\begin{tikzpicture}[scale=1]
    \tikzset{
        %>=latex
        cir/.style={circle, draw, minimum size=1.8cm, text centered, font=\large,line width=0.5mm},
        point/.style={circle, fill, inner sep=1.5pt},
        box/.style={draw, minimum width=1.6cm, minimum height=1.6cm, text centered,line width=0.5mm,font=\large},
        ->-/.style={decoration={
        markings,mark=at position #1 with {\arrow[scale=1.5]{>}}},postaction={decorate},line width=0.5mm},
        -<-/.style={decoration={
        markings,
        mark=at position #1 with {\arrow[scale=1.5]{<}}},postaction={decorate},line width=0.5mm},
        dashed/.style={dash pattern=on 8pt off 5pt,line width=0.5mm},
        elli/.style={ellipse,draw, minimum width=2cm, minimum height=1.8cm,line width=0.5mm,font=\normalsize
    }}
%Sp with N1 AF-chiral and N2 Fermi

\begin{scope}[xshift=7cm]
\node[cir] (Sp) at (0,0) {$\mathrm{Sp}(N)$};
\node[box] (N1) at (-2.5,-3.5) {$N_{1}$};
\node[box] (N2) at (2.5,-3.5) {$N_{2}$};
\draw[-<-=0.5] (Sp) -- (N1.north) node[midway,yshift=0.3cm,xshift=-0.7cm] {\LARGE$-\mathsf{y}_1$};
\draw[->-=0.5,dashed] (Sp) -- (N2.north) node[midway,yshift=0.3cm,xshift=0.7cm] {\LARGE$\mathsf{y}_2$};
\node[scale=2,rotate=90] () at (-5,-1.75) {$\Big\Updownarrow$};
\end{scope}
%N1-N1m2-N1m4-..-N2p4-N2p2-N2
\begin{scope}[xshift=-13cm]
    \node[box] (N1) at (0,-3.5) {$N_{1}$};
    \node[cir] (N1m2) at (0,0) {$N_{1}-2$};
    \node[cir] (N1m4) at (3.5,0) {$N_{1}-4$};
    \node[] (dots) at (6,0) {$\ldots $};
    \node[cir] (N2p4) at (8.5,0) {$N_{2}+4$};
    \node[cir] (N2p2) at (12,0) {$N_{2}+2$};
    \node[box] (N2) at (12,-3.5) {$N_{2}$};
    \draw[-<-=0.5] (N1)  --  (N1m2) node[midway,xshift=1.5cm] {\Large$\mathsf{y}_1-\frac{N_{1}\mathsf{y}_1}{N_{1}-2}$};
    \draw[-<-=0.5] (N1m2) -- (N1m4) node[midway,yshift=1.5cm] {\Large$\frac{N_{1}\mathsf{y}_1}{N_{1}-2}-\frac{N_{1}\mathsf{y}_1}{N_{1}-4}$};
    \draw[-<-=0.5] (N1m4) -- (dots);
    \draw[-<-=0.5] (dots) -- (N2p4);
    \draw[-<-=0.5] (N2p4) -- (N2p2) node[midway,yshift=1.5cm] {\Large$\frac{N_{1}\mathsf{y}_1}{N_{2}+4}-\frac{N_{1}\mathsf{y}_1}{N_{2}+2}$};
    \draw[-<-=0.5] (N2p2) -- (N2) node[midway,xshift=-1.4cm] {\Large$\frac{N_{1}\mathsf{y}_1}{N_{2}+2}-\mathsf{y}_2$};
    \draw[-<-=0.5,dashed] (N2) -- (N1) node[midway,yshift=-0.7cm] {\Large$\mathsf{y}_2-\mathsf{y}_1$};
\end{scope}
\end{tikzpicture}

%% file: figures/SU2-SU2k-Sp2k.tex
\begin{tikzpicture}[scale=1]
    \tikzset{
        cir/.style={circle, draw, minimum size=2cm, text centered, font=\Large,line width=0.5mm},
        point/.style={circle, fill, inner sep=1.5pt},
        box/.style={draw, minimum width=1.6cm, minimum height=1.6cm, text centered,line width=0.5mm,font=\Large},
        ->-/.style={decoration={
        markings,mark=at position #1 with {\arrow[scale=1.5]{>}}},postaction={decorate},line width=0.5mm},
        -<-/.style={decoration={
        markings,
        mark=at position #1 with {\arrow[scale=1.5]{<}}},postaction={decorate},line width=0.5mm},
        dashed/.style={dash pattern=on 8pt off 5pt,line width=0.5mm} 
    }
\node[cir] (Sp2m) at (0,0) {$\mathrm{Sp}(2m)$};
\node[cir] (SU2mp2) at (3.5,0) {$2m$+$2$};
\node[cir] (SU2k) at (9.5,0) {$N$};
\node[box] (Ul)  at (12,2) {$N_2$};
\node[box] (Ukl)  at (12,-3) {$N_1$};

\draw[->-=.5] (Sp2m.east) -- (SU2mp2.west) node[midway, yshift=0.6cm] {\LARGE$\mathsf{x}$}  ;
\draw[->-=0.5] (SU2mp2.east) --(6,0) node[midway, yshift=0.6cm] {\LARGE$\mathsf{x}$};
\node[] () at (6.5,0) {$\ldots $} ;
\draw[->-=0.5] (7,0) -- (SU2k.west) node[midway, yshift=0.6cm] {\LARGE$\mathsf{x}$};
\draw[-<-=0.5,dashed] (SU2k.60) -- (Ul.west) ;
\draw[->-=0.5] (SU2k.-60) -- (Ukl.north);

\begin{scope}[xshift=18cm]
\node[cir] (Sp2k) at (0,0) {$\mathrm{Sp}(N)$};
\node[box] (Ul)  at (2.5,2) {$N_2$};
\node[box] (Ukl)  at (2.5,-3) {$N_1$};
\draw[-<-=0.5,dashed] (Sp2k.60) -- (Ul.west);
\draw[->-=0.5] (Sp2k.-60) -- (Ukl.north);
\end{scope}

\node[scale=2,rotate=90] (arrow) at (15,0) {$\Big\Downarrow$};
\node[]  ()  at ($(arrow) + (0,0.6)$) {\large$\mathsf{x}\rightarrow 0$};

\end{tikzpicture}

%% file: figures/SU-AS-N-n-m.tex
\begin{tikzpicture}[scale=1]
    \tikzset{
        cir/.style={circle, draw, minimum size=1.8cm, text centered, font=\Large,line width=1.5pt},
        point/.style={circle, fill, inner sep=1.5pt},
        box/.style={draw, minimum width=1.6cm, minimum height=1.6cm, text centered,line width=1.5pt,font=\Large},
        ->-/.style={decoration={
        markings,mark=at position #1 with {\arrow[scale=1.3]{>}}},postaction={decorate},line width=1.5pt},
        -<-/.style={decoration={
        markings,
        mark=at position #1 with {\arrow[scale=1.3]{<}}},postaction={decorate},line width=1.5pt},
        dashed/.style={dash pattern=on 8pt off 5pt,line width=1.5pt},
        elli/.style={ellipse,draw, minimum width=2.5cm, minimum height=2cm,line width=1.5pt,font=\large
    }}    

%figure1
\coordinate (origin) at (0,0);
\draw[line width=1.5pt] ($(origin)+(0,-1.3)$) circle(0.7cm) node[yshift=-0.2cm]{\Large AS} node[yshift=-1cm] {\Large$Y$};
\node[cir,fill=white] (SUA) at (origin) {$N_{A}$} ;
\node[box] (n1) at (-3,-3) {${n_{1}}$};
\draw[line width=1.5pt,dashed] ($(n1)+(0,-1.3)$) circle(0.7cm) node[yshift=-0.2cm]{\Large $\overline{\mbox{AS}}$} node[yshift=-1.2cm] {\Large$\Lambda$};
\node[box,fill=white] () at (n1) {${n_{1}}$};
\node[box] (N) at (3,-3) {$N$};
\node[box] (m1) at (0,3.5) {$\mathrm{U}(m_{1})$};
\draw[->-=0.5] (SUA) -- (n1.north) node[midway,above,yshift=0.2cm] {\Large$X_1$};
\draw[-<-=0.5] (SUA) -- (N.north) node[midway,above,yshift=0.2cm] {\Large$X_2$};
\draw[->-=0.5,dashed] (SUA)  -- (m1) node[midway,left] {\Large$\Psi$};
\draw[-<-=0.5,dashed] (N) -- (n1) node[midway,below,yshift=-0.2cm] {\Large$\Gamma$};
\draw[->-=0.5] (m1) -- (N.55) node[midway,right,xshift=0.2cm,yshift=0.2cm] {\Large$M_1$};
\draw[-<-=0.2] (m1) -- (n1.125) node[midway,left,xshift=-0.2cm,yshift=0.2cm] {\Large$M_2$};
\draw[->-=0.8] (m1) -- (n1.125);

\node (A) at (0,-5) {\Large\textbf{Theory A}};

\node[] (arr1) at (0,-6.5) {\scalebox{2}{$\Big\Downarrow$}};
\node[xshift=1cm,yshift=0.3cm] () at (arr1) {\ref{fig:seiberg-Sp}};
\node[xshift=1cm,yshift=-0.3cm] () at (arr1) {Fig \ref{fig:Sp-ASodd}};

%figure2
\begin{scope}[yshift=-13cm]
    \coordinate (origin) at (0,0);
    \node[cir,fill=white] (SUA) at (origin) {$N_{A}$} ;
    \node[box] (n1) at (-3.5,-2) {${n_{1}}$};
    \draw[line width=1.5pt,dashed] ($(n1)+(0,-1.3)$) circle(0.7cm) node[yshift=-1.2cm]{\Large $\overline{\mbox{AS}}$} ;
    \node[box,fill=white] () at (n1) {${n_{1}}$};
    \node[box] (N) at (0,-4.2) {$N$};
    \node[box] (m1) at (3.5,-2) {$\mathrm{U}(m_{1})$};
    \node[box] (m2) at (-3.5,2) {$\mathrm{U}(m_{2})$};
    \draw[->-=0.5] (SUA) -- (n1);
    \draw[-<-=0.5] (SUA) -- (N);
    \node[elli] (Sp) at (0,4.2) {$\mathrm{Sp}(N_{A}$+$n_{2}$-$m_{2}$-2)};
    \draw[->-=0.5] (Sp) -- (SUA);
    \draw[dashed] (3.5,3.3) circle(0.7cm) node[yshift=1.3cm]{\Large $\overline{\mbox{AS}}$};
    \node[box,fill=white] (n2) at (3.5,2) {$n_{2}$};
    \draw[->-=0.5] (n2) -- (Sp);
    \draw[-<-=0.5,dashed] (n2.205) -- (SUA);
    \draw[-<-=0.5,dashed] (m1) -- (SUA);
    \draw[-<-=0.5] (m2) -- (SUA);
    \draw[->-=0.5,dashed] (m2) -- (Sp);
    \draw[-<-=0.5,dashed] (N) -- (n1);
    \draw[->-=0.5] (m1) -- (N);
    \draw[-<-=0.2] (m1) -- (n1);
    \draw[->-=0.8] (m1) -- (n1);
    \draw[-<-=0.2] (m2) -- (n2);
    \draw[->-=0.8] (m2) -- (n2);

 \node[rotate=90] (arr3) at (7.5,0) {\scalebox{2}{$\Big\Downarrow$}} ;
 \node[yshift=0.5cm] () at (arr3) {\ref{fig:seiberg-SU-N1-N2-N3}};
\end{scope}

%figure3
\begin{scope}[xshift=15cm,yshift=-13cm]
    \coordinate (origin) at (0,0);
    \node[cir,fill=white] (SUB) at (origin) {$N_{B}$} ;

    \node[box] (n1) at (3.5,2) {${n_{1}}$};
    \draw[line width=1.5pt,dashed] ($(n1)+(0,1.3)$) circle(0.7cm) node[yshift=1.3cm]{\Large $\overline{\mbox{AS}}$};
    \node[box,fill=white] () at (n1) {${n_{1}}$};
    \node[box] (N) at (0,-4.2) {$N$};
    \node[box] (m1) at (-3.5,2) {$\mathrm{U}(m_{1})$};
    \node[box] (m2) at (3.5,-2) {$\mathrm{U}(m_{2})$};
    \draw[-<-=0.5,dashed] (SUB) -- (n1);
    \draw[->-=0.5] (SUB) -- (N.north);
    \node[elli] (Sp) at (0,4.2) {$\mathrm{Sp}(N_{B}$+$n_{1}$-$m_{1}$-2)};
    \draw[-<-=0.5] (Sp) -- (SUB);
    \draw[dashed] (-3.5,-3.3) circle(0.7cm) node[yshift=-1.3cm]{\Large $\overline{\mbox{AS}}$};
    \node[box,fill=white] (n2) at (-3.5,-2) {$n_{2}$};
    \draw[-<-=0.5,dashed] (n2) -- (N);
    \draw[->-=0.5] (n2) -- (SUB);
    \draw[->-=0.5] (m1) -- (SUB);
    \draw[->-=0.5,dashed] (m2) -- (SUB);
    \draw[-<-=0.5] (m2) -- (N);
    \draw[-<-=0.5,dashed] (m1) -- (Sp);
    \draw[->-=0.6] (Sp) -- (n1);
    \draw[-<-=0.2] (m1) -- (n1);
    \draw[->-=0.8] (m1) -- (n1);
    \draw[-<-=0.2] (m2) -- (n2);
    \draw[->-=0.8] (m2) -- (n2);
\node[] (arr2) at (0,6.5) {\scalebox{2}{$\Big\Uparrow $}};
\node[xshift=1cm,yshift=0.3cm] () at (arr2) {\ref{fig:seiberg-Sp}};
\node[xshift=1cm,yshift=-0.3cm] () at (arr2) {Fig \ref{fig:Sp-ASodd}};

\end{scope}

%figure 4
\begin{scope}[xshift=15cm]
\coordinate (origin) at (0,0);
\draw[line width=1.5pt] ($(origin)+(0,-1.3)$) circle(0.7cm) node[yshift=-0.2cm]{\Large AS} node[yshift=-1.1cm] {\Large$\widetilde{Y}$};
\node[cir,fill=white] (SUB) at (origin) {$N_{B}$} ;
\node[box] (n2) at (-3,-3) {${n_{2}}$};
\draw[line width=1.5pt,dashed] ($(n2)+(0,-1.3)$) circle(0.7cm) node[yshift=-0.2cm]{\Large $\overline{\mbox{AS}}$} node[yshift=-1.2cm] {\Large$\widetilde{\Lambda}$};
\node[box,fill=white] () at (n2) {${n_{2}}$};
\node[box] (N) at (3,-3) {$N$};
\node[box] (m2) at (0,3.5) {$\mathrm{U}(m_{2})$};
\draw[->-=0.5] (SUB) -- (n2.north) node[midway,above,yshift=0.2cm] {\Large$\widetilde{X}_1$};
\draw[-<-=0.5] (SUB) -- (N.north) node[midway,above,yshift=0.2cm] {\Large$\widetilde{X}_2$};
\draw[->-=0.5,dashed] (SUB)  -- (m2) node[midway,left] {\Large$\widetilde{\Psi}$};
\draw[-<-=0.5,dashed] (N) -- (n2) node[midway,below,yshift=-0.2cm] {\Large$\widetilde{\Gamma}$};
\draw[->-=0.5] (m2) -- (N.55) node[midway,right,xshift=0.2cm,yshift=0.2cm] {\Large$\widetilde{M}_1$};
\draw[-<-=0.2] (m2) -- (n2.125) node[midway,left,xshift=-0.2cm,yshift=0.2cm] {\Large$\widetilde{M}_2$};
\draw[->-=0.8] (m2) -- (n2.125);
\node (B) at (0,-5) {\Large\textbf{Theory B}};

\end{scope}

\end{tikzpicture}

%% file: figures/SU-Sym-N-n-m.tex
\begin{tikzpicture}[scale=1]
    \tikzset{
        cir/.style={circle, draw, minimum size=1.9cm, text centered, font=\Large,line width=1.5pt},
        point/.style={circle, fill, inner sep=1.5pt},
        box/.style={draw, minimum width=1.6cm, minimum height=1.6cm, text centered,line width=1.5pt,font=\Large},
        ->-/.style={decoration={
        markings,mark=at position #1 with {\arrow[scale=1.3]{>}}},postaction={decorate},line width=1.5pt},
        -<-/.style={decoration={
        markings,
        mark=at position #1 with {\arrow[scale=1.3]{<}}},postaction={decorate},line width=1.5pt},
        dashed/.style={dash pattern=on 8pt off 5pt,line width=1.5pt},
        elli/.style={ellipse,draw, minimum width=2.5cm, minimum height=2cm,line width=1.5pt,font=\large
    }}    

%figure1
\coordinate (origin) at (0,0);
\draw[line width=1.5pt] ($(origin)+(0,-1.5)$) circle(0.7cm) node[yshift=-0.2cm]{\Large Sym} node[yshift=-1cm] {\Large$Y$};
\node[cir,fill=white] (SUA) at (origin) {$N_{A}$} ;
\node[box] (n1) at (-3,-3) {${n_{1}}$};
\draw[line width=1.5pt,dashed] ($(n1)+(0,-1.3)$) circle(0.7cm) node[yshift=-0.2cm]{\Large $\overline{\mbox{Sym}}$} node[yshift=-1.2cm] {\Large$\Lambda$};
\node[box,fill=white] () at (n1) {$n_{1}$};
\node[box] (N) at (3,-3) {$N$};
\node[box] (m1) at (0,3.5) {$\mathrm{U}(m_{1})$};
\draw[->-=0.5] (SUA) -- (n1.north) node[midway,above,yshift=0.2cm] {\Large$X_1$};
\draw[-<-=0.5] (SUA) -- (N.north) node[midway,above,yshift=0.2cm] {\Large$X_2$};
\draw[->-=0.5,dashed] (SUA)  -- (m1) node[midway,left] {\Large$\Psi$};
\draw[-<-=0.5,dashed] (N) -- (n1) node[midway,below,yshift=-0.2cm] {\Large$\Gamma$};
\draw[->-=0.5] (m1) -- (N.55) node[midway,right,xshift=0.2cm,yshift=0.2cm] {\Large$M_1$};
\draw[-<-=0.2] (m1) -- (n1.125) node[midway,left,xshift=-0.2cm,yshift=0.2cm] {\Large$M_2$};
\draw[->-=0.8] (m1) -- (n1.125);

\node (A) at (0,-5) {\Large\textbf{Theory A}};

%figure 2
\begin{scope}[xshift=15cm]
\coordinate (origin) at (0,0);
\draw[line width=1.5pt] ($(origin)+(0,-1.5)$) circle(0.7cm) node[yshift=-0.2cm]{\Large Sym} node[yshift=-1.1cm] {\Large$\widetilde{Y}$};
\node[cir,fill=white] (SUB) at (origin) {$N_{B}$} ;
\node[box] (n2) at (-3,-3) {$n_{2}$};
\draw[line width=1.5pt,dashed] ($(n2)+(0,-1.3)$) circle(0.7cm) node[yshift=-0.2cm]{\Large $\overline{\mbox{Sym}}$} node[yshift=-1.2cm] {\Large$\widetilde{\Lambda}$};
\node[box,fill=white] () at (n2) {$n_{2}$};
\node[box] (N) at (3,-3) {$N$};
\node[box] (m2) at (0,3.5) {$\mathrm{U}(m_{2})$};
\draw[->-=0.5] (SUB) -- (n2.north) node[midway,above,yshift=0.2cm] {\Large$\widetilde{X}_1$};
\draw[-<-=0.5] (SUB) -- (N.north) node[midway,above,yshift=0.2cm] {\Large$\widetilde{X}_2$};
\draw[->-=0.5,dashed] (SUB)  -- (m2) node[midway,left] {\Large$\widetilde{\Psi}$};
\draw[-<-=0.5,dashed] (N) -- (n2) node[midway,below,yshift=-0.2cm] {\Large$\widetilde{\Gamma}$};
\draw[->-=0.5] (m2) -- (N.55) node[midway,right,xshift=0.2cm,yshift=0.2cm] {\Large$\widetilde{M}_1$};
\draw[-<-=0.2] (m2) -- (n2.125) node[midway,left,xshift=-0.2cm,yshift=0.2cm] {\Large$\widetilde{M}_2$};
\draw[->-=0.8] (m2) -- (n2.125);
\node (B) at (0,-5) {\Large\textbf{Theory B}};
\end{scope}

\end{tikzpicture}

%% file: figures/Sp-adj.tex
\begin{tikzpicture}[scale=1]
    \tikzset{
        cir/.style={circle, draw, minimum size=2.2cm, text centered, font=\large,line width=0.5mm},
        point/.style={circle, fill, inner sep=1.5pt},
        box/.style={draw, minimum width=1.6cm, minimum height=1.6cm, text centered,line width=0.5mm,font=\Large},
        ->-/.style={decoration={
        markings,mark=at position #1 with {\arrow[scale=1.5]{>}}},postaction={decorate},line width=0.5mm},
        -<-/.style={decoration={
        markings,
        mark=at position #1 with {\arrow[scale=1.5]{<}}},postaction={decorate},line width=0.5mm},
        dashed/.style={dash pattern=on 8pt off 5pt,line width=0.5mm},
        elli/.style={ellipse,draw, minimum width=2cm, minimum height=2cm,line width=0.5mm,font=\large
    }}
%1
\node[cir] (Sp2N) at (0,-3) {};
\draw[line width=0.5mm,yshift=1.3cm] (0,-3) circle(0.8cm) node[yshift=1.3cm]{\Large Adj};
\node[cir,fill=white] () at (Sp2N) {$\mathrm{Sp}(2N)$};
\node[rotate=90] (arrow) at (2.8,-3) {\scalebox{2}{$\Big\Updownarrow$}};
\node[yshift=0.5cm] () at (arrow) {\ref{fig:seiberg-Sp}};

%2
\begin{scope}[xshift=5cm]
    \node[cir] (Sp2N) at (0,0) {$\mathrm{Sp}(2N)$};
    \node[scale=1.3] () at ($(Sp2N)+(0,2.5)$) {$X$} node[yshift=1.8cm,scale=1.3] {$+$};
    \node[cir] (SO2Np2) at (0,-6) {\normalsize$\mathrm{SO}(2N+2)$};
    \draw[line width=0.5mm] (SO2Np2.north) -- (Sp2N.south);
    \node[rotate=90] (arrow) at (2.5,-3) {\scalebox{2}{$\Big\Updownarrow$}};
    \node[yshift=0.5cm] () at (arrow) {Fig. \ref{fig:SU-Sp}};
\end{scope}

%3
\begin{scope}[xshift=11cm]
    \node[cir] (Sp2Nm2) at (0,1) {\normalsize$\mathrm{Sp}(2N-2)$};
    \node[cir] (SU2N) at (0,-3) {$2N$};
    \node[scale=1.3] () at ($(SU2N)+(-1.5,2)$) {$X+$} ;
    \node[cir] (SO2Np2) at (0,-7) {\normalsize$\mathrm{SO}(2N+2)$};
    \draw[-<-=.5] (SO2Np2.north) -- (SU2N.south);
    \draw[-<-=.5] (SU2N.north) -- (Sp2Nm2.south);
    \node[rotate=90] (arrow) at (3,-3) {\scalebox{2}{$\Big\Updownarrow$}};
    \node[yshift=0.5cm] () at (arrow) {\ref{fig:seiberg-SU-N1-N2-N3}};
\end{scope}

%4
\begin{scope}[xshift=18cm]
    \node[cir] (Sp2Nm2) at (0,1) {\normalsize$\mathrm{Sp}(2N-2)$};
    \node[cir] (SU2N) at (0,-3) {$2$};
    \node[scale=1.3] () at ($(SU2N)+(1.5,2)$) {$+X$} ;

    \node[cir] (SO2Np2) at (0,-7) {\normalsize$\mathrm{SO}(2N+2)$};
    \draw[->-=.5] (SO2Np2.north) -- (SU2N.south);
    \draw[->-=.5,dashed] (SU2N.north) -- (Sp2Nm2.south);
    \draw[->-=.5,bend right=30] (Sp2Nm2.215) to (SO2Np2.145);
    \node[rotate=90] (arrow) at (3,-3) {\scalebox{2}{$\Big\Updownarrow$}};
    \node[yshift=0.5cm] () at (arrow) {\ref{fig:seiberg-SO}};
\end{scope}

%5
\begin{scope}[xshift=24cm]

\coordinate (Sp2Nm2) at (0,-1) {};
\draw[line width=0.5mm] ($(Sp2Nm2)+(0,1.5)$) circle(0.8cm) node[yshift=1.3cm]{\Large Adj};
\node[cir,fill=white] () at (Sp2Nm2) {\normalsize$\mathrm{Sp}(2N-2)$};

\node[cir,fill=white] (SU2) at (0,-6) {$2$};
\draw[line width=0.5mm] ($(SU2)+(0,1.5)$) circle(0.8cm) node[yshift=1.3cm]{\Large Sym};
\node[cir,fill=white] () at (SU2) {$2$};

\node[scale=1.3] () at ($(Sp2Nm2)+(2,-1)$) {$+X$} ;
\node[scale=1.3] () at ($(Sp2Nm2)+(2,-2)$) {$+\Psi$} ;
\end{scope}

\end{tikzpicture}

%% file: figures/SO-adj-even.tex
\begin{tikzpicture}[scale=1]
    \tikzset{
        cir/.style={circle, draw, minimum size=2.2cm, text centered, font=\Large,line width=0.5mm},
        point/.style={circle, fill, inner sep=1.5pt},
        box/.style={draw, minimum width=1.6cm, minimum height=1.6cm, text centered,line width=0.5mm,font=\Large},
        ->-/.style={decoration={
        markings,mark=at position #1 with {\arrow[scale=1.5]{>}}},postaction={decorate},line width=0.5mm},
        -<-/.style={decoration={
        markings,
        mark=at position #1 with {\arrow[scale=1.5]{<}}},postaction={decorate},line width=0.5mm},
        dashed/.style={dash pattern=on 8pt off 5pt,line width=0.5mm},
        elli/.style={ellipse,draw, minimum width=2cm, minimum height=2cm,line width=0.5mm,font=\large
    }}
%1
\node[cir] (SO2N) at (0,-3) {};
\draw[line width=0.5mm,yshift=1.3cm] (0,-3) circle(0.8cm) node[yshift=1.3cm]{\Large Adj};
\node[cir,fill=white] () at (SO2N) {\large$\mathrm{SO}(2N)$};
\node[rotate=90] (arrow) at (3.5,-3) {\scalebox{2}{$\Big\Updownarrow$}};
\node[yshift=0.5cm] () at (arrow) {\ref{fig:seiberg-Sp}};

%2
\begin{scope}[xshift=8.5cm]
    \node[cir] (Sp2Nm2) at (0,-0.5) {\normalsize$\mathrm{Sp}(2N-2)$};
    \node[scale=1.3] () at (-2,-3) {$X+$} ;
    \node[cir] (SO2N) at (0,-5.5) {\large$\mathrm{SO}(2N)$};
    \draw[line width=0.5mm] (SO2N.north) -- (Sp2Nm2.south);
    \node[rotate=90] (arrow) at (3.5,-3) {\scalebox{2}{$\Big\Updownarrow$}};
    \node[yshift=0.5cm] () at (arrow) {\ref{fig:seiberg-SO}};
\end{scope}

\begin{scope}[xshift=15.5cm]
\node[cir] (Sp2Nm2) at (0,-3) {};
\draw[line width=0.5mm,yshift=1.5cm] (0,-3) circle(0.8cm) node[yshift=1.5cm]{\Large Adj};
\node[cir,fill=white] () at (Sp2Nm2) {\normalsize$\mathrm{Sp}(2N-2)$};

    \node[scale=1.3] () at (2,-2.5) {$+X$} ;
    \node[scale=1.3] () at (2,-3.5) {$+\Psi$} ;

\end{scope}

\end{tikzpicture}

%% file: figures/SO-adj-odd.tex
\begin{tikzpicture}[scale=1]
    \tikzset{
        cir/.style={circle, draw, minimum size=2.2cm, text centered, font=\Large,line width=0.5mm},
        point/.style={circle, fill, inner sep=1.5pt},
        box/.style={draw, minimum width=1.6cm, minimum height=1.6cm, text centered,line width=0.5mm,font=\Large},
        ->-/.style={decoration={
        markings,mark=at position #1 with {\arrow[scale=1.5]{>}}},postaction={decorate},line width=0.5mm},
        -<-/.style={decoration={
        markings,
        mark=at position #1 with {\arrow[scale=1.5]{<}}},postaction={decorate},line width=0.5mm},
        dashed/.style={dash pattern=on 8pt off 5pt,line width=0.5mm},
        elli/.style={ellipse,draw, minimum width=2.5cm, minimum height=2cm,line width=0.5mm,font=\large}
    }
        \coordinate (SO2Np1) at (0,0);
        \draw[line width=0.5mm] ($(SO2Np1)+(0,1.4)$) circle(0.7cm) node[yshift=2.5cm,midway]{\Large Adj};
        \node[cir,fill=white] () at (SO2Np1) {\normalsize$\mathrm{SO}(2N+1)$};
            \node[rotate=90] (arrow) at (3.7,0) {\scalebox{2}{$\Big\Updownarrow$}};
    \node[yshift=0.5cm] () at (arrow) {Fig. \ref{fig:Sp-ASodd}};

    \begin{scope}[xshift=7cm]
    \node[cir] (Sp2Nm2) at (0,0) {\normalsize$\mathrm{Sp}(2N-2)$};
    \node[cir]  (SO2Np1)    at (4,2.5) {\normalsize$\mathrm{SO}(2N+1)$};
    \node[box]  (U1)    at (4,-2.5) {$\mathrm{U}(1)$};
    \draw[->-=0.5] (Sp2Nm2.45) -- (SO2Np1.200);
    \draw[-<-=0.5,dashed] (Sp2Nm2.-45) -- (U1.west);
    \draw[->-=0.5] (SO2Np1.south) -- (U1.north);
    \node[rotate=90] (arrow) at (7,-0) {\scalebox{2}{$\Big\Updownarrow$}};
    \node[yshift=0.5cm] () at (arrow) {\ref{fig:seiberg-SO}};
    \end{scope}

    \begin{scope}[xshift=17cm]
        \coordinate (Sp2Nm2) at (0,0);
        \draw[line width=0.5mm] ($(Sp2Nm2)+(0,1.4)$) circle(0.7cm) node[yshift=2.5cm,midway]{\Large Adj};
        \node[cir,fill=white] () at (Sp2Nm2) {\normalsize$\mathrm{Sp}(2N-2)$};
        \node[scale=1.3] () at (2,0.5) {$+X$} ;
        \node[scale=1.3] () at (2,-0.5) {$+\Psi$} ;
    \end{scope}
\end{tikzpicture}

%% file: figures/SU-Sym-AS-even.tex
\begin{tikzpicture}[scale=1]
    \tikzset{
        cir/.style={circle, draw, minimum size=2.1cm, text centered, font=\large,line width=0.5mm},
        point/.style={circle, fill, inner sep=1.5pt},
        box/.style={draw, minimum width=1.6cm, minimum height=1.6cm, text centered,line width=0.5mm,font=\Large},
        ->-/.style={decoration={
        markings,mark=at position #1 with {\arrow[scale=1.3]{>}}},postaction={decorate},line width=0.5mm},
        -<-/.style={decoration={
        markings,
        mark=at position #1 with {\arrow[scale=1.3]{<}}},postaction={decorate},line width=0.5mm},
        dashed/.style={dash pattern=on 8pt off 5pt,line width=0.5mm},
        elli/.style={ellipse,draw, minimum width=2.5cm, minimum height=2cm,line width=0.5mm,font=\Large
    }}
%1
\coordinate (SUN) at (0,-4.5);
\draw[line width=0.5mm] ($(SUN)+(0,1.3)$) circle(0.8cm) node[yshift=1.3cm]{\Large Sym};
\draw[line width=0.5mm] ($(SUN)+(0,-1.3)$) circle(0.8cm) node[yshift=-1.3cm]{\Large AS};
\node[cir,fill=white] () at (SUN) {$N$};
\node[rotate=90] (arr1) at (2.25,-4.5) {\scalebox{2}{$\Big\Updownarrow$}};
\node[scale=1.3] () at (0,-8.5) {$+\Psi_{1}^{(0)}+\Psi_{2}^{(0)}$};
\node[scale=1.3] () at (0,-9.5) {$+\Psi_{3}^{(0)}+Z^{(0)}$};

\node[yshift=0.5cm] () at (arr1) {\ref{fig:seiberg-SO}};
\node[yshift=-0.5cm] () at (arr1) {\ref{fig:seiberg-Sp}};

%2
\begin{scope}[xshift=4.5cm]
\node[scale=1.3] () at (0,-10) {$+Z^{(1)}+\Psi_{1}^{(0)}$};
\node[scale=1.3] () at (0,-11) {$+\Psi_{3}^{(0)}+Z^{(0)}$};

\node[scale=1.3] () at (0,1) {$+\Psi_2^{(0)}+P^{(1)}$};
\draw[line width=0.4mm] (-1.5,1.5) rectangle (1.5,0.5);

\node[cir] (SUN) at (0,-4.5) {$N$};
\node[cir] (SONp2) at ($(SUN)+(0,3.5)$) {\normalsize$\mathrm{SO}(N+2)$};
\node[cir] (SpNm2) at ($(SUN)+(0,-3.5)$) {\normalsize$\mathrm{Sp}(N-2)$};
\draw[->-=0.5] (SONp2) -- (SUN);
\draw[->-=0.5] (SpNm2) -- (SUN);
\node[rotate=90] (arr1) at (2.25,-4.5) {\scalebox{2}{$\Big\Updownarrow $}};
\node[yshift=0.5cm] () at (arr1) {Fig \ref{fig:SU-Sp}};
\end{scope}

%3
\begin{scope}[xshift=9.25cm]
\node[scale=1.3] () at (2,-3) {$+\Psi_{1}^{(0)}$};
\node[scale=1.3] () at (2,-4) {$+Z^{(0)}~$};
\node[scale=1.3] () at (2,-5) {$+\Psi_{3}^{(0)}$};
\node[scale=1.3] () at (2,-6) {$+Z^{(1)}$};

\node[cir] (SUN) at (0,-3) {$N$};
\node[cir] (SONp2) at ($(SUN)+(0,3.5)$) {\normalsize$\mathrm{SO}(N+2)$};
\node[cir] (SUNm2) at ($(SUN)+(0,-3.5)$) {\normalsize$N-2$};
\node[cir] (SpNm4) at ($(SUNm2)+(0,-3.5)$) {\normalsize$\mathrm{Sp}(N-4)$};

\draw[->-=0.5] (SONp2) -- (SUN);
\draw[->-=0.5] (SUNm2) -- (SUN);
\draw[->-=0.5] (SpNm4) -- (SUNm2);

\node[rotate=90] (arr1) at (3.75,-4.5) {\scalebox{2}{$\Big\Updownarrow $}};
\node[yshift=0.5cm] () at (arr1) {\ref{fig:seiberg-SU-N1-N2-N3}};
\end{scope}

% \begin{scope}[xshift=16cm]
% \node[cir,draw=none]  (SUN) at (0,-3) {};
% \node[cir] () at (SUN) {$N$};
% \node[elli] (SONp2) at ($(SUN)+(0,3.5)$) {\large$\mathrm{SO}(N+2)$};
% \node[cir] (SU2) at ($(SUN)+(0,-3.5)$) {$2$};
% \node[elli] (SpNm4) at ($(SU2)+(0,-3.5)$) {\large$\mathrm{Sp}(N-4)$};

% \draw[->-=0.5] (SONp2) -- (SUN);
% \draw[-<-=0.5] (SU2) -- (SUN);
% \draw[-<-=0.5,dashed] (SpNm4) -- (SU2);
% \draw[->-=0.5,bend right=30] (SpNm4.25) to (SUN.-25);

% \draw[->,line width=0.3mm] (3,-4.5) -- (4,-4.5) node[midway,yshift=0.5cm]{\ref{fig:seiberg-SU-N1-N2-N3}};
% \end{scope}

%4
\begin{scope}[xshift=16.25cm]
\node[scale=1.3] () at (2,-2.5) {$+\Psi_{3}^{(0)}$};
\node[scale=1.3] () at (2,-1.5) {$+Z^{(0)}~$};
\node[scale=1.3] () at (2,-0.5) {$+\Psi_{1}^{(0)}$};
\node[scale=1.3] () at (2,0.5) {$+Z^{(1)}$};
\node[cir,draw=none]  (SUNm2) at (0,-3) {\normalsize$N-2$};
\node[cir] () at (SUNm2) {\normalsize$N-2$};
\node[cir] (SONp2) at ($(SUNm2)+(0,3.5)$) {\normalsize$\mathrm{SO}(N+2)$};
\node[cir] (SU2) at ($(SUNm2)+(0,-3.5)$) {$2$};
\node[cir] (SpNm4) at ($(SU2)+(0,-3.5)$) {\normalsize$\mathrm{Sp}(N-4)$};

\draw[-<-=0.5] (SONp2) -- (SUNm2);
\draw[->-=0.5,dashed] (SU2) -- (SUNm2);
\draw[-<-=0.5,bend right=30] (SpNm4.25) to (SUNm2.-25);
\draw[-<-=0.5,bend left=40] (SU2.155) to (SONp2.-155);

\node[rotate=90] (arr1) at (3.25,-4.5) {\scalebox{2}{$\Big\Updownarrow $}};
\node[yshift=0.5cm] () at (arr1) {\ref{fig:seiberg-SO}};
\node[yshift=-0.5cm] () at (arr1) {\ref{fig:seiberg-Sp}};
\end{scope}

%5
\begin{scope}[xshift=21.5cm]
\node[scale=1.3] () at (2,1) {$+\Psi_{1}^{(1)}$};
\node[scale=1.3] () at (2,0) {$+\Psi_{2}^{(1)}$};

\node[scale=1.3] () at (2.75,-1) {$+\Psi_{3}^{(1)}~(=\Psi_{1}^{(0)})$};
\node[scale=1.3] () at (2,-2) {$+Z^{(1)}$};
\node[scale=1.3] () at (2,-3.5) {$+\Psi_{4}^{(1)}$};
\node[scale=1.3] () at (2,-4.5) {$+\widetilde{Z}^{(1)}$};

\coordinate (SUNm2) at (0,-1.5);
\draw[line width=0.5mm] ($(SUNm2)+(0,1.3)$) circle(0.7cm) node[yshift=1.3cm]{\Large Sym};
\draw[line width=0.5mm] ($(SUNm2)+(0,-1.3)$) circle(0.7cm) node[yshift=-1.3cm]{\Large AS};
\node[cir,fill=white] () at (SUNm2) {\normalsize$N-2$};
\node[scale=1.3] () at (2,-8.5) {$+\Psi_{3}^{(0)}$};
\node[scale=1.3] () at (2,-9.5) {$+Z^{(0)}$};
\coordinate (SU2) at (0,-9);
\draw[line width=0.5mm] ($(SU2)+(0,1.3)$) circle(0.8cm) node[yshift=1.3cm]{\Large Sym};
\node[cir,fill=white] () at (SU2) {$2$};
\end{scope}

\end{tikzpicture}

%% file: figures/SU-Sym-AS-EG-even.tex
\begin{tikzpicture}[scale=1]
    \tikzset{
        cir/.style={circle, draw, minimum size=2.1cm, text centered, font=\Large,line width=0.5mm},
        point/.style={circle, fill, inner sep=1.5pt},
        box/.style={draw, minimum width=1.6cm, minimum height=1.6cm, text centered,line width=0.5mm,font=\Large},
        ->-/.style={decoration={
        markings,mark=at position #1 with {\arrow[scale=1.3]{>}}},postaction={decorate},line width=0.5mm},
        -<-/.style={decoration={
        markings,
        mark=at position #1 with {\arrow[scale=1.3]{<}}},postaction={decorate},line width=0.5mm},
        dashed/.style={dash pattern=on 8pt off 5pt,line width=0.5mm},
        elli/.style={ellipse,draw, minimum width=2.5cm, minimum height=2cm,line width=0.5mm,font=\Large
    }}
%1
\coordinate (SU2) at (0,-4.5);
\draw[line width=0.5mm] ($(SU2)+(0,1.3)$) circle(0.8cm) node[yshift=1.3cm]{\Large Sym} node[yshift=0.5cm,xshift=-1.3cm]{\Large $A^{(i)}$}  ;
\node[cir,fill=white] () at (SU2) {$2$};
\node[] (arr1) at (-3.5,-4.5) {\scalebox{2}{$\sum_{i=0}^{\frac{N-4}{2}}$}};
\draw[line width=0.4mm] (-2,-1) rectangle (2,-7.5);
\node[scale=1.3] () at (0,-6.5) {$+\Psi_{3}^{(i)}+Z^{(i)}$};

\node[scale=2] (plus) at (3,-4.5) {$+$}; 
\node[scale=2] (plus) at (5,-4.3) {$\sum_{i=1}^{\frac{N-2}{2}}$};

\begin{scope}[xshift=5.5cm]
    \node[scale=1.3] () at (1.5,-3.5) {$+\Psi_{4}^{(i)}$};
    \node[scale=1.3] () at (1.5,-5) {$+\widetilde{Z}^{(i)}$};
    \draw[line width=0.4mm] (0.7,-3) rectangle (2.3,-5.5);
    \node[scale=2] (plus) at (3,-4.5) {$+$};
\end{scope}

\begin{scope}[xshift=13cm]
\coordinate (SU2) at (0,-4.5);
\draw[line width=0.5mm] ($(SU2)+(0,1.3)$) circle(0.8cm) node[yshift=1.3cm]{\Large Sym} node[yshift=0.5cm,xshift=1.6cm]{\Large $X^{(\frac{N-2}{2})}$};
\draw[line width=0.5mm] ($(SU2)+(0,-1.3)$) circle(0.8cm) node[yshift=-1.3cm]{\Large AS} node[yshift=-0.5cm,xshift=1.6cm]{\Large $Y^{(\frac{N-2}{2})}$};
    \node[scale=1.3] () at (-2.5,-3.5) {$+\Psi_{1,2,3}^{(\frac{N-2}{2})}$};
    \node[scale=1.3] () at (-2.5,-5) {$+Z^{(\frac{N-2}{2})}$};
\draw[line width=0.4mm] (-3.7,-1) rectangle (2.5,-7.5);

\node[cir,fill=white] () at (SU2) {$2$};

\end{scope}

\end{tikzpicture}

%% file: figures/SU-Sym-AS-odd.tex
\begin{tikzpicture}[scale=1]
    \tikzset{
        cir/.style={circle, draw, minimum size=2cm, text centered, font=\large,line width=0.5mm},
        point/.style={circle, fill, inner sep=1.5pt},
        box/.style={draw, minimum width=1.6cm, minimum height=1.6cm, text centered,line width=0.5mm,font=\large},
        ->-/.style={decoration={
        markings,mark=at position #1 with {\arrow[scale=1.5]{>}}},postaction={decorate},line width=0.5mm},
        -<-/.style={decoration={
        markings,
        mark=at position #1 with {\arrow[scale=1.5]{<}}},postaction={decorate},line width=0.5mm},
        dashed/.style={dash pattern=on 8pt off 5pt,line width=0.5mm},
        elli/.style={ellipse,draw, minimum width=2.5cm, minimum height=2cm,line width=0.5mm,font=\Large
    }}
%1
\coordinate (SUN) at (-0.5,-4.5);
\draw[line width=0.5mm] ($(SUN)+(0,1.3)$) circle(0.8cm) node[yshift=1.3cm]{\Large Sym};
\draw[line width=0.5mm] ($(SUN)+(0,-1.3)$) circle(0.8cm) node[yshift=-1.3cm]{\Large AS};
\node[cir,fill=white] () at (SUN) {$N$};
\node[rotate=90] (arr1) at (2.5,-4.5) {\scalebox{2}{$\Big\Updownarrow $}};
\node[scale=1.3] () at (-0.5,-8.5) {$+\Psi_{1} +\Psi_{2} $};
\node[scale=1.3] () at (-0.5,-9.5) {$+\Psi_{3} +Z $};

\node[yshift=0.5cm] () at (arr1) {\ref{fig:seiberg-SO}};
\node[yshift=-0.5cm]() at (arr1) {Fig. \ref{fig:Sp-ASodd}};

%2
\begin{scope}[xshift=5.5cm]

\node[cir,draw=none]  (SUN) at (0,-4.5) {};
\node[cir] () at (SUN) {$N$};
\node[scale=1.3] () at ($(SUN)+(3,3.5)$) {$+\Psi_{1} +\Psi_{2} $};
\node[scale=1.3] () at ($(SUN)+(3,2.5)$) {$+\Psi_{3} +Z $};
\node[scale=1.3] () at ($(SUN)+(3.6,1.5)$) {$+Z^ {(0)}$};

\node[cir] (SONp2) at ($(SUN)+(0,3.5)$) {\normalsize$\mathrm{SO}(N+2)$};
\node[cir] (SpNm3) at ($(SUN)+(0,-3.5)$) {\normalsize$\mathrm{Sp}(N-3)$};
\node[box]  (U1) at ($(SUN)+(3.6,0)$) {$\mathrm{U}(1)$};
\draw[->-=0.5] (SONp2) -- (SUN);
\draw[->-=0.5] (SpNm3) -- (SUN);
\draw[-<-=0.5] (U1) -- (SUN);
\draw[->-=0.5,dashed] (U1) -- (SpNm3);
\node[rotate=90] (arr2) at (6.5,-4.5) {\scalebox{2}{$\Big\Updownarrow $}};
\node[yshift=0.5cm] () at (arr2) {\ref{fig:seiberg-SU-N1-N2-N3}};
\end{scope}

%3
\begin{scope}[xshift=15cm]
\node[cir] (SUNm1) at (0,-4.5) {\normalsize$N-1$};
\node[scale=1.3] () at ($(SUNm1)+(3,3.5)$) {$+\Psi_{1} +\Psi_{2} $};
\node[scale=1.3] () at ($(SUNm1)+(3,2.5)$) {$+\Psi_{3} +Z $};
\node[scale=1.3] () at ($(SUNm1)+(3.6,1.5)$) {$+Z^ {(0)}$};

\node[cir] (SONp2) at ($(SUNm1)+(0,3.5)$) {\normalsize$\mathrm{SO}(N+2)$};
\node[cir] (SpNm3) at ($(SUNm1)+(0,-3.5)$) {\normalsize$\mathrm{Sp}(N-3)$};
\node[box]  (U1) at ($(SUNm1)+(3.5,0)$) {$\mathrm{U}(1)$};
\draw[-<-=0.5] (SONp2) -- (SUNm1);
\draw[-<-=0.5] (SpNm3) -- (SUNm1);
\draw[->-=0.5,dashed] (U1) -- (SUNm1);
\draw[-<-=0.5] (U1) -- (SONp2);

\node[rotate=90] (arr3) at (6.5,-4.5) {\scalebox{2}{$\Big\Updownarrow $}};
\node[yshift=0.5cm] () at (arr3) {\ref{fig:seiberg-SO}};
\node[yshift=-0.5cm]() at (arr3) {\ref{fig:seiberg-Sp}};
\end{scope}

\begin{scope}[xshift=25cm]

\coordinate (SUNm1) at (0,-4.5);
\node[scale=1.3] () at ($(SUNm1)+(3,1.5)$) {$+\Psi_1^ {(0)}(=\Psi_{3} )$};
\node[scale=1.3] () at ($(SUNm1)+(2.4,0.5)$) {$+\Psi_2^ {(0)} \quad$}; %from Sp
\node[scale=1.3] () at ($(SUNm1)+(3,-0.5)$) {$+\Psi_3^ {(0)}(=\Psi_{1} )$};
\node[scale=1.3] () at ($(SUNm1)+(2.4,-1.5)$) {$+Z^ {(0)}\quad$};

\node[scale=1.3] () at ($(SUNm1)+(0,-4)$) {$+\Psi_2 +P^ {(0)}$};
\node[scale=1.3] () at ($(SUNm1)+(0,-5)$) {$+Z +\Gamma^ {(0)}$};

\draw[line width=0.4mm] (-1.5,-8) rectangle (1.5,-10);

\draw[line width=0.5mm] ($(SUNm1)+(0,1.3)$) circle(0.8cm) node[yshift=1.3cm]{\Large Sym};
\draw[line width=0.5mm] ($(SUNm1)+(0,-1.3)$) circle(0.8cm) node[yshift=-1.3cm]{\Large AS};
\node[cir,fill=white] () at (SUNm1) {$N-1$};
\end{scope}
\end{tikzpicture}

%% file: figures/Sp-SU-AS-AS-odd.tex
\begin{tikzpicture}[scale=1]
    \tikzset{
        %>=latex
        cir/.style={circle, draw, minimum size=1.8cm, text centered, font=\Large,line width=0.5mm},
        point/.style={circle, fill, inner sep=1.5pt},
        box/.style={draw, minimum width=1.6cm, minimum height=1.6cm, text centered,line width=0.5mm,font=\Large},
        ->-/.style={decoration={
        markings,mark=at position #1 with {\arrow[scale=1.5]{>}}},postaction={decorate},line width=0.5mm},
        -<-/.style={decoration={
        markings,
        mark=at position #1 with {\arrow[scale=1.5]{<}}},postaction={decorate},line width=0.5mm},
        dashed/.style={dash pattern=on 8pt off 5pt,line width=0.5mm},
        elli/.style={ellipse,draw, minimum width=2.5cm, minimum height=2cm,line width=0.5mm,font=\Large
    }}

% Sp with F 
    \node[fill=blue!10!,scale=1.7]  at (0,-5) {AS-1'};
    \node[cir,scale=0.85] (SpNm3) at (0,0) {Sp($N$-$1$)};
    \node[box]  (UN)    at (-1.5,-3) {$N$};
    \node[box]  (U1)    at (1.5,-3) {$\mathrm{U}(1)$} node[right, xshift=2.5cm, yshift=-3.5cm] {\Large$\mathsf{y}$};
    \draw[->-=0.5] (SpNm3.-135) -- (UN.north);
    \draw[-<-=0.5] (SpNm3.-45) -- (U1.north);
    % \draw[->-=0.5] (UN.east) -- (U1.west);
    \node[rotate=90] () at (4.3,-2.5) {\scalebox{2}{$\Big\Updownarrow$}};

    \begin{scope}[xshift=7cm,yshift=-2cm]
    \node[fill=blue!10!,scale=1.7]  at (1.5,-3) {AS-3'};

        \node[box] (UN) at (0,-1) {$N$};
        \node[box]  (U1)    at (3,-1) {$\mathrm{U}(1)$} node[right, xshift=4cm,yshift=-1.7cm] {\Large$\mathsf{y}$};
        \draw[-<-=0.5] (UN.east) -- (U1.west);
        \node[font=\Large,scale=1.3] (Gamma) at ($(U1)+(2,0)$) {$+\Psi$};
        \draw[line width=0.5mm] (UN.120) arc[start angle=135+90,end angle=45-90, radius=0.7cm] node[yshift=0.5cm,midway]{\Large AS};
    \end{scope}

% SU with AS and F
\begin{scope}[yshift=-8cm,xshift=4cm]
\node[fill=blue!10!,scale=1.7]  at (0,-5) {AS-2'};
    \node[cir] (gauge) at (0,0) {$N$-$2$};
         \draw[line width=0.5mm] (gauge.120) arc[start angle=135+90,end angle=45-90, radius=0.7cm] node[yshift=0.5cm,midway]{\large AS};
    \node[box]  (UN)    at (-1.5,-3) {$N$};
    \node[box]  (U1)    at (1.5,-3) {$\mathrm{U}(1)$} node[right, xshift=2.5cm, yshift=-3.5cm] {\Large$\mathsf{y}$};
    \draw[-<-=0.5] (gauge.-135) -- (UN.north);
    % \draw[line width=0.5mm] (gauge.-45) -- (U1.north);
\draw[-<-=0.5] (UN.east) -- (U1.west);
\node[font=\Large,scale=1.3] (Gamma) at ($(gauge)+(3.5,-1.3)$) {$+\Psi$};
\node[rotate=30] () at (-2.5,2) {\scalebox{2}{$\Big\Updownarrow$}};
\node[rotate=-30] () at (3,2) {\scalebox{2}{$\Big\Updownarrow$}};
\end{scope}
\end{tikzpicture}

%% file: figures/02-SU-as-N1-N2-N2=1.tex
\begin{tikzpicture}[scale=1]
    \tikzset{
        cir/.style={circle, draw, minimum size=1.9cm, text centered, font=\Large,line width=0.5mm},
        point/.style={circle, fill, inner sep=1.5pt},
        box/.style={draw, minimum width=1.6cm, minimum height=1.6cm, text centered,line width=0.5mm,font=\Large},
        ->-/.style={decoration={
        markings,mark=at position #1 with {\arrow[scale=1.5]{>}}},postaction={decorate},line width=0.5mm},
        -<-/.style={decoration={
        markings,
        mark=at position #1 with {\arrow[scale=1.5]{<}}},postaction={decorate},line width=0.5mm},
        elli/.style={ellipse,draw, minimum width=1.6cm, minimum height=1.8cm,line width=0.5mm,font=\large},
        dashed/.style={dash pattern=on 8pt off 5pt,line width=0.5mm}    
    }

%theory A
\node[scale=2] () at (0,4) {$+\Psi$};

\node[font=\Large] (AS1) at (0,2.3)  {AS} node[yshift=1.4cm,xshift=1.4cm] {\LARGE$2\mathsf{x}_{0}$} node[yshift=1.4cm,xshift=-1.4cm] {\LARGE$Y$};
\node[font=\Large]  at (0,-6)  {\textbf{Theory A}};    

\draw[line width=0.5mm] (0,1.2) circle (0.7);
\node[cir,fill=white] (SU) at (0,0) {$n-1$};
\node[box] (U1)   at (3,-3.5)  {$\mathrm{U}(1)$};
\node[box] (n1)   at (-3,-3.5)  {$n$};
%N2
\draw[->-=.5]  (U1.north) -- (SU) node[midway, yshift=0.5cm,xshift=0.5cm] {\LARGE$\mathsf{x_0}$} node[midway, xshift=-0.5cm, yshift=-0.5cm] {\LARGE$X_2$}  ;
%N1
\draw[->-=.5]  (SU) -- (n1.north) node[midway,yshift=0.5cm,xshift=-0.8cm]{\LARGE$\mathsf{x}_{2}$-$\mathsf{x}_0$} node[midway,xshift=0.5cm, yshift=-0.5cm] {\LARGE$X_1$};
\draw[->-=0.5,dashed] (n1) -- (U1) node[midway,yshift=0.5cm] {\LARGE$-\mathsf{x}_2$} node[midway,yshift=-0.7cm] {\LARGE$\Gamma$};

\node[rotate=90,scale=2] (arrow) at (5.5,0) {$\Big\Updownarrow$};
\node[yshift=0.5cm] () at (arrow) {\ref{fig:seiberg-Sp}};
\node[yshift=-0.7cm] () at (arrow) {\Large$n+m$ odd};
\node[yshift=-1.5cm] () at (arrow) {\Large$m\in \{0,1\}$};

% picture 2
\begin{scope}[xshift=12cm]
\node[elli] (SP) at (0,3.5)  {$\mathrm{Sp}(n$+$m$-$3)$};
\node[cir] (SUn1m1) at (0,0) {$n-1$};
\node[box] (U1mm)   at (3,-3.5)  {$\mathrm{U}(1-m)$};
\node[box] (n1)   at (-3,-3.5)  {$n$};
\node[box] (m)   at (-4,3.5) {$\mathrm{U}(m)$};

\draw[->-=.5]  (U1mm.north) -- (SUn1m1) node[midway, yshift=0.5cm,xshift=0.5cm] {\LARGE$\mathsf{x_0}$};
\draw[->-=.5]  (SUn1m1) -- (n1.north) node[midway,yshift=0.5cm,xshift=-0.5cm]{\LARGE$\mathsf{x}_{2}$-$\mathsf{x_0}$};
\draw[->-=.5] (SP.south) -- (SUn1m1.north) node[midway,xshift=0.6cm] {\LARGE$\mathsf{x}_{0}$};
\draw[-<-=.6] (SP.west) -- (m.east) ;

\draw[->-=0.5,dashed] (n1) -- (U1mm) node[midway,yshift=0.5cm] {\LARGE$-\mathsf{x}_2$};
\draw[->-=0.5,dashed] (n1.120) -- (m.-65) node[midway,xshift=-0.9cm] {\LARGE$-\mathsf{x}_2$};

\node[rotate=90,scale=2] (arrow) at (5.5,0) {$\Big\Updownarrow$};
\node[yshift=0.5cm] () at (arrow) {\ref{fig:seiberg-SU-N1-N2-N3}};
\end{scope}

%picture3
\begin{scope}[xshift=24cm]
\node[elli] (Sp) at (0,3.5)  {$\mathrm{Sp}(n$+$m$-$3)$};
\node[box] (U1) at (0,0) {$\mathrm{U(1)}$} node[xshift=1.3cm,yshift=-0.5cm] {\LARGE$\mathsf{y}$};
%\draw[line width=0.5mm, dashed] ($\mathrm{(SPN)+(2.5,0)}$) circle (0.7);
\draw[dashed,-<-=0.5] (Sp) -- (U1) node[midway,xshift=0.7cm] {\LARGE$\Lambda$};
\node[elli,fill=white] () at (Sp) {$\mathrm{Sp}(n$+$m$-$3)$};
\node[box] (U1mm)   at (3.5,-3.5)  {$\mathrm{U}(1-m)$};
\node[box] (n1)   at (-3.5,-3.5)  {$n$};
\node[box,fill=white] (m)   at (-4,3.5) {};
\node[box,fill=white] ()   at (m) {$\mathrm{U}(m)$};

\draw[-<-=.5] (Sp) -- (m) node[midway,yshift=0.7cm] {\LARGE$\widetilde{X}_{2}$};
\draw[->-=.5] (Sp.220)  -- (n1.north) node[midway,xshift=-0.6cm,yshift=-0.2cm] {\LARGE$\mathsf{x}_{2}$} node[midway,xshift=-0.3cm, yshift=1cm] {\LARGE$\widetilde{X}_{1}$};
%\draw[->-=.5] (U1mm)  -- (n1) node[midway,yshift=-0.5cm] {\LARGE$\mathsf{x}_{1}$};
\draw[->-=.5] (n1) -- (U1) node[midway,yshift=-0.7cm,xshift=0cm]{\LARGE-$\mathsf{x}_2$} node[midway,yshift=0.1cm,xshift=0.75cm]{\LARGE$Z$} ;
\draw[dashed,->-=0.5] (U1) -- (U1mm)  node[midway,xshift=-0.7cm,yshift=-0.2cm] {\LARGE$\widetilde{\Psi}$};
\draw[->-=.5,dashed] (n1.120) -- (m) node[midway,xshift=-0.8cm,yshift=-0.5cm] {\LARGE$-\mathsf{x}_2$} node[midway,xshift=-0.6cm,yshift=0.5cm] {\LARGE$\widetilde{\Gamma}$};
%\node[font=\Large] (F) at ($\mathrm{(SPN)+(3.5,0)}$)  {F};
\node[font=\Large] (AS) at (0,-6)  {\textbf{Theory B}};
\end{scope}

\end{tikzpicture}

%% file: figures/SU-AS-AF-even.tex
\begin{tikzpicture}[scale=1]
    \tikzset{
        %>=latex
        cir/.style={circle, draw, minimum size=2cm, text centered, font=\Large,line width=0.5mm},
        point/.style={circle, fill, inner sep=1.5pt},
        box/.style={draw, minimum width=1.8cm, minimum height=1.8cm, text centered,line width=0.5mm,font=\Large},
        ->-/.style={decoration={
        markings,mark=at position #1 with {\arrow[scale=1.3]{>}}},postaction={decorate},line width=0.5mm},
        -<-/.style={decoration={
        markings,
        mark=at position #1 with {\arrow[scale=1.3]{<}}},postaction={decorate},line width=0.5mm},
        dashed/.style={dash pattern=on 8pt off 5pt,line width=0.5mm},
        elli/.style={ellipse,draw, minimum width=2.5cm, minimum height=2cm,line width=0.5mm,font=\large,
    }}

    \coordinate (origin) at (0,2);
    \draw[line width=1.5pt] ($(origin)+(0,1.4)$) circle(0.7cm) node[yshift=1.1cm]{\Large AS};
    \node[cir,fill=white] (SU) at (origin) {$n-2$};
    \node[box]  (n1) at  ($(origin)-(0,4)$) {$n$};
    \draw[->-=0.5] (SU) -- (n1);
    \node[rotate=90] (arr1) at (2.5,0) {\scalebox{2}{$\Big\Updownarrow $}}; 
    \node[yshift=0.5cm] () at (arr1) {\ref{fig:seiberg-Sp}};
    \node[yshift=-0.5cm] () at (arr1) {\large even $n$};
    \node[scale=1.5] (det) at (0,-4) {$+\Psi$};

    \begin{scope}[xshift=5.5cm]
    \coordinate (origin) at (0,0);
    \node[cir] (Sp)  at ($(origin)+(0,4)$) {\normalsize$\mathrm{Sp}(n-4)$};
    \node[cir] (SU) at (origin) {$n-2$};
    \node[box]  (n1) at  ($(origin)-(0,4)$) {$n$};
    \draw[->-=0.5] (SU) -- (n1);
    \draw[->-=0.5] (Sp) -- (SU);
    \node[rotate=90] (arr1) at (2.5,0) {\scalebox{2}{$\Big\Updownarrow $}}; 
    \node[yshift=0.5cm] () at (arr1)  {\ref{fig:seiberg-SU-N1-N2-N3}};
    \end{scope}

    \begin{scope}[xshift=10.5cm]
    
    \coordinate (origin) at (0,0);
    \node[cir] (Sp)  at ($(origin)+(0,4)$) {\normalsize$\mathrm{Sp}(n-4)$};
    \node[cir] (SU) at (origin) {$2$};
    \node[box]  (n1) at  ($(origin)-(0,4)$) {$n$};
    \draw[-<-=0.5] (SU) -- (n1);
    \draw[dashed] (Sp) -- (SU);
    \draw[->-=0.5,bend left=30] (Sp.-40) to (n1.44);
    \node[rotate=90] (arr1) at (3.5,0) {\scalebox{2}{$\Big\Downarrow $}}; 
    \node[yshift=1.4cm] () at (arr1)  {\Large$\substack{ \text {Merge}
    \\
    \\ \mathrm{SU}(2)
    \\
    \\\text {and}
    \\
    \\ \mathrm{Sp}(n-4)}$};
    \end{scope}

    \begin{scope}[xshift=16.5cm]
    \coordinate (origin) at (0,2);
    \node[cir] (Sp) at (origin) {\normalsize$\mathrm{Sp}(n-2)$};
    \node[box]  (n1) at  ($(origin)-(0,4)$) {$n$};
    \draw[->-=0.5] (Sp) -- (n1);
    \node[rotate=90] (arr1) at (2.5,0) {\scalebox{2}{$\Big\Updownarrow $}}; 
    \node[yshift=0.5cm] () at (arr1) {\ref{fig:seiberg-Sp}};
    \node[yshift=1.8cm,xshift=1.5cm,scale=1.5] at (Sp) {$\bigoplus \frac{n-2}2$};
    \end{scope}

    \begin{scope}[xshift=22cm]

    \coordinate (origin) at (0,0);
    \draw[line width=1.5pt] ($(origin)+(0,1.4)$) circle(0.7cm) node[yshift=1.2cm]{\Large AS};
    \node[box,fill=white]  (n1) at  (origin) {$n$};
    \node[scale=1.3] (det) at (0,-2) {$+{\Psi}$};
    \node[yshift=3.8cm,xshift=1.5cm,scale=1.5] at (origin) {$\bigoplus \frac{n-2}2$};
    \end{scope}
\end{tikzpicture}

%% file: figures/SU-AS-AF-odd.tex
\begin{tikzpicture}[scale=1]
    \tikzset{
        %>=latex
        cir/.style={circle, draw, minimum size=2cm, text centered, font=\Large,line width=0.5mm},
        point/.style={circle, fill, inner sep=1.5pt},
        box/.style={draw, minimum width=1.8cm, minimum height=1.8cm, text centered,line width=0.5mm,font=\Large},
        ->-/.style={decoration={
        markings,mark=at position #1 with {\arrow[scale=1.3]{>}}},postaction={decorate},line width=0.5mm},
        -<-/.style={decoration={
        markings,
        mark=at position #1 with {\arrow[scale=1.3]{<}}},postaction={decorate},line width=0.5mm},
        dashed/.style={dash pattern=on 8pt off 5pt,line width=0.5mm},
        elli/.style={ellipse,draw, minimum width=2.5cm, minimum height=2cm,line width=0.5mm,font=\large,
    }}
    %1
    \begin{scope}[xshift=1.5cm]

    \coordinate (origin) at (-1.5,2);
    \draw[line width=1.5pt] ($(origin)+(0,1.4)$) circle(0.7cm) node[yshift=1.1cm]{\Large AS};
    \node[cir,fill=white] (SU) at (origin) {$n-2$};
    \node[box]  (n1) at  ($(origin)-(0,4)$) {$n$};
%    \node[box] (U1) at ($(origin)+(2,-4)$) {$\mathrm{U}(1)$} node[xshift=1.7cm,yshift=-2.5cm] {\LARGE$\mathsf{y}$};
%    \draw[-<-=0.5] (n1) -- (U1);
    \draw[->-=0.5] (SU) -- (n1);
    \node[rotate=90] (arr1) at (1,0) {\scalebox{2}{$\Big\Updownarrow $}}; 
    \node[scale=1.5] (det) at (-1.5,-4) {$+\Psi$};
    
    \node[yshift=0.5cm] () at (arr1) {Fig. \ref{fig:seiberg-Sp-2}};
    \node[yshift=-0.5cm] () at (arr1) {\large odd $n$};
    \end{scope}

    %2
    \begin{scope}[xshift=5.5cm]
    \coordinate (origin) at (0,0);

    \node[cir] (Sp)  at ($(origin)+(-2,4)$) {\normalsize$\mathrm{Sp}(n-3)$};
    \node[box] (U1)  at ($(origin)+(2,4)$) {$\mathrm{U}(1)$};
    \node[cir] (SU) at (origin) {$n-2$};
    \node[box]  (n1) at  ($(origin)-(0,4)$) {$n$};
    \draw[->-=0.5] (SU) -- (n1);
    \draw[->-=0.5] (Sp) -- (SU);
    \draw[-<-=0.5] (Sp) -- (U1);
    %\draw[-<-=0.3,dashed] (SU) -- (U1);
    \draw[->-=0.5,dashed] (SU) -- (U1);

    \node[rotate=90] (arr2) at (3.,0) {\scalebox{2}{$\Big\Updownarrow $}}; 
    \node[yshift=0.5cm] () at (arr2)  {\ref{fig:seiberg-SU-N1-N2-N3}};
    \end{scope}
    %3
    \begin{scope}[xshift=13cm]
    \coordinate (origin) at (0,0);

    \node[cir] (Sp)  at ($(origin)+(-2,4)$) {\normalsize$\mathrm{Sp}(n-3)$};
    \node[box] (U1)  at ($(origin)+(2,4)$) {$\mathrm{U}(1)$};
    \node[cir] (SU) at (origin) {$2$};
    \node[box]  (n1) at  ($(origin)-(0,4)$) {$n$};
    \draw[-<-=0.5] (SU) -- (n1);
    \draw[dashed,-<-=0.5] (Sp) -- (SU);
    \draw[-<-=0.5] (Sp) -- (U1);
    \draw[->-=0.5] (SU) -- (U1);

    \draw[->-=0.5,bend right=30] (Sp) to (n1.135);
    \draw[->-=0.5,bend left=30,dashed] (U1) to (n1.45);
    
    \node[rotate=90] (arr3) at (4,0) {\scalebox{2}{$\Big\Downarrow $}}; 
    \node[yshift=1.5cm] () at (arr3) {\Large$\substack{ \text {Merge}
    \\
    \\ \mathrm{SU}(2)
    \\
    \\\text {and}
    \\
    \\ \mathrm{Sp}(n-3)}$};
    \end{scope}

    \begin{scope}[xshift=19cm]
    \coordinate (origin) at (0,0);

    \node[box] (U1)  at ($(origin)+(0,3.5)$) {$\mathrm{U}(1)$};
    \node[cir] (Sp) at (origin) {\normalsize$\mathrm{Sp}(n-1)$};
    \node[box]  (n1) at  ($(origin)-(0,3.5)$) {$n$};
    \draw[-<-=0.5,bend left=30,dashed] (U1.-45) to (n1.45);
    \draw[->-=0.5] (Sp) -- (n1);
    \draw[-<-=0.5] (Sp) -- (U1);
    \node[rotate=90] (arr3) at (3,0) {\scalebox{2}{$\Big\Updownarrow $}}; 
    \node[yshift=0.5cm] () at (arr3)   {Fig. \ref{fig:seiberg-Sp-2}};
    \node[yshift=1.8cm,xshift=1.5cm,scale=1.5] at (U1) {$\bigoplus \frac{n-1}2$};
    \end{scope}

    \begin{scope}[xshift=24cm]
    \coordinate (origin) at (0,0);
    \draw[line width=1.5pt] ($(origin)+(0,1.3)$) circle(0.7cm) node[yshift=1.1cm]{\Large AS};
    \node[box,fill=white]  (n) at  (origin) {$n$};
    %\node[box] (U1) at (3.5,0) {$\mathrm{U}(1)$}  node[xshift=4.8cm,yshift=-0.5cm] {\LARGE$\mathsf{y}$};
    %\draw[-<-=0.5] (n) -- (U1);

    \node[scale=1.5] (det) at (0,-2) {$+{\Psi}$};
        \node[yshift=3.8cm,xshift=1.5cm,scale=1.5] at (origin) {$\bigoplus \frac{n-1}2$};
    \end{scope}
\end{tikzpicture}

%% file: 2d-duality.bbl
\providecommand{\href}[2]{#2}\begingroup\raggedright\begin{thebibliography}{10}

\bibitem{Witten:1993yc}
E.~Witten, \emph{{Phases of $N=2$ theories in two-dimensions}},
  \href{https://doi.org/10.1016/0550-3213(93)90033-L}{\emph{Nucl.Phys.}
  {\bfseries B403} (1993) 159}
  [\href{https://arxiv.org/abs/hep-th/9301042}{{\ttfamily hep-th/9301042}}].

\bibitem{Hori:2006dk}
K.~Hori and D.~Tong, \emph{{Aspects of Non-Abelian Gauge Dynamics in
  Two-Dimensional N=(2,2) Theories}},
  \href{https://doi.org/10.1088/1126-6708/2007/05/079}{\emph{JHEP} {\bfseries
  05} (2007) 079} [\href{https://arxiv.org/abs/hep-th/0609032}{{\ttfamily
  hep-th/0609032}}].

\bibitem{Hori:2011pd}
K.~Hori, \emph{{Duality In Two-Dimensional (2,2) Supersymmetric Non-Abelian
  Gauge Theories}}, \href{https://doi.org/10.1007/JHEP10(2013)121}{\emph{JHEP}
  {\bfseries 10} (2013) 121} [\href{https://arxiv.org/abs/1104.2853}{{\ttfamily
  1104.2853}}].

\bibitem{Benini:2016qnm}
F.~Benini and B.~Le~Floch, \emph{{Supersymmetric localization in two
  dimensions}}, \href{https://doi.org/10.1088/1751-8121/aa77bb}{\emph{J. Phys.
  A} {\bfseries 50} (2017) 443003}
  [\href{https://arxiv.org/abs/1608.02955}{{\ttfamily 1608.02955}}].

\bibitem{Rennemo:2016oiu}
J.V.~Rennemo and E.~Segal, \emph{{Hori-mological projective duality}},
  \href{https://doi.org/10.1215/00127094-2019-0014}{\emph{Duke Math. J.}
  {\bfseries 168} (2019) 2127}
  [\href{https://arxiv.org/abs/1609.04045}{{\ttfamily 1609.04045}}].

\bibitem{Gadde:2013lxa}
A.~Gadde, S.~Gukov and P.~Putrov, \emph{{(0, 2) trialities}},
  \href{https://doi.org/10.1007/JHEP03(2014)076}{\emph{JHEP} {\bfseries 03}
  (2014) 076} [\href{https://arxiv.org/abs/1310.0818}{{\ttfamily 1310.0818}}].

\bibitem{Nawata:2023aoq}
S.~Nawata, Y.~Pan and J.~Zheng, \emph{{Class S theories on $S^2$}},
  \href{https://doi.org/10.1103/PhysRevD.109.105015}{\emph{Phys. Rev. D}
  {\bfseries 109} (2024) 105015}
  [\href{https://arxiv.org/abs/2310.07965}{{\ttfamily 2310.07965}}].

\bibitem{Gaiotto:2009we}
D.~Gaiotto, \emph{{$N=2$ dualities}},
  \href{https://doi.org/10.1007/JHEP08(2012)034}{\emph{JHEP} {\bfseries 1208}
  (2012) 034} [\href{https://arxiv.org/abs/0904.2715}{{\ttfamily 0904.2715}}].

\bibitem{Gadde:2015wta}
A.~Gadde, S.S.~Razamat and B.~Willett, \emph{{On the reduction of 4d $
  \mathcal{N}=1 $ theories on $ {\mathbb{S}}^2 $}},
  \href{https://doi.org/10.1007/JHEP11(2015)163}{\emph{JHEP} {\bfseries 11}
  (2015) 163} [\href{https://arxiv.org/abs/1506.08795}{{\ttfamily
  1506.08795}}].

\bibitem{Closset:2013sxa}
C.~Closset and I.~Shamir, \emph{{The $\mathcal{N}=1$ Chiral Multiplet on
  $T^2\times S^2$ and Supersymmetric Localization}},
  \href{https://doi.org/10.1007/JHEP03(2014)040}{\emph{JHEP} {\bfseries 03}
  (2014) 040} [\href{https://arxiv.org/abs/1311.2430}{{\ttfamily 1311.2430}}].

\bibitem{Putrov:2015jpa}
P.~Putrov, J.~Song and W.~Yan, \emph{{(0,4) dualities}},
  \href{https://doi.org/10.1007/JHEP03(2016)185}{\emph{JHEP} {\bfseries 03}
  (2016) 185} [\href{https://arxiv.org/abs/1505.07110}{{\ttfamily
  1505.07110}}].

\bibitem{Gomis:2012wy}
J.~Gomis and S.~Lee, \emph{{Exact Kahler Potential from Gauge Theory and Mirror
  Symmetry}}, \href{https://doi.org/10.1007/JHEP04(2013)019}{\emph{JHEP}
  {\bfseries 1304} (2013) 019}
  [\href{https://arxiv.org/abs/1210.6022}{{\ttfamily 1210.6022}}].

\bibitem{Hori:2000kt}
K.~Hori and C.~Vafa, \emph{{Mirror symmetry}},
  \href{https://arxiv.org/abs/hep-th/0002222}{{\ttfamily hep-th/0002222}}.

\bibitem{Benini:2012ui}
F.~Benini and S.~Cremonesi, \emph{{Partition Functions of ${\mathcal{N}=(2,2)}$
  Gauge Theories on S$^{2}$ and Vortices}},
  \href{https://doi.org/10.1007/s00220-014-2112-z}{\emph{Commun. Math. Phys.}
  {\bfseries 334} (2015) 1483}
  [\href{https://arxiv.org/abs/1206.2356}{{\ttfamily 1206.2356}}].

\bibitem{Doroud:2012xw}
N.~Doroud, J.~Gomis, B.~Le~Floch and S.~Lee, \emph{{Exact Results in D=2
  Supersymmetric Gauge Theories}},
  \href{https://doi.org/10.1007/JHEP05(2013)093}{\emph{JHEP} {\bfseries 1305}
  (2013) 093} [\href{https://arxiv.org/abs/1206.2606}{{\ttfamily 1206.2606}}].

\bibitem{Dedushenko:2017osi}
M.~Dedushenko and S.~Gukov, \emph{{IR duality in 2D $N=(0,2)$ gauge theory with
  noncompact dynamics}},
  \href{https://doi.org/10.1103/PhysRevD.99.066005}{\emph{Phys. Rev. D}
  {\bfseries 99} (2019) 066005}
  [\href{https://arxiv.org/abs/1712.07659}{{\ttfamily 1712.07659}}].

\bibitem{Sacchi:2020pet}
M.~Sacchi, \emph{{New 2d $ \mathcal{N} $ = (0, 2) dualities from four
  dimensions}}, \href{https://doi.org/10.1007/JHEP12(2020)009}{\emph{JHEP}
  {\bfseries 12} (2020) 009}
  [\href{https://arxiv.org/abs/2004.13672}{{\ttfamily 2004.13672}}].

\bibitem{Seiberg:1994pq}
N.~Seiberg, \emph{{Electric-magnetic duality in supersymmetric nonAbelian gauge
  theories}},
  \href{https://doi.org/10.1016/0550-3213(94)00023-8}{\emph{Nucl.Phys.}
  {\bfseries B435} (1995) 129}
  [\href{https://arxiv.org/abs/hep-th/9411149}{{\ttfamily hep-th/9411149}}].

\bibitem{Jia:2014ffa}
B.~Jia, E.~Sharpe and R.~Wu, \emph{{Notes on nonabelian (0,2) theories and
  dualities}}, \href{https://doi.org/10.1007/JHEP08(2014)017}{\emph{JHEP}
  {\bfseries 08} (2014) 017} [\href{https://arxiv.org/abs/1401.1511}{{\ttfamily
  1401.1511}}].

\bibitem{Gadde:2014ppa}
A.~Gadde, S.~Gukov and P.~Putrov, \emph{{Exact Solutions of 2d Supersymmetric
  Gauge Theories}}, \href{https://doi.org/10.1007/JHEP11(2019)174}{\emph{JHEP}
  {\bfseries 11} (2019) 174} [\href{https://arxiv.org/abs/1404.5314}{{\ttfamily
  1404.5314}}].

\bibitem{Benini:2012cz}
F.~Benini and N.~Bobev, \emph{{Exact two-dimensional superconformal R-symmetry
  and c-extremization}},
  \href{https://doi.org/10.1103/PhysRevLett.110.061601}{\emph{Phys. Rev. Lett.}
  {\bfseries 110} (2013) 061601}
  [\href{https://arxiv.org/abs/1211.4030}{{\ttfamily 1211.4030}}].

\bibitem{Benini:2013cda}
F.~Benini and N.~Bobev, \emph{{Two-dimensional SCFTs from wrapped branes and
  c-extremization}}, \href{https://doi.org/10.1007/JHEP06(2013)005}{\emph{JHEP}
  {\bfseries 06} (2013) 005} [\href{https://arxiv.org/abs/1302.4451}{{\ttfamily
  1302.4451}}].

\bibitem{Intriligator_1995}
K.~Intriligator and P.~Pouliot, \emph{Exact superpotentials, quantum vacua and
  duality in supersymmetric {SP}(nc) gauge theories},
  \href{https://doi.org/10.1016/0370-2693(95)00618-u}{\emph{Physics Letters B}
  {\bfseries 353} (1995) 471}.

\bibitem{Intriligator:1995id}
K.A.~Intriligator and N.~Seiberg, \emph{{Duality, monopoles, dyons, confinement
  and oblique confinement in supersymmetric SO(N(c)) gauge theories}},
  \href{https://doi.org/10.1016/0550-3213(95)00159-P}{\emph{Nucl. Phys. B}
  {\bfseries 444} (1995) 125}
  [\href{https://arxiv.org/abs/hep-th/9503179}{{\ttfamily hep-th/9503179}}].

\bibitem{Kim:2024vci}
S.-S.~Kim, X.~Li, S.~Nawata and F.~Yagi, \emph{{Freezing and BPS jumping}},
  \href{https://doi.org/10.1007/JHEP05(2024)340}{\emph{JHEP} {\bfseries 05}
  (2024) 340} [\href{https://arxiv.org/abs/2403.12525}{{\ttfamily
  2403.12525}}].

\bibitem{Amariti:2024usp}
A.~Amariti, P.~Glorioso, F.~Mantegazza, D.~Morgante and A.~Zanetti,
  \emph{{Dualities from dualities in 2d $\mathcal{N}=(0,2)$}},
  \href{https://arxiv.org/abs/2410.12453}{{\ttfamily 2410.12453}}.

\bibitem{Aharony:2016jki}
O.~Aharony, S.S.~Razamat, N.~Seiberg and B.~Willett, \emph{{The long flow to
  freedom}}, \href{https://doi.org/10.1007/JHEP02(2017)056}{\emph{JHEP}
  {\bfseries 02} (2017) 056}
  [\href{https://arxiv.org/abs/1611.02763}{{\ttfamily 1611.02763}}].

\bibitem{Bonetti:2018fqz}
F.~Bonetti, C.~Meneghelli and L.~Rastelli, \emph{{VOAs labelled by complex
  reflection groups and 4d SCFTs}},
  \href{https://doi.org/10.1007/JHEP05(2019)155}{\emph{JHEP} {\bfseries 05}
  (2019) 155} [\href{https://arxiv.org/abs/1810.03612}{{\ttfamily
  1810.03612}}].

\bibitem{Pan:2021ulr}
Y.~Pan, Y.~Wang and H.~Zheng, \emph{{Defects, modular differential equations,
  and free field realization of N=4 vertex operator algebras}},
  \href{https://doi.org/10.1103/PhysRevD.105.085005}{\emph{Phys. Rev. D}
  {\bfseries 105} (2022) 085005}
  [\href{https://arxiv.org/abs/2104.12180}{{\ttfamily 2104.12180}}].

\bibitem{Beem:2013sza}
C.~Beem, M.~Lemos, P.~Liendo, W.~Peelaers, L.~Rastelli and B.C.~van Rees,
  \emph{{Infinite Chiral Symmetry in Four Dimensions}},
  \href{https://doi.org/10.1007/s00220-014-2272-x}{\emph{Commun. Math. Phys.}
  {\bfseries 336} (2015) 1359}
  [\href{https://arxiv.org/abs/1312.5344}{{\ttfamily 1312.5344}}].

\bibitem{Eager:2019zrc}
R.~Eager, G.~Lockhart and E.~Sharpe, \emph{{Hidden exceptional symmetry in the
  pure spinor superstring}},
  \href{https://doi.org/10.1103/PhysRevD.101.026006}{\emph{Phys. Rev. D}
  {\bfseries 101} (2020) 026006}
  [\href{https://arxiv.org/abs/1902.09504}{{\ttfamily 1902.09504}}].

\bibitem{Mohri:1997ef}
K.~Mohri, \emph{{D-branes and quotient singularities of Calabi-Yau fourfolds}},
  \href{https://doi.org/10.1016/S0550-3213(98)00085-6}{\emph{Nucl. Phys. B}
  {\bfseries 521} (1998) 161}
  [\href{https://arxiv.org/abs/hep-th/9707012}{{\ttfamily hep-th/9707012}}].

\bibitem{Garcia-Compean:1998sla}
H.~Garcia-Compean and A.M.~Uranga, \emph{{Brane box realization of chiral gauge
  theories in two-dimensions}},
  \href{https://doi.org/10.1016/S0550-3213(98)00725-1}{\emph{Nucl. Phys. B}
  {\bfseries 539} (1999) 329}
  [\href{https://arxiv.org/abs/hep-th/9806177}{{\ttfamily hep-th/9806177}}].

\bibitem{Franco:2015tna}
S.~Franco, D.~Ghim, S.~Lee, R.-K.~Seong and D.~Yokoyama, \emph{{2d (0,2) Quiver
  Gauge Theories and D-Branes}},
  \href{https://doi.org/10.1007/JHEP09(2015)072}{\emph{JHEP} {\bfseries 09}
  (2015) 072} [\href{https://arxiv.org/abs/1506.03818}{{\ttfamily
  1506.03818}}].

\bibitem{Franco:2015tya}
S.~Franco, S.~Lee and R.-K.~Seong, \emph{{Brane Brick Models, Toric Calabi-Yau
  4-Folds and 2d (0,2) Quivers}},
  \href{https://doi.org/10.1007/JHEP02(2016)047}{\emph{JHEP} {\bfseries 02}
  (2016) 047} [\href{https://arxiv.org/abs/1510.01744}{{\ttfamily
  1510.01744}}].

\bibitem{Benini:2014mia}
F.~Benini, D.S.~Park and P.~Zhao, \emph{{Cluster Algebras from Dualities of 2d
  ${\mathcal{N}}$ = (2, 2) Quiver Gauge Theories}},
  \href{https://doi.org/10.1007/s00220-015-2452-3}{\emph{Commun. Math. Phys.}
  {\bfseries 340} (2015) 47} [\href{https://arxiv.org/abs/1406.2699}{{\ttfamily
  1406.2699}}].

\bibitem{Yagi:2015lha}
J.~Yagi, \emph{{Quiver gauge theories and integrable lattice models}},
  \href{https://doi.org/10.1007/JHEP10(2015)065}{\emph{JHEP} {\bfseries 10}
  (2015) 065} [\href{https://arxiv.org/abs/1504.04055}{{\ttfamily
  1504.04055}}].

\bibitem{de-la-Cruz-Moreno:2020xop}
J.~de-la Cruz-Moreno and H.~Garc\'\i{}a-Compe\'an, \emph{{Star-triangle type
  relations from $2d$ $\mathcal{N}=(0,2)$ $USp(2N)$ dualities}},
  \href{https://doi.org/10.1007/JHEP01(2021)023}{\emph{JHEP} {\bfseries 01}
  (2021) 023} [\href{https://arxiv.org/abs/2008.02419}{{\ttfamily
  2008.02419}}].

\bibitem{Benini:2013nda}
F.~Benini, R.~Eager, K.~Hori and Y.~Tachikawa, \emph{{Elliptic genera of
  two-dimensional N=2 gauge theories with rank-one gauge groups}},
  \href{https://doi.org/10.1007/s11005-013-0673-y}{\emph{Lett. Math. Phys.}
  {\bfseries 104} (2014) 465}
  [\href{https://arxiv.org/abs/1305.0533}{{\ttfamily 1305.0533}}].

\bibitem{Benini:2013xpa}
F.~Benini, R.~Eager, K.~Hori and Y.~Tachikawa, \emph{{Elliptic Genera of 2d
  ${\mathcal{N}}$ = 2 Gauge Theories}},
  \href{https://doi.org/10.1007/s00220-014-2210-y}{\emph{Commun. Math. Phys.}
  {\bfseries 333} (2015) 1241}
  [\href{https://arxiv.org/abs/1308.4896}{{\ttfamily 1308.4896}}].

\bibitem{jeffrey1995localization}
L.C.~Jeffrey and F.C.~Kirwan, \emph{Localization for nonabelian group actions},
  {\emph{Topology} {\bfseries 34} (1995) 291}
  [\href{https://arxiv.org/abs/alg-geom/9307001}{{\ttfamily
  alg-geom/9307001}}].

\bibitem{brion1999arrangement}
M.~Brion and M.~Vergne, \emph{{Arrangement of hyperplanes. I. Rational
  functions and Jeffrey-Kirwan residue}}, {\emph{Annales scientifiques de
  l'Ecole normale sup{\'e}rieure} {\bfseries 32} (1999) 715}
  [\href{https://arxiv.org/abs/math/9903178}{{\ttfamily math/9903178}}].

\bibitem{szenes2003toric}
A.~Szenes and M.~Vergne, \emph{{Toric reduction and a conjecture of Batyrev and
  Materov}}, {\emph{Invent. Math.} {\bfseries 158} (2004) 453–495}
  [\href{https://arxiv.org/abs/math/0306311}{{\ttfamily math/0306311}}].

\end{thebibliography}\endgroup
